\documentclass[fleqn,usenatbib]{mnras}

\usepackage{newtxtext,newtxmath}
\usepackage[T1]{fontenc}

\DeclareRobustCommand{\VAN}[3]{#2}
\let\VANthebibliography\thebibliography
\def\thebibliography{\DeclareRobustCommand{\VAN}[3]{##3}\VANthebibliography}

\usepackage{graphicx}	% Including figure files
\usepackage{amsmath}	% Advanced maths commands
\usepackage{hyperref}

\newcommand{\plato}{{\it PLATO}}
\newcommand{\HWO}{{\it HWO}}

\newcommand{\harpsn}{HARPS-N}

\newcommand{\logrhk}{$\log\,\mathrm{R}^\prime_\mathrm{HK}$}

\newcommand{\phoenix}{PHOENIX}
\newcommand{\muram}{MURaM}

\newcommand{\chisq}{$\chi^2$}

\title[Fe\,{\sc i} 4377 \AA\ Line as a Solar Faculae Indicator]{The Fe\,{\sc i} 4377 \AA\ Line as a Solar Faculae Indicator: Insights from Spectral Ratio Analysis}

\author[K. L. Hobbs et al.]{\parbox{\textwidth}{\Large
Katlyn L. Hobbs,$^{1}$
Christopher A. Watson,$^{1}$
Jean C. Costes,$^{1}$
Yvonne Unruh,$^{2}$
Dana Clarice Yaptangco,$^{2}$
Krishnamurthy Sowmya,$^{3}$
Mitchell E. Young,$^{1}$
Ernst J. W. de Mooij,$^{1}$
Alexander G. M. Pietrow,$^{4}$
\\
Pál Váradi Nagy,$^{5}$
Alexander I. Shapiro,$^{3,6}$
Veronika Witzke,$^{3}$
Federica Rescigno,$^{7}$
Ryan A. Rubenzahl,$^{8}$
Megan Bedell,$^{8}$
Andrew Collier Cameron,$^{9}$
Xavier Dumusque,$^{10}$
Sean M. O'Brien,$^{1}$
\\
Benjamin M. J. Cadell,$^{1}$
Baptiste Klein,$^{11}$
Niamh Mallaghan,$^{1}$
Niamh K. O'Sullivan,$^{11}$
Toby Rodel$^{1}$
\\
and
Sara Tavella$^{10}$
}
\vspace{0.3cm}
\\
$^{1}$Queen's University Belfast, Belfast, University Road, BT7 1NN, UK\\
$^{2}$Department of Physics, Imperial College London, UK\\
$^{3}$University of Graz, Institute of Physics, Universitätsplatz 5, 8010 Graz, Austria\\
$^{4}$Leibniz-Institut für Astrophysik Potsdam (AIP), An der Sternwarte 16, 14482 Potsdam, Germany\\
$^{5}$Independent Researcher, Cluj-Napoca, Romania\\
$^{6}$Max-Planck-Institut für Sonnensystemforschung, Justus-von-Liebig-Weg 3, 37077 Göttingen, Germany\\
$^{7}$School of Physics \& Astronomy, University of Birmingham, Edgbaston, Birmingham B15 2TT, UK\\
$^{8}$Center for Computational Astrophysics, Flatiron Institute, Simons Foundation, 162 Fifth Avenue, New York, NY 10010, USA\\
$^{9}$Centre for Exoplanet Science, SUPA School of Physics and Astronomy, University of St Andrews, North Haugh, St Andrews KY16 9SS, UK\\
$^{10}$Astronomy Department of the University of Geneva, 51 ch. des Maillettes, 1290 Versoix, Switzerland\\
$^{11}$Denys Wilkinson Building, Department of Physics, University of Oxford, OX1 3RH, UK
}

\date{Accepted XXX. Received YYY; in original form ZZZ}

\pubyear{\the\year{}}

\begin{document}
\label{firstpage}
\pagerange{\pageref{firstpage}--\pageref{lastpage}}
\maketitle{}

% Abstract of the paper
\begin{abstract}
Faculae are a dominant source of stellar activity noise in radial velocity measurements, yet their low contrast and broad surface distribution make them difficult to track in disc-integrated observations. We apply Spectral Ratio Analysis (SRA) to HARPS-N Sun-as-a-star observations to isolate and characterize the spectral imprint of facular regions over rotational timescales. The resulting SRA spectra show coherent, line-dependent variability sensitive to surface magnetic activity, with the \mbox{Fe\,{\sc i} 4377 \AA} line exhibiting a particularly strong diagnostic response to facular coverage. We interpret the observed signatures using two complementary synthetic frameworks: composite PHOENIX spectra, from which we derive best-fit facular temperature contrasts in the range 200–400 K, and MPS-ATLAS spectra synthesized using MURaM simulations of the quiet Sun including a small-scale dynamo and magnetically-enhanced facular analogues with initial mean vertical magnetic fields of 100G, 200G, and 300G. Both approaches are benchmarked against facular filling factors measured from Solar Dynamics Observatory (SDO) disc-resolved images. We find good agreement between SDO-measured and SRA-inferred filling factors using the \mbox{Fe\,{\sc i} 4377 \AA} line, with Pearson R coefficients of 0.587–0.927 across models and timescales. The  estimated filling factors track the solar activity cycle, rising from $\sim$ 1.5 \%\ at lower activity to $\sim$ 5.5 \% at higher activity, consistent with SDO-measured filling factors. These results demonstrate that SRA offers a means to reliably track surface magnetic activity in disc-resolved spectra, which is necessary for mitigating the effects of activity on RV characterization of exoplanet masses and atmospheres at modern precision.

% We caution that degeneracies between temperature, magnetic field strength, and filling factor limit the absolute accuracy of these parameters. 

% We find good agreement between SDO-measured filling factors and those inferred from modelling SRA residuals. Our results demonstrate sensitivity to modest changes in plage coverage even during low solar activity.

% \muram\ simulations incorporating quiet-Sun small-scale dynamo models alongside magnetically enhanced ($100\,\mathrm{G} \leq B \leq 300\,\mathrm{G}$) faculae analogues. 

%Faculae exhibit low contrast with respect to the surrounding photosphere and are uniformly distributed across the stellar surface, making them difficult to detect and quantify observationally. As a result, their contribution to disc-integrated spectral variability remains poorly constrained, posing a fundamental challenge for disentangling stellar activity from true planetary signals.

\end{abstract}

% Select between one and six entries from the list of approved keywords.
% Don't make up new ones.

%%%%%%%%%%%%%%%%%%%%%%%%%%%%%%%%%%%%%%%%%%%%%%%%%%

%%%%%%%%%%%%%%%%% BODY OF PAPER %%%%%%%%%%%%%%%%%%

\section{Introduction}\label{intro}
The discovery of Earth-sized exoplanets orbiting at $\sim$1 AU around Sun-like stars remains one of the key goals of the exoplanet community. Although such planets have yet to be found, they will provide critical benchmarks for understanding the processes of planet formation, the diversity of planetary systems, and the conditions required for potentially habitable environments. Multiple surveys are targeting long-period, low-mass exoplanets orbiting bright nearby stars, including: the High Accuracy Radial velocity Planet Searcher for the Northern hemisphere (HARPS-N) Rocky Planet Search \citep[][]{Cosentino2012, Motalebi2015}; the Echelle SPectrograph for Rocky Exoplanets and Stable Spectroscopic Observations (ESPRESSO) blind radial velocity survey \citep[][]{Hojjatpanah2019,Pepe2021}; the EXtreme PREcision Spectrograph (EXPRES) 100 Earths survey \citep[][]{Jurgenson2016, Brewer2020, Brewer2022}; and the NN-EXPLORE Exoplanet Investigations with Doppler spectroscopy (NEID) Earth twin survey \citep[][]{Schwab2016, Gupta2021, Gupta2026}. Future efforts such as the Terra Hunting Experiment \citep{Thompson2016} and the \textit{PLAnetary Transits and Oscillations of stars} \citep[\plato;][]{Rauer2014, Rauer2025} will further expand this search. The exoplanet discoveries that emerge from these surveys will provide key targets for further characterization with future direct imaging missions such as the \textit{Habitable Worlds Observatory} \citep[\HWO;][]{Hartman2018}.

Among the primary exoplanet detection techniques currently available, the radial velocity (RV) method is fundamental to the characterization of Earth-analogues. While the transit method is valuable for measuring the radii of planetary bodies, ground-based RV follow-up is typically required to measure the mass of any detected planetary candidates.
Achieving such mass measurements is particularly challenging for Earth-analogues since they induce extremely small Doppler signals on their host stars (for context, Earth induces a $\sim$10 cm s$^{-1}$ RV semi-amplitude reflex motion on the Sun). Current state-of-the-art instruments can, in principle, reach the instrumental precision required to detect such signals \citep[see][for review]{Fischer2016, Crass2021, Burt2025}, depending on survey design and a sufficient number of observations \citep[see][]{Langellier2021, Luhn2023, Gupta2023}. Nevertheless, low-mass exoplanets with long ($\sim$ 1 year) orbital periods remain largely undetected, primarily because stellar variability induces RV signals that are typically an order of magnitude larger than those generated by such planets. This stellar variability is primarily driven by surface convection, magnetic activity, and their interactions. Active regions (such as spots \citep{Huerta2008_starspotref}, plages or faculae \citep{Meunier2010}), granulation and supergranulation \citep{Lakeland2024, OSullivan2025}, oscillations \citep{Luhn2024_oscillations}, and flares \citep{Pietrow2024} distort spectral line profiles, producing apparent RV shifts that can mask or mimic planetary signals. These stochastic variations can be of the order of 0.5 -- 10\,m\,s$^{-1}$ or more, occurring on timescales ranging from minutes (e.g., oscillations) to decades (e.g., magnetic cycles). In some cases, this has led to spurious or disputed planet detections (e.g. Barnard's star, \citealp{Ribas2018, Lubin2022}, Kapetyn's star, \citealp{Anglada2014, Robertson2015, Ji2019}, and HD\,41248, \citealp{Jenkins2013, Santos2014, Jenkins2014, Feng2017b, Faria2020}). Resolving small planetary signals thus requires disentangling these diverse sources of stellar variability signals across both amplitude and time.

In particular, plages and faculae are large impediments to the detection of Earth-analogues. These features are bright, magnetically active regions that form when strong magnetic fields suppress the underlying convection in the stellar chromosphere and photosphere, thereby suppressing the overall convective blueshift \citep{Meunier2010, Lakeland2024}. This is the dominant RV effect in stars exhibiting solar-like activity levels \citep{Haywood2016, Milbourne2019, Meunier2024}. 
Although related, faculae and plage refer to magnetic structures in different layers of the solar atmosphere, with faculae referring to bright, magnetically active regions in the photosphere, while plage refers to their chromospheric counterparts \citep{Cretignier2024}. Since RV measurements are predominantly derived from photospheric absorption lines, they are primarily sensitive to the photospheric manifestation of magnetic activity \citep{Dravins1981}. While literature often uses ``plage'' and ``faculae'' interchangeably, we adopt the term faculae henceforth. 
On the Sun, faculae-driven variability drives RV amplitudes over rotational timescales of $\sim$0.4 -- 1 m s$^{-1}$ depending on the activity level, while net changes in facular filling factor across the solar cycle create variations of up to $\sim$8 -- 10 m s$^{-1}$ \citep{Meunier2010, Cretignier2024}.
The facular filling factor can reach up to $\sim$10\% at solar maximum, while their relatively diffuse distributions across the stellar surface, and lower contrast with the surrounding photosphere compared to spots make these regions difficult to track in disc-integrated measurements. Furthermore, the lifetime of faculae is on the order of weeks to months, making them longer living than other variability features (i.e. spots with lifetimes of days to weeks), and often persist across multiple rotations of the Sun \citep{Cameron2019, Kontogiannis2025, Finley2025}.
Thus, these bright, magnetically active regions represent a major barrier to the detection of low-amplitude planetary signals.

Stellar faculae also complicate the characterization of exoplanet atmospheres. Transmission spectroscopy (which observes starlight passing through a planet's atmosphere during transit) is sensitive to such stellar features. \citet{Oshagh2014} demonstrated that occultations of stellar faculae during transit can bias the inferred planet-to-star radius ratio, producing overestimates of up to 10\% at short wavelengths in M-dwarf systems. Similarly, \citet{Rackham2018} showed that unocculted faculae regions introduce wavelength-dependent contamination in transmission spectra, affecting inferred planetary radii and atmospheric characterization. These effects of stellar contamination highlight the need to understand facular regions and their effects on stellar spectra, both for RV measurement and transmission spectroscopy. 

A number of techniques and observing strategies have been developed to mitigate the effect of stellar noise on RV signals, however, a unified way of addressing the stellar activity problem has not been identified, and the individual contributions of particular active regions remain poorly understood \citep[e.g.,][]{Zhao2022}. This underscores the need to develop a deeper understanding of how stellar activity imprints on the observed spectra, and motivates the development of more sensitive and physically-motivated activity indicators. Recent work has also shown that commonly used solar reference spectra are contaminated with activity, highlighting the need to understand activity indicators and develop truly quiet variability references \citep{Hanassi-Savari2025}. Particularly, indicators that probe photospheric changes, where spectral lines that are used for RV extraction are typically formed, may prove more effective than traditional proxies such as \logrhk\ %Ca\,\textsc{ii} H\&K lines (\logrhk) 
that instead track chromospheric activity.
%\citep{Cretignier2024}.

Ultimately, these novel indicators can be used in tandem with existing diagnostics that measure the line-profile asymmetries caused by stellar activity. These include, for example, the $\Delta \alpha B^2$ indicator, a proxy for the differential unsigned magnetic flux that is measured directly from the profiles of absorption lines \citep{Lienhard2023}, or the Full Width at Half-Maximum \citep[FWHM;][]{Boisse2011} and the Bisector Inverse Slope \citep[BIS;][]{Queloz2001} that are derived from the shape of the cross-correlation function \citep[CCF;][]{Baranne1996CCF,Pepe2002CCF} used to measure RVs. These can be integrated into advanced
mitigation techniques \citep[e.g.,][]{Rajpaul2015, John2021, Klein2024,
Liang2024, Gilbertson2024}, improving both the interpretation of RV
modelling and the effectiveness of noise reduction, while providing
deeper insights into the stellar processes driving the observed RV
variability.

To further understand the spectral impact of activity and develop new indicators, \citet{Thompson2017} introduced an approach using ratios of high- and low-activity spectra to examine how the spectra of $\alpha$~Cen~B varied between magnetically quiet and active states. Following \citet{Costes2026}, we refer to this approach as Spectral Ratio Analysis (SRA). Using spectra taken with the High Accuracy Radial Velocity Planet Searcher \citep[HARPS:][] {Pepe2000}, \citet{Thompson2017} uncovered hundreds of residual spectral features within the 4300 -- 5300 \AA\ range with flux variability that correlated with \logrhk. Additionally, many of these features showed velocity shifts that were modulated by the rotation period of the star and by the stellar activity levels.
To understand how particular surface activity features correlated with these spectral variations, \citet{Thompson2020magneticactivityharpsnpaper2} used HARPS-N disc-integrated Solar spectra and found that many of these spectral signatures correlated with facular filling factors derived from disc-resolved images from the
Helioseismic and Magnetic Imager \citep[HMI;][]{Scherrer2012} instrument on the Solar Dynamics Observatory \citep[SDO;][]{Pesnell2012}.
They further showed that these spectral features varied in amplitude with the solar rotation period, and in phase with a significant faculae region visible in SDO images. 

In this study, we build on the known correlation between facular filling factor and the strength of SRA residual features established by \citet{Thompson2017, Thompson2020magneticactivityharpsnpaper2} and \citet{Costes2026} to develop a framework that can infer facular areal coverage directly from high-resolution spectra. We focus on the \ion{Fe}{I} 4377~\AA\ line, which exhibits the most pronounced response to facular coverage and thus provides the strongest diagnostic leverage \citep{Thompson2020magneticactivityharpsnpaper2}. In brief, we compute model SRA residuals using synthetic spectra from \phoenix\ \citep{Husser2013} and \muram\ \citep{Vogler2005}, and determine the best-fitting models to HARPS-N SRA solar observations to characterize faculae-driven spectral variability. Crucially, this approach enables facular filling factors to be inferred directly from spectral observations. This framework can thus be applied to high-resolution spectral observations of other stars.

In Section \ref{sec:Harpsn} we discuss the HARPS-N Solar Telescope observations used in this study. In Section~\ref{sec:feline} we motivate our choice of the Fe\,{\sc I} 4377 ~\AA~feature as a tracer of faculae further. In Section \ref{stellarmodels} we describe our process of using \phoenix\ and \muram\ models to fit the observations and estimate the areal coverage of facular regions; in Section \ref{comparisonswSDO}, we compare our estimations to faculae filling factor values measured from disc-resolved images taken by SDO. Finally, in Section~\ref{sec:conclusions} we outline our conclusions.

\section{HARPS-N Solar Telescope Observations}\label{sec:Harpsn}
% \subsection{Instruments}
The HARPS-N solar telescope \citep{Dumusque2015, Phillips2016} is a 3-inch guided telescope mounted on the exterior of the 3.58m Telescope Nationale Galileo (TNG) enclosure at the Roque de Los Muchachos observatory in La Palma, Spain. The instrument has been observing the Sun daily since 2015, with 5-minute integration times. It operates by feeding sunlight via an integrating sphere along an optical fibre into the HARPS-N spectrograph to obtain disc-integrated Sun-as-a-star observations. HARPS-N itself is a wavelength-stabilized echelle spectrograph with a long-term instrumental precision of $\sim$ 0.5 m/s, a wavelength coverage spanning 3830 \AA\ to 6900 \AA, and a resolution of 115,000 \citep{HARPSN}. We refer the reader to \citep{Dumusque2015} for more details about the solar telescope. 

\subsection{Selected Data}\label{sec:selected data}

\begin{figure*}
	% To include a figure from a file named example.*
	% Allowable file formats are eps or ps if compiling using latex
	% or pdf, png, jpg if compiling using pdflatex
	\includegraphics[width=\textwidth]{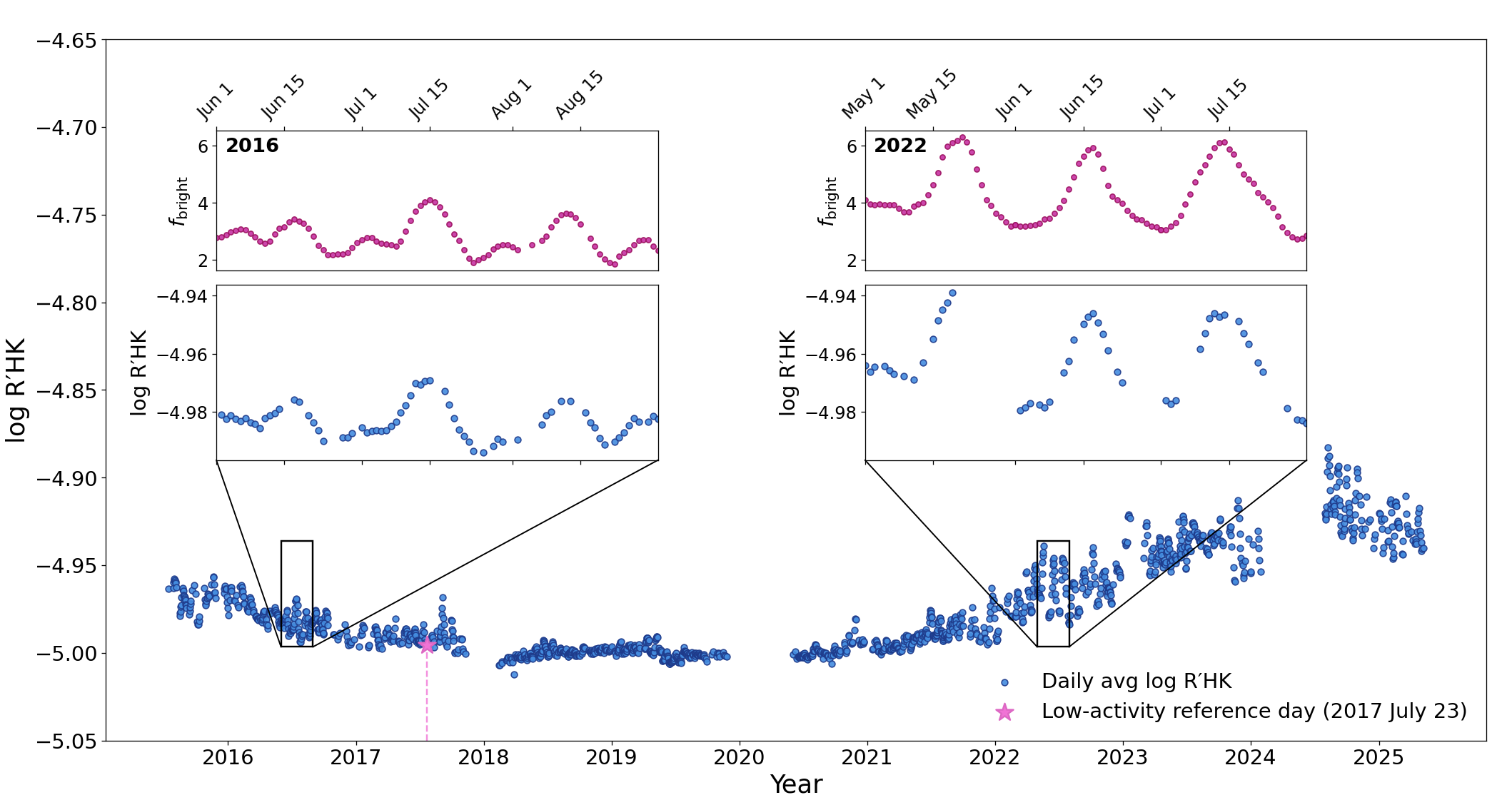}
    \caption{Time series of the disc-integrated solar chromospheric activity $\log R'_{\rm HK}$ index, measured from HARPS-N Sun-as-a-star observations between 2015 and 2025. Blue points show daily averaged values, while the pink star indicates the low-activity reference day used throughout this work. The long-term evolution traces the solar magnetic cycle, with a clear minimum around 2019–2020 and rising activity toward the subsequent maximum. The inset panels show variation in facular filling factor (measured using SDO, see Section~\ref{comparisonswSDO}) in pink in the top panel, and variation in \logrhk\ in the bottom panel. These insets highlight the shorter timescale variability over the June-August 2016 and May-July 2022 periods studied in this paper (corresponding to low- and high-activity phases, respectively) and show clear rotational modulation and active-region evolution in both cases. }
    \label{fig:logrhk}
\end{figure*}

All data used in this study were processed with the Data Reduction Software (DRS) version 3.2.0 \citep{Dumusque2026}. For this path-finding study, we focused on two time intervals, each covering approximately three solar rotations: 2016 June 01 through 2016 August 31, and 2022 May 01 through 2022 July 31. These intervals probe different activity levels during descending and ascending phases of the 11-year solar cycle, respectively, as seen in Figure \ref{fig:logrhk}. Rather than the quietest or most active available epochs, these data were chosen to better reflect the possible activity states when observing other stars and to test the sensitivity of our method away from the activity extremes.  Each interval was selected to span multiple rotations in order to capture the full lifecycle of active regions as they traverse the visible solar disc: emerging at the limb, transiting disc centre, and receding at the opposite limb. This allows us to track how the emergence, evolution, and disappearance of features such as faculae affect the spectral line profiles studied. Both intervals exhibit clear periodic signals in the \logrhk\ chromospheric activity index and faculae filling factor (measured using SDO, see Section~\ref{comparisonswSDO}) across multiple rotations (see Figure \ref{fig:logrhk}), as well as pronounced rotational modulation in sunspot coverage, with sunspot numbers varying significantly within individual rotation cycles\footnote{\url{https://lasp.colorado.edu/lisird/data/international_sunspot_number}} \citep{SILSO_Sunspot_Number}. Periods of large variation in activity across individual rotation cycles were chosen such that contrast could be seen in the features over time.
These two periods also allow us to test the method across an instrumental intervention. The 2016 data share the same instrumental configuration as our chosen low-activity template, whereas the 2022 data follow a 2018 instrumental change. Including the post-intervention period therefore tests whether the method can reliably recover active-region areal coverage despite changes to the instrument, providing a more stringent demonstration of its robustness.

We began with the full set of S2D (Science 2D, order-by-order extracted) solar spectra. We chose these over one-dimensional spectra to preserve the pixel sampling and instrumental line-spread function within each echelle order. We applied a series of quality cuts to ensure reliable and homogeneous measurements across all observations. \cite{Cameron2019} define a quality flag representing the probability that an observation is free from extinction-related contamination. Following the recommendations of this work, we only selected data with quality flags $\ge 0.99$. This removes spectra affected by partial obscuration of the solar disc (e.g., due to the passage of clouds, calima, high levels of differential extinction, or temporary tracking errors), which can distort the rotational line profile and introduce systematic errors in the measured radial velocities \citep{Cameron2019}. Our analysis is, however, performed directly on the spectra rather than on the derived radial velocities. While the quality flag ensures that extinction-related systematics are adequately corrected at the RV level, the spectra themselves are not corrected for these effects and may still retain wavelength-dependent distortions arising from differential extinction across the solar disc. To minimize atmospheric extinction gradients and differential refraction effects, we further restrict the sample to observations obtained at an airmass less than 1.5, corresponding to times when the Sun is sufficiently elevated above the horizon. 
Furthermore, large extinction correction values (${\rm RV}_{\rm ext}$) are indicative of unstable observing conditions and can imprint systematic trends in the derived radial velocities and residual spectra \citep{Cameron2019}, therefore we exclude spectra with radial velocity extinction corrections where $|{\rm RV}_{\rm ext}| > 10~\mathrm{cm\,s^{-1}}$.

Finally, the features seen in SRA (see Section \ref{sec:dataprocsra}) typically have amplitudes around $1\%$, thus we require high signal-to-noise, ensuring that photon noise is subdominant to activity-driven variability in the wavelength region of interest. Following \citet{Thompson2020magneticactivityharpsnpaper2}, we require a mean signal-to-noise ratio $\ge 100$ within echelle orders 40–49. We further increase signal-to-noise by combining daily spectra (see Section \ref{sec:dataprocsra} for details).

\subsection{Data Processing and SRA}\label{sec:dataprocsra}
For the analysis in this paper, we follow the data processing steps of \citet{Thompson2017, Thompson2020magneticactivityharpsnpaper2}, and \citet{Costes2026}. For clarity, and to highlight the differences in our analysis, we outline the steps here. 

Following the motivation outlined in Section \ref{sec:selected data}, we identify 2017 July 23 as a low-activity day from our filtered data set prior to the instrument interference in 2018, selected on the basis of its low $\log R'_{\rm HK}$ value (pink in Figure \ref{fig:logrhk}). As mentioned previously, we choose an intermediate activity level day as opposed to a Solar minimum reference to simulate differences in activity levels similar to those that might be observed in other stars. Additionally, images from the Helioseismic and Magnetic Imager (HMI) on SDO \citep{Schou2012, Pesnell2012, Couvidat2016} confirm a nearly spotless solar surface with minimal presence of faculae, making this date a close representation of an almost immaculate solar photosphere. All spectra from this day are used, and by following the recommended procedure on the \textit{Data \&\ Analysis Centre for Exoplanets} (DACE)\footnote{\href{https://dace.unige.ch/dashboard/}{https://dace.unige.ch/dashboard/}}, we began by dividing the flux per pixel by the pixel size to remove any remaining effects of dispersion from the continuum. While standard barycentric corrections applied by the DRS account for the motion of the observatory relative to the Solar System barycentre, Sun-as-a-star analyses require a heliocentric reference frame as the Solar System barycentre is offset from the centre of the Sun due to planetary gravitational effects. Thus, spectra were additionally adjusted using the barycentric to heliocentric correction from DACE. This adjustment shifts the wavelength grid of each spectrum. We took the spectrum with the highest signal-to-noise ratio (SNR) across echelle orders 41-50 for the day as our template wavelength grid and interpolated each heliocentric-corrected spectrum onto this grid using a cubic spline interpolation. Finally, we combined all the spectra passing our data quality cuts into a single low-activity template via a weighted average, where each spectrum was weighted by the inverse square of its mean SNR across echelle orders 41–50. When binning per day, the SNR is generally limited by the number of flat fields taken daily which are used to reduce the data; as a result, we obtained an averaged template spectrum with SNR $\sim$900.

High-activity days are treated analogously: spectra from each day were combined into weighted averages following the same procedure -- including interpolating heliocentric-corrected spectra to the same reference wavelength grid defined by the low activity day -- yielding spectra with typical signal-to-noise ratios of $\sim$600 -- 900. These averaged spectra were then compared to the low-activity template using SRA, constructed by dividing each high-activity spectrum by the low activity template. Large-scale continuum and blaze profile variations largely cancel out when taking the ratio of the spectra. We thus obtain a residual spectrum for each day, highlighting line profile changes relative to the low-activity reference as a function of activity level.

Similar to \citet{Thompson2020magneticactivityharpsnpaper2} and \citet{Costes2026}, we find a characteristic ripple pattern present in the residual spectra. These features appear across all spectral orders with varying amplitude and wavelength scales. Changing the low-activity template used to generate the residuals had no effect on the presence of these ripples. Similar patterns have been reported in other instruments, such as ESPRESSO, which exhibit wiggle-like structures that are likely caused by optical interference \citep{Tabernero2021}. In \citet{Thompson2020magneticactivityharpsnpaper2}, they postulate that etaloning is the suspected cause of this pattern.

To isolate and remove these ripples,  we applied a Generalized Lomb-Scargle periodogram to each residual spectrum to identify the dominant frequency of the ripple. However, we observe that the frequency of the ripple varies as a function of wavelength. In general, ripple frequencies tended to decrease at longer wavelengths. Rather than fitting multiple sine waves -- as done by \cite{Thompson2020magneticactivityharpsnpaper2} -- we fit a wavelength-scaled sinusoidal function whose frequency changes smoothly with wavelength, which more effectively suppressed the ripple patterns. The fitted function is defined as:
\begin{equation}
M_\mathrm{ripple}(\lambda) = A \cdot \sin\left(2\pi f \cdot (\lambda - T_0) \cdot \frac{\lambda_{\text{med}}}{\lambda} \right) + k
\end{equation}
where $M\mathrm{_{ripple}}\left(\lambda \right)$ is the model amplitude of the ripple at a given wavelength, $A$ is the amplitude, $f$ is the frequency identified from the Lomb-Scargle periodogram,  $T_0$ is the phase offset, $k$ is a constant vertical offset, $\lambda$ is the wavelength array, and $\lambda_{\text{med}}$ is the median wavelength of the spectrum \citep{Costes2026}. The fitted model was subsequently subtracted from each residual spectrum, and allowed the activity-induced features to be more visible. 

\begin{figure*}
	% To include a figure from a file named example.*
	% Allowable file formats are eps or ps if compiling using latex
	% or pdf, png, jpg if compiling using pdflatex
	\includegraphics[width=\textwidth]{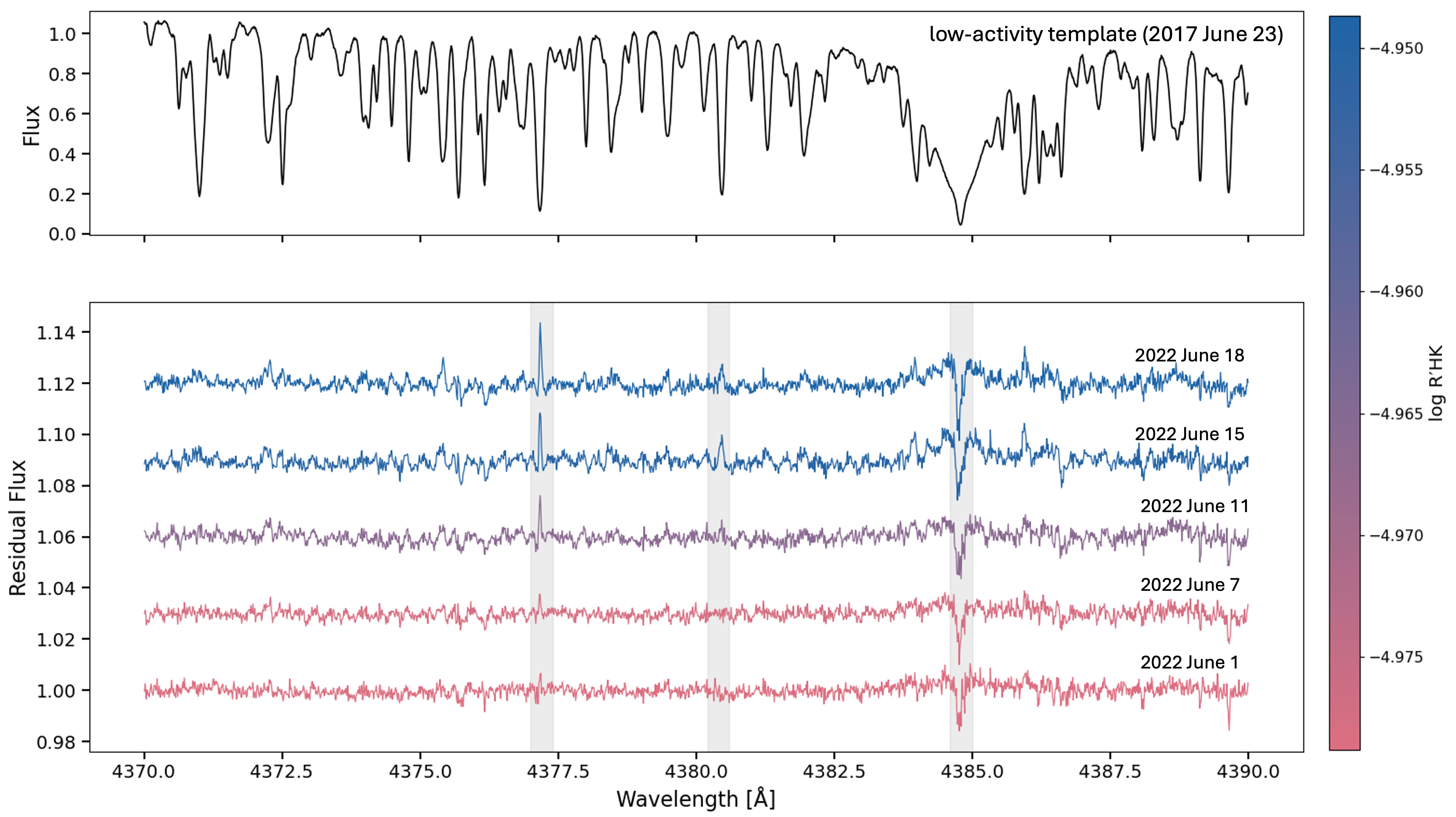}
    \caption{Top panel: observed low activity template Sun-as-a-star HARPS-N Solar spectrum, continuum normalized and shown over a select wavelength range for clarity of individual lines. Bottom: the ratio spectra across five different activity levels, over the same selected wavelength range. Various prominent features are highlighted here: The Fe\,{\sc i} lines at 4377 \AA\ and 4381 \AA\ change in amplitude, with the first showing dramatic differences across activity. Variation can also be seen in the 4381 \AA\  Fe\,{\sc i} line, where the change is primarily seen in the wings of the feature.}
    % \label{fig:2016bestfitex}
    \label{fig:features}
\end{figure*}

After applying the filters and processing steps outlined above, we obtain the final residual spectra used in our analysis. Examples of these reduced residual spectra are shown in Figure \ref{fig:features}, in which we reproduce the results from \citet{Thompson2020magneticactivityharpsnpaper2} in a small region of the solar spectrum as an example. The top panel shows the low-activity template spectrum as a reference for the line spectroscopic line profiles. The bottom panel of the figure shows five random selected daily SRA residual spectra with varying values of \logrhk, plotted in increasing \logrhk\ from bottom to top. Note that all spectra in this study are plotted in vacuum wavelengths. 

\section{The Fe\,{\sc I} 4377 ~\AA~feature as a tracer of faculae}\label{sec:feline}

In Figure \ref{fig:logrhk} we note that \logrhk\ is strongly correlated with the faculae filling-factor variations. However, \logrhk\ is unlikely to provide a reliable characterisation of facular features specifically, which are the primary drivers of RV variations (see, e.g., \citealt{Meunier2026}). Ca\,\textsc{ii} H\&K emission responds to changes in plage regions, the chromospheric counterparts of photospheric faculae (see Section \ref{intro}), rather than to the faculae themselves. Plage area coverages derived from Ca\,\textsc{ii} measurements are at least three times larger than photospheric facular coverages (see, e.g., \citealt{Chatzistergos2017}; \citealt{Sowmya2021}), because the flux tubes that form magnetic features expand with height in the atmosphere. Additionally, the retrieval of plage area coverages from Ca\,\textsc{ii} emission is affected by contamination from spots. As demonstrated by \citet{Sowmya2023}, spots also lead to an increase in Ca\,\textsc{ii} H\&K emission, similar to plages. The flux tubes forming spots expand with height, leading to the formation of superpenumbrae in the chromosphere, which appear brighter in the Ca\,\textsc{ii} line cores than the surrounding quiet regions. Consequently, the observed chromospheric Ca\,\textsc{ii} emission is a combination of contributions from both plages and spots, making it challenging to separate their individual effects. Although this is not a major concern for the Sun, where spot coverages are much smaller than faculae/plage coverages, it may become an issue for stars that are more active than the Sun and have larger spot coverages. Information gathered from the lines that form \logrhk\ therefore cannot be directly mapped to faculae. Ca\,\textsc{ii} H\&K lines additionally require intensive non-LTE modelling that limits their use for other spectral types.

These limitations motivate a more direct diagnostic of photospheric faculae. In addition, such studies specifically targeting photospheric faculae may be applicable to better constraining their impact on, for example, transmission spectroscopy (as discussed in Section \ref{intro} -- see \citealt{Rackham2018} and \citealt{Oshagh2014}). Here, we demonstrate that SRA features directly trace faculae. SRA primarily arise from photospheric rather than chromospheric variations, and are well captured by standard \phoenix\ and \muram\ models (see Sections \ref{Phoenix} and \ref{muram}), thus these offer a more direct and broadly applicable route to characterising faculae on other stars.

SRA residuals in fact reveal hundreds of spectral features that vary with stellar activity. \citet{Costes2026} showed that these features better capture RV variability than classical proxies, and that SRA-based indicators outperform conventional ones in planetary-injection-recovery tests. Because individual features vary independently and can behave quite differently from one another, we focus in this pilot study on a single line to interpret its variations. We selected Fe\,{\sc i} 4377 \AA\ as it shows one of the strongest and most coherent activity-related signals in the time series, in both amplitude and response, as found by \citet{Thompson2017, Thompson2020magneticactivityharpsnpaper2} and \citet{Costes2026}. While further investiation of SRA features and the impact of spots is needed, \citet{Thompson2020magneticactivityharpsnpaper2} notes that the 4377 \AA\ feature is more strongly correlated with faculae filling factor than spot filling factor. This is thus a more reliable diagnostic of faculae using their spectral imprints. It is also one of the few such features covered by the currently available MP-ATLAS \muram\ models, making it a natural choice for this study. The exploration of further lines is deferred to future work.

\begin{figure}
	% To include a figure from a file named example.*
	% Allowable file formats are eps or ps if compiling using latex
	% or pdf, png, jpg if compiling using pdflatex
	\includegraphics[width=0.5\textwidth]{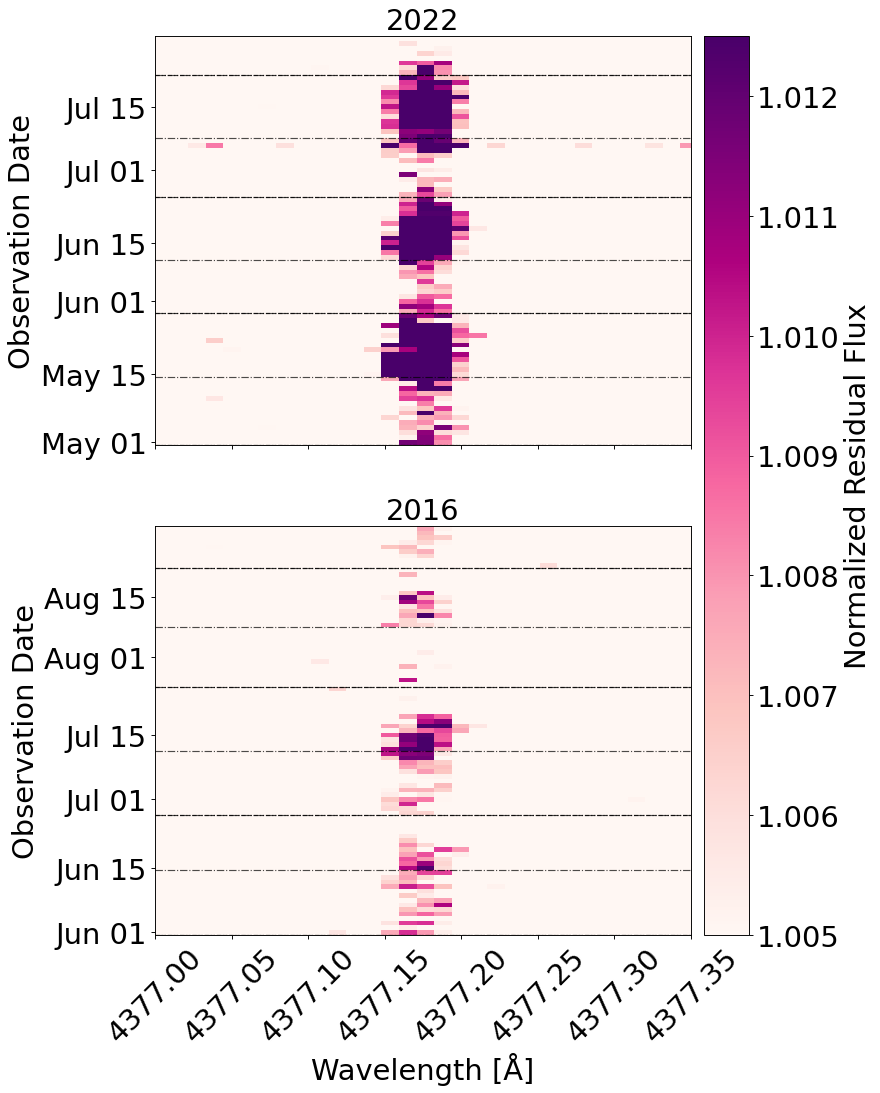}
    \caption{Trailed spectrum of the Fe\,{\sc i} 4377 \AA\ feature as a function of time. Carrington rotation periods and half-periods are indicated by dashed and dotted lines, respectively. The colour scale traces the evolution of the residual line profile and the variation in its amplitude is consistent with rotational modulation. The slanted morphology of each emergent residual feature reflects the expected blue-to-red shift of the feature as the prominent active region rotates across the stellar disc.}
    \label{fig:trailedspectrum}
\end{figure}

Figure \ref{fig:trailedspectrum} highlights the time modulation of this feature, which exhibits changes in line depth consistent with the presence of faculae regions on the solar surface \citep{Thompson2020magneticactivityharpsnpaper2}. In this figure, we see the intensity of the residual flux change over the visible half of the solar rotation period, confirming that the feature is linked to rotationally-modulated activity on the visible disc of the Sun. 

We attribute the strong sensitivity of the Fe\,{\sc i} 4377 \AA\ line to the following underlying physical drivers. First, solar quiet photospheric plasma temperatures are such that iron is predominantly singly ionized. Fe\,{\sc i} populations are thus a minority species relative to Fe\,{\sc ii} \citep[see e.g.][]{Mihalas1978,Gehren2001,Rutten2003RT}. This results in a higher sensitivity of Fe\,{\sc i} to temperature \citep[see e.g. the discussion in Section 3.3 of][]{Sowmya2026}. The temperature enhancement associated with facular magnetic fields \citep[e.g.][]{VAL1981,Fontenla1999} increases the degree of ionization, thereby reducing the Fe\,{\sc i} population. Second, Fe\,{\sc i} 4377 \AA\ line is a strong spectral line that forms in the middle-upper photosphere where the temperature enhancement caused by facular magnetic fields is more pronounced. Third, Fe\,{\sc i} 4377 \AA\ line is a resonance line with the ground state as the lower level ($E_{\mathrm{low}} = 0.0$\,eV; as recorded in the VALD3 database \citealt{VALD,Ryabchikova2015}). An increase in temperature redistributes the remaining Fe\,{\sc i} population among excited levels according to the Boltzmann distribution \citep{Mihalas1978,Rutten2003RT,Gray2005}, thereby decreasing the population of the ground state that serves as the lower level of the transition and reducing the line opacity. The combined action of these effects leads to a substantial weakening of the Fe\,{\sc i} 4377 \AA\ line in facular regions compared to the quiet Sun, making it an excellent tracer of changes in facular regions on the Solar disc.

To further contextualize this behaviour, we turn to spatially resolved observations of the Solar surface at the Fe\,{\sc i} 4377 wavelength. These data allow us to directly connect variations in the line profile with the distribution and viewing geometry of facular regions. We obtained observations with a modified Solar Explorer \citep[Sol’Ex;][]{Buil2023} spectroheliograph located in Cluj-Napoca, Romania. The instrument was connected to a 62/400 mm refractor with a blue bandpass filter placed before the aperture as an energy rejection filter (ERF), effectively reducing its diameter down to 4~cm. The spectroheliograph has an effective spectral resolving power of ${\cal R} \approx 20\,000$, and can be tuned to a wide range of wavelengths when the ERF is changed \citep{Pal2025}. The setup can scan the full solar disc in about 1~min, during which time it takes around 4500 slit positions. The solar disc was reconstructed and stored as a FITS cube using the JSol’Ex\footnote{\href{https://github.com/melix/astro4j/tree/main/jsolex}{github.com/melix/astro4j/tree/main/jsolex}} open source software \citep{Champeau2026} as described in \cite{Pal2025}. In this case, Sol'Ex was tuned to a spectral window extending from approximately 4373~\AA\ to 4386~\AA\ and the data were calibrated using the atlas from \citet{Pietrow2026}.

\begin{figure*}
	% To include a figure from a file named example.*
	% Allowable file formats are eps or ps if compiling using latex
	% or pdf, png, jpg if compiling using pdflatex
	\includegraphics[width=\textwidth]{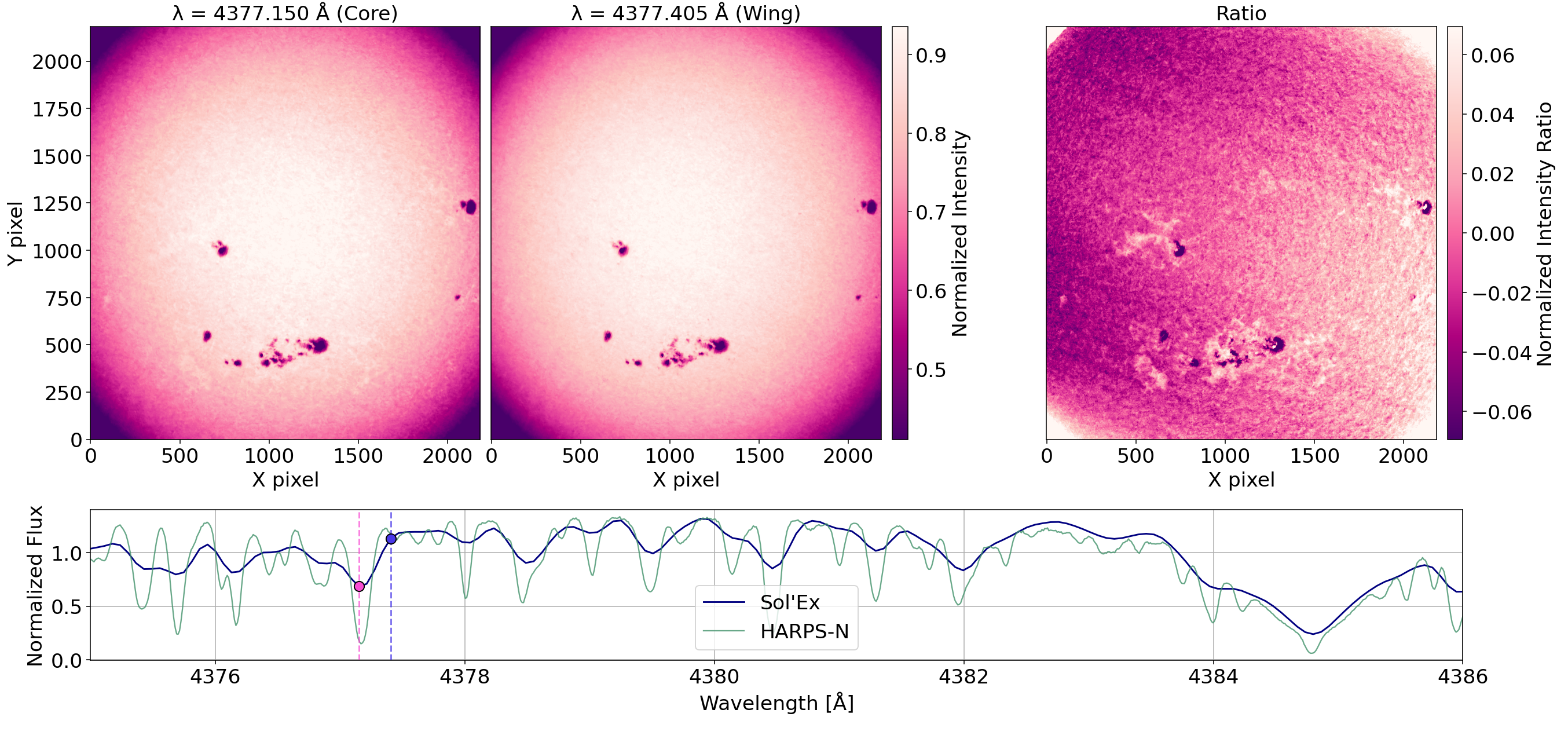}
    \caption{Top row: solar disc observed with Sol’Ex tuned to the line core of the 4377 \AA\, feature (left), the outer wing of the same line (middle), and their ratio (right), which highlights enhanced contrast in facular regions. Bottom row: comparison between the Sol’Ex spectrum and the corresponding HARPS-N spectrum. The pink and purple markers indicate the wavelengths corresponding to the line-core and wing images shown in the top row, respectively.}
    \label{fig:solarimage}
\end{figure*}

A total of 19 scans were recorded between 2025-08-15 and 2025-08-30 to sample different faculae filling factors; however, in the current work, only the best seeing data from day 2025-08-28 was used as a context image, shown in Figure \ref{fig:solarimage}. 
The top row of this figure shows two spectroheliograms of the solar surface constructed from different parts of the spectral line: the line core, and a continuum region at the red-edge of the line-wing. The final panel in the top row shows the ratio between the two images. In the image taken in the core of the Fe\,{\sc i}~4377~\AA\ line (top left), the facular regions appear brighter compared to surrounding inactive regions. In contrast, the images taken in the continuum at the red-edge of the line (top middle), shows little to no enhancement in these same regions. The ratio image (top right) further emphasizes that the Fe\,{\sc i}~4377~\AA\ (and in particular its core) is indeed sensitive to the presence of faculae. 

We thus focus on this line as a diagnostic of faculae and use it to infer properties of these regions. Our primary goal was to determine whether HARPS-N spectra could be used to determine the fractional area of the solar disc covered by faculae at different activity levels. To investigate this, we compared the SRA residuals to synthetic spectra generated from stellar atmosphere models, allowing us to adjust relevant parameters and evaluate how different filling-factors of quiet-Sun and faculae-like components reproduce the observed solar variations.

\section{Stellar Models and Faculae Estimations}\label{stellarmodels}

To quantitatively interpret the observed modulation of the Fe\,{\sc i} 4377 \AA\,line and relate it to faculae coverage on the solar surface, we model the SRA-derived residuals using synthetic stellar spectra. We adopt two complementary modelling frameworks that differ in both physical assumptions and dimensionality: one based on one-dimensional PHOENIX stellar atmosphere models, and another using three-dimensional magnetohydrodynamic (MHD) simulations from \muram. 

\subsection{PHOENIX}\label{Phoenix}

\phoenix\ is a general-purpose stellar atmosphere code that computes self-consistent model atmospheres and synthetic spectra for stars across a broad range of temperatures, gravities, and compositions \citep{hauschildt1999}. The code solves the radiative transfer equation in hydrostatic equilibrium for either plane-parallel or spherical geometry, incorporating detailed atomic and molecular opacities treated line by line. \phoenix\ includes comprehensive treatments of molecular and non-equilibrium chemistries and convection, enabling accurate modelling of both hot and cool stellar atmospheres. The resulting temperature–pressure structures and emergent spectra are routinely used to interpret stellar observations and derive fundamental stellar parameters.

The \phoenix\ models used in this work are drawn from the Göttingen Spectral Library \citep{Husser2013}, and for this study we restrict our analysis to Local Thermodynamic Equilibrium (LTE) effects. Despite not including the physics necessary to self-consistently model faculae regions and thus being an incomplete representation of their complex structure \citep{Witzke2022}, approximating faculae as slightly hotter than the quiet photosphere has been used in both solar and stellar activity modelling \citep[e.g.,][]{Fontenla1999,Dumusque2014}. Libraries of \phoenix\ synthetic spectra have been used in numerous studies to simulate or approximate the effects of stellar activity from spots, plage and faculae. Typically, spectra computed at different effective temperatures are combined according to prescribed filling factors and surface distributions to represent the quiet photosphere and active regions. \cite{herrero2016} applied this method to construct composite stellar spectra and quantify the impact of activity on line profiles and radial velocities. The SOAP-GPU framework uses high-resolution PHOENIX spectra as templates for the quiet photosphere, spots, and faculae in efficient forward models of stellar activity \citep{zhao2023,zhao2025}. A similar philosophy underlies the PAStar model of \citet{petralia2025}, which employs \phoenix\ libraries to synthesize spectra of heterogeneous stellar photospheres for activity studies. 

\begin{figure}
	% To include a figure from a file named example.*
	% Allowable file formats are eps or ps if compiling using latex
	% or pdf, png, jpg if compiling using pdflatex
	\includegraphics[width=0.5\textwidth]{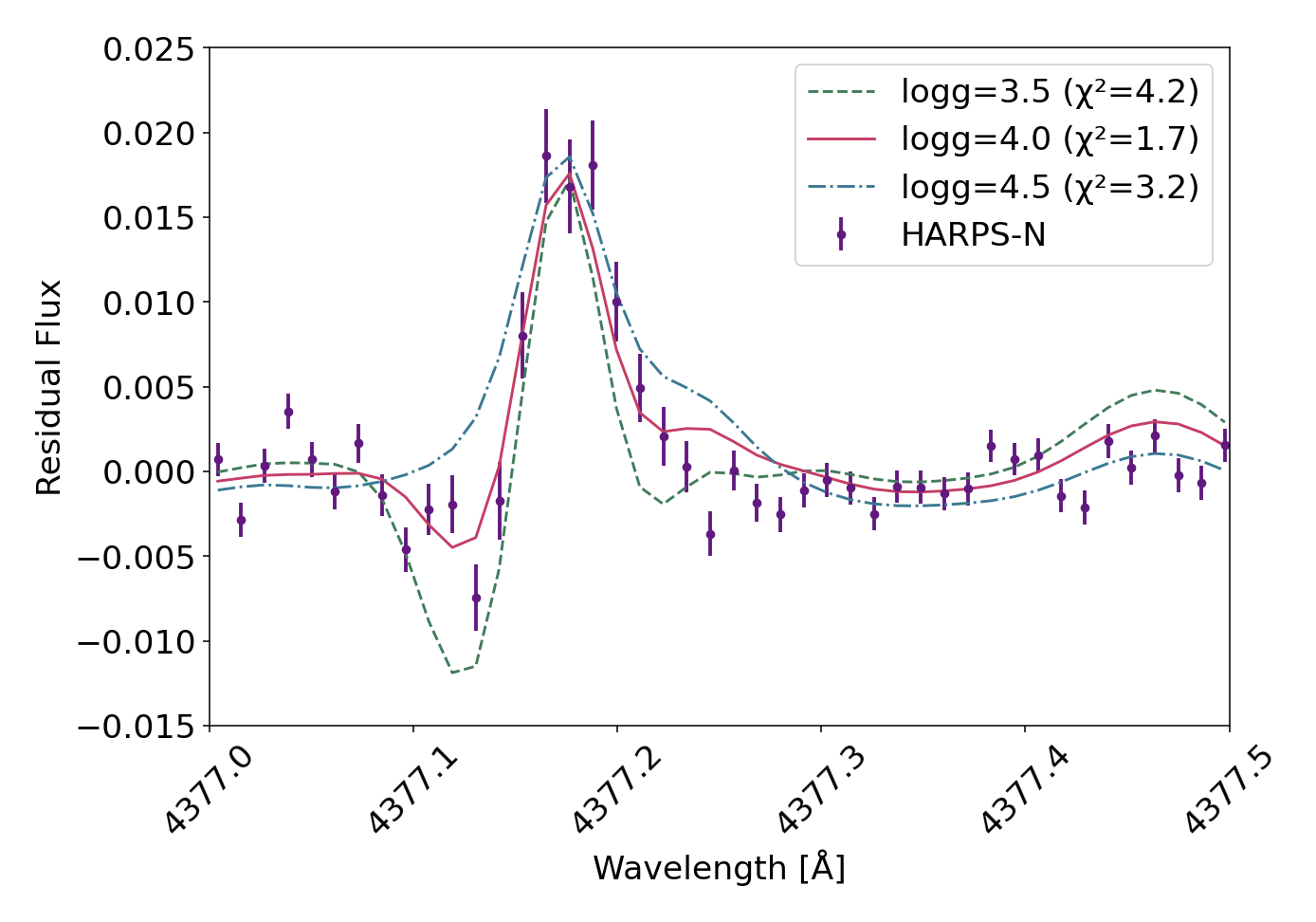}
    \caption{Example residual spectra of the HARPS-N data (purple points) in comparison to \phoenix\ models of varying surface gravity values (3.5 in green, 4.0 in pink, 4.5 in blue.) Modelled residual spectra are created by dividing a linear combination of quiet Sun spectrum plus faculae spectrum (normalized by filling factor, here set to 4.5\%) by a low activity  reference (1\% filling factor). See text for further details. Faculae coverage of 4.5 $\%$ was chosen to match SDO-measured values for this particular day (2022-06-13). Variation in the residuals seen here are representative of changes caused by varying surface gravity, and the best-fit reduced \chisq\ values for each model are given in the legend, with surface gravity of 4.0 producing the best fit to the data.}
    \label{fig:loggexamples}
\end{figure}

For this study, we adopt solar metallicity ([Fe/H] = 0.0). We take a 5800 K synthetic spectrum as our “quiet Sun” reference, and we adopt a surface gravity of $\log g = 4.5$. Higher-temperature models (5800–6300 K) are used as proxies for facular regions. To approximate the modified opacity structure in facular regions, we experimented with a range of surface gravities, \mbox{$\log g$ = 3.5,} 4.0, and 4.5. We find that the resulting faculae estimates (see Section \ref{comparisonswSDO}) are relatively insensitive to $\log g$ over the range tested, however minor differences are seen in the line profile at 4377 \AA\ -- we show this in Figure \ref{fig:loggexamples}. A surface gravity value of 4.0 gives the lowest reduced \chisq\ values, likely because the decrease in surface gravity leads to the appearance of a trough on the left side of the feature, similar to the trough seen in the observed SRA residuals. Thus, we adopt $\log g = 4.0$\ as a representative value for facular regions. This lower surface gravity serves as a proxy for the change in pressure broadening within the facular regions \citep{Spruit1976, Fontenla1993}. 

The \phoenix\ models are tabulated in discrete 100 K steps, thus we linearly interpolated between grid points (in step sizes of 50 K) to create spectra at intermediate temperatures, allowing finer control over the contrast between the faculae temperature and quiet Sun. All synthetic spectra were continuum-normalized by fitting a fourth-order polynomial and dividing by the resulting fit. This order of polynomial provides sufficient flexibility to model broad continuum variations without overfitting spectral line structure. The process removes large-scale variations in the continuum shape and ensures consistency with the normalization applied to the observed spectra, allowing direct comparison of line profiles. 

\phoenix\ synthetic spectra are provided in the stellar rest frame at high intrinsic resolution \citep[R $\sim$ 500,000 in the optical][]{Husser2013} and do not include rotational broadening (i.e. effectively $v\sin i = 0~\mathrm{km\,s^{-1}}$). To approximate disc-integrated observations, the quiet-Sun spectrum was rotationally broadened using the PyAstronomy routine \texttt{fastRotBroad}, adopting a projected rotational velocity of $v\sin i = 2~\mathrm{km\,s^{-1}}$ and a linear limb-darkening coefficient of 0.4 \citep{Claret2000}. The facular spectrum was treated as a spatially-localized component and was therefore not rotationally broadened, representing emission from a small region near the disc centre where projected rotational effects are negligible. Instrumental broadening was applied by convolving all synthetic spectra with a Gaussian line spread function with width $2.54\,\mathrm{km\,s^{-1}}$, corresponding to resolving power of \harpsn\ (115,000). All synthetic spectra were subsequently interpolated onto the reference HARPS-N wavelength grid defined in Section~\ref{sec:Harpsn} using cubic spline interpolation for direct comparison with observations.

The quiet Sun and faculae spectra were then linearly combined by adjusting their relative contributions, simulating different fractional surface coverages. Equation \ref{eqn:plagecombinations} describes these calculations,
\begin{equation}\label{eqn:plagecombinations}
F_{\rm model} = \, (1 - \alpha) F_{\rm quiet}  + \alpha F_{\rm faculae},
\end{equation}

\noindent where $F_{\rm quiet}$ is the flux of the quiet Sun spectrum, $F_{\rm faculae}$ is flux of the faculae proxy spectrum, and $\alpha$ is the fractional area coverage of faculae on the visible solar disc.

We began by creating a low-activity model spectrum, where the faculae coverage was set to 1\%, similar to the solar faculae coverage as determined from SDO images (see Section \ref{comparisonswSDO} for details) of our low activity template on the day of 2017 July 23. We then created the `high' activity spectra by allowing the combinations of quiet Sun and faculae proxy spectra to vary. Specifically, we varied the faculae filling factor, $\alpha$, from 1–10\%, with step sizes of 0.2\%. A radial velocity shift applied to all facular proxy spectra was also treated as a free parameter in the grid search, allowed to vary over $\pm 2.5\,\mathrm{km\,s^{-1}}$, with step sizes of $ 0.25\,\mathrm{km\,s^{-1}}$ to account for Doppler shifts arising from the movement of active regions across the solar disc. For each of these combinations, SRA was then applied to create residuals between high-activity and low-activity spectra that highlights the relative changes in line profiles.

We determined the best-fit parameters (temperature and faculae coverage) for each observing day using a grid search over synthetic models. We include the radial velocity of the facular proxy spectrum as a free parameter in the grid search; however, the best-fit RV values recovered from fitting the synthetic models to the HARPS-N residual spectra show no significant or coherent signal. We therefore do not analyze the recovered RV values further and instead focus on the best-fit temperature and faculae coverage. For each grid point, we computed the reduced \chisq\ statistic between the synthetic residual spectrum and the HARPS-N residual spectrum in a narrow wavelength window of 4377.0 to 4377.5\,\AA. The best-fitting model was selected as the parameter combination yielding the minimum reduced \chisq\ value. This fit is repeated for each day, and we present the results in Section \ref{comparisonswSDO} along with fits using \muram\ models (Section~\ref{muram}).

\subsection{MURaM}\label{muram}

For the second set of stellar models, we used 3D radiative magnetohydrodynamic (MHD) simulations of the solar photosphere and upper convection zone performed with the \muram\ code {\footnote{MURaM: The Max-Planck-Institute for Aeronomy/ University of Chicago Radiation Magneto-hydrodynamics code.} \citep{Vogler2005}. It solves the full set of compressible magnetohydrodynamic equations, coupled with non-grey radiative transfer and a realistic equation of state that includes partial ionization effects. \muram\ simulations including a small-scale dynamo \citep[SSD;][]{VoeglerandSchuessler2007,SchuesslerandVoegler2008,Rempel2014}, but no additional magnetic fields, are used to represent the quiet Sun \citep[see e.g.][for details]{Witzke2024} while simulations with average vertical magnetic fields of 100, 200, and 300 G imposed on the SSD setup are used to represent faculae \citep[see e.g.][]{Witzke2022,Kostogryz2024}. The imposed magnetic fields reorganize, leading to the formation of kG flux concentrations in the intergranular lanes \citep[see e.g.][]{Kostogryz2024,Bhatiaetal2026}. Synthetic spectra are then calculated from the simulated \muram\ atmospheres using MPS-ATLAS \citep{Witzke2021} radiative transfer code. The spectra are computed for 10 viewing geometries, parametrized by $\mu = \cos\theta$ where $\theta$ is the heliocentric angle. $\mu = 1$ corresponds to disc centre, and $\mu$ decreases in steps of 0.1 down to $\mu = 0.1$ at the limb. The spectra are synthesized for at least 40 snapshots at a time cadence of 90\,s from each set of \muram\ simulations described here.

We take a similar approach to creating our models with these synthetic spectra as we do with the PHOENIX spectra, with some differences. First, the spectra for each limb angle combination are averaged by stacking across multiple `snapshots' run by the \muram\ code. To construct a synthetic quiet Sun spectrum representative of the full stellar disc, we follow a disc-integration procedure. We assume a normalized solar radius of 1, and the SSD spectra for each $\mu$ angle are then weighted by the sky projected area of their corresponding annulus. The resulting spectrum is a weighted sum over all $\mu$ bins and represents the disc-integrated quiet Sun. Rotational broadening is then applied to this spectrum to simulate the net rotational effect of the Sun viewed as a star, and instrumental broadening is applied.  

The synthetic spectra with injected magnetic fields of 100, 200, and 300~G were taken as the faculae proxy. We use the same method to produce model residuals as done for the PHOENIX spectra (see Section \ref{Phoenix}). In summary, we again created a low activity template with a 1\% faculae filling factor, and created a grid of high activity models by varying the facular filling factors from 1\% to 10\%, while also varying the $\mu$ angle. We add this parameter to account for the centre-to-limb variation of facular contrasts and line profile shapes, with the goal of tracking faculae regions as they move across the solar disc. We normalize both the low activity and high activity synthetic spectra using the same polynomial normalization process. The synthetic SRA residuals are then created by dividing the active synthetic spectrum by the inactive one, and these are then compared to the HARPS-N observations and the reduced \chisq\ between the two are calculated across the full grid of generated synthetic SRA spectra.

An example day of the resulting best-fit model residuals to the data are shown in Figure \ref{fig:Bcomparison}. The 100~G models tend to underestimate the amplitude of the Fe\,{\sc i} feature, leading to overestimation of faculae areal coverage as the models favours a high filling factor to reach the amplitude of the observed residuals. Due to this, 100~G was removed as a parameter in further investigations. Additionally, while 300~G models provide the lowest reduced \chisq\ value, little difference is seen between the 200 G and 300 G models. For the remainder of this study we focus on the 200 G models for further analysis as \citet{SchrijverandHarvey1994, Carlsson2019, Schrijver2020} cite this as the canonical value for the magnetic field of faculae regions.
\begin{figure}
	% To include a figure from a file named example.*
	% Allowable file formats are eps or ps if compiling using latex
	% or pdf, png, jpg if compiling using pdflatex
	\includegraphics[width=0.5\textwidth]{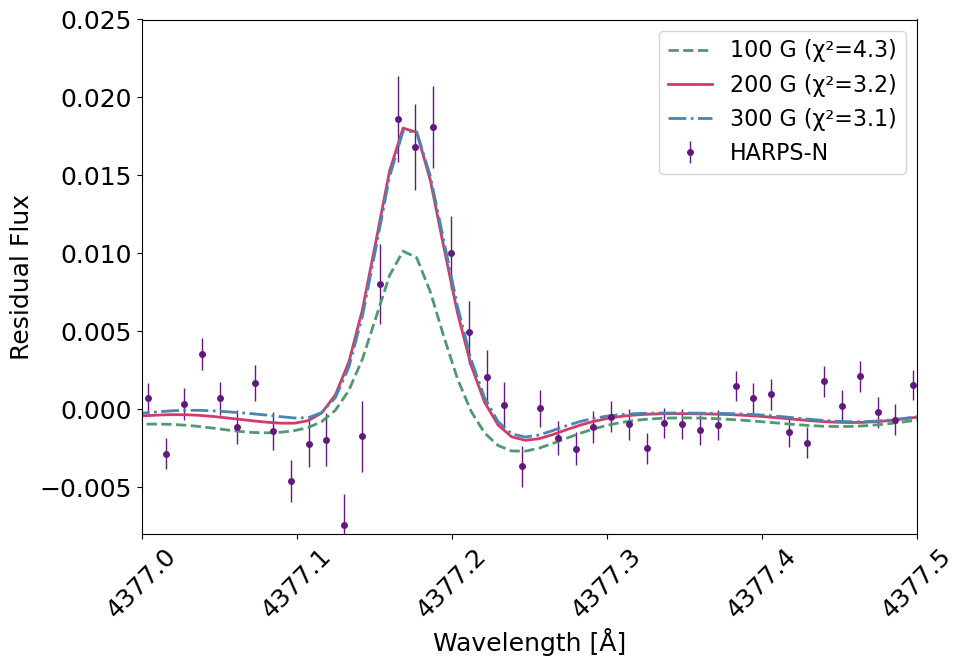}
    \caption{Example spectra of the HARPS-N data (purple) in comparison to \muram\ models of varying magnetic field strengths. 100~G model as a green dashed line, 200~G as a pink solid line, and 300~G as a dash-dotted blue line. Faculae coverage level is set to 4.5\%, matching SDO values for this particular day (2022 June 13). Variations in the residuals seen here are representative of changes caused by varying magnetic field strengths, where the reduced \chisq\ of each model to the residuals are given in the legend. Here, we can see that the 100~G model has a lower amplitude than the other models.}
    \label{fig:Bcomparison}
\end{figure}

Figure \ref{fig:bestfits} shows examples of best fit models for both \phoenix\ and \muram\ compared with the HARPS-N observations. The most prominent difference between the models is the presence of a pre-feature trough, which is reproduced by \phoenix, but is absent in the \muram\ fits. The pre-feature trough is seen in all observed spectra, however its origin, and thus the discrepancy between synthetic models, is beyond the scope of this work. 
\begin{figure*}
	% To include a figure from a file named example.*
	% Allowable file formats are eps or ps if compiling using latex
	% or pdf, png, jpg if compiling using pdflatex
	\includegraphics[width=1\textwidth]{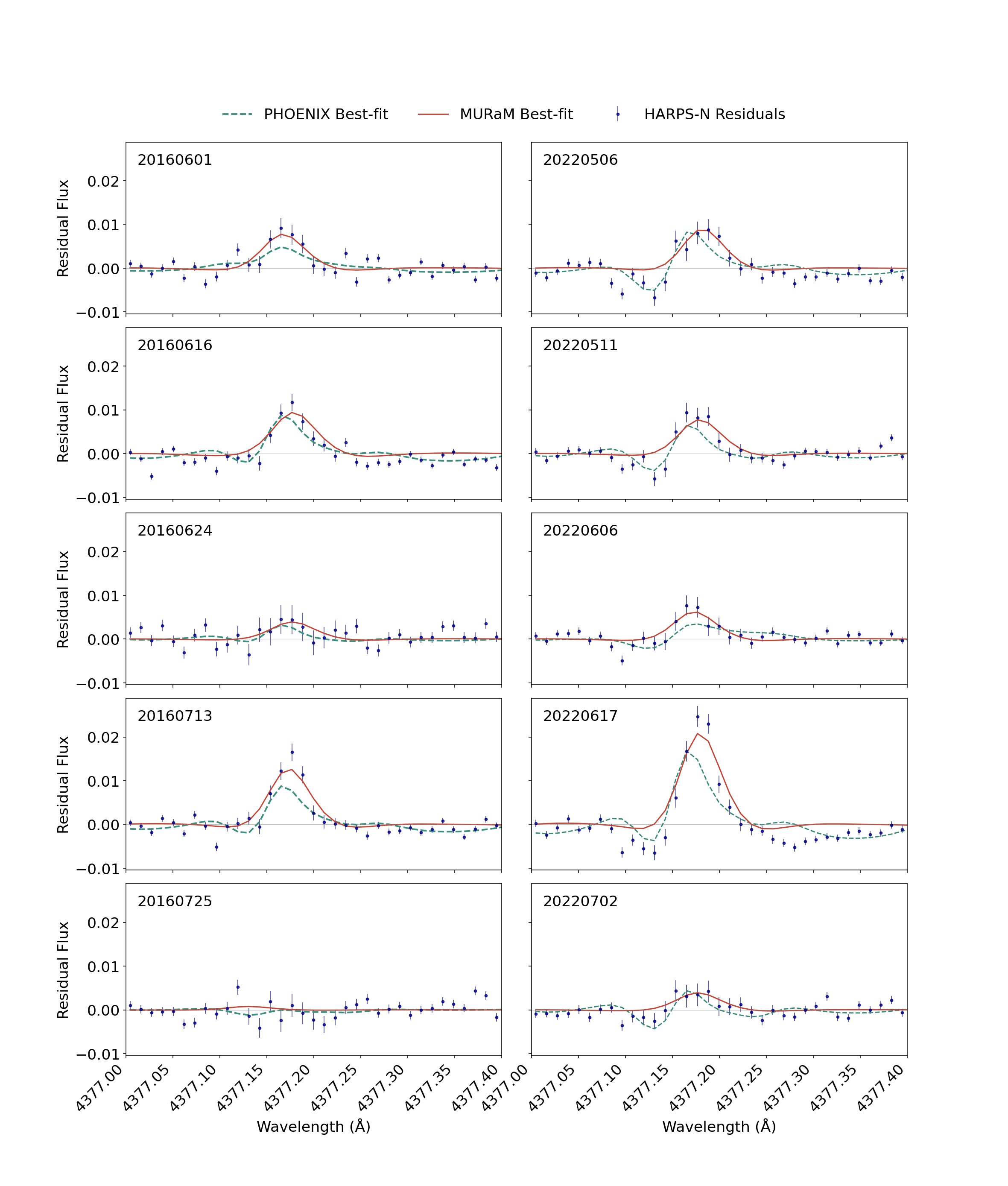}
    \caption{Various examples of the Fe\,{\sc i} 4377\,\AA\ feature are shown for different observation days, with HARPS-N data displayed as dark blue points, \phoenix\ best-fit models as green dashed lines, and \muram\ models in as a red solid line. The left column corresponds to dates during the 2016 period, and the right column shows examples from the 2022 period. Dates were chosen to demonstrate the variation in amplitude of the SRA residuals across a rotation period, and to show the difference in amplitude of the feature between 2016 and 2022.}
    \label{fig:bestfits}
\end{figure*}

\section{Comparison with SDO/HMI Filling Factors}\label{comparisonswSDO}
In this section, we compare the facular filling factors inferred from our HARPS-N spectral fitting with independent, disc-resolved measurements of the Sun from SDO/HMI. The goal is to assess whether the spectroscopically derived filling factors recover the true temporal variability of surface magnetic regions as seen in spatially resolved observations.

SDO observes the solar surface with continuous full-disc coverage using two 4096x4096 pixel cameras -- allowing for resolution at the granule level. It has three scientific instruments: the Atmospheric Imaging Assembly (AIA), Extreme Ultraviolet Variability Experiment, and the Helioseismic and Magnetic Imager (HMI) \citep{Schou2012, Pesnell2012, Couvidat2016}. 

For the dates analysed in this work, we use SDO/HMI magnetogram and continuum intensitygram data to derive spatial information about faculae regions of the Solar disc. Using SolAster \citep{Ervin2022}, we followed the method outlined by \citet{Haywood2016} and \citet{Milbourne2019} to calculate faculae filling factors. Using HMI intensity and magnetogram images, we set a magnetic field strength threshold value to differentiate between active and quiet pixels; this is set to 24 G. To differentiate between spots and faculae within the active pixels, we apply an intensity threshold of $I_\mathrm{thresh} > 0.89\hat{I}_{\mathrm{quiet}}$, where $\hat{I}_{\mathrm{quiet}}$ is the mean intensity of the quiet Sun regions. 

We compute facular filling factors both with and without an additional area threshold. In the thresholded case (hereafter denoted by `T'), we apply an area threshold of 2 microhemispheres ($\mu$Hem), corresponding to $6.08 \times 10^{6}\ \mathrm{km}^2$ on the solar surface. This threshold is applied to exclude small, network level brightness variations. The non-thresholded filling factors (hereafter denoted as `NT') include all pixels satisfying the magnetic and intensity criteria, regardless of spatial extent. Figure \ref{fig:sdo_mu_hist_combined} shows active regions for two example days in 2016 and 2022, and the distribution of limb angles for these active regions.

\begin{figure*}
    \includegraphics[width=\textwidth]{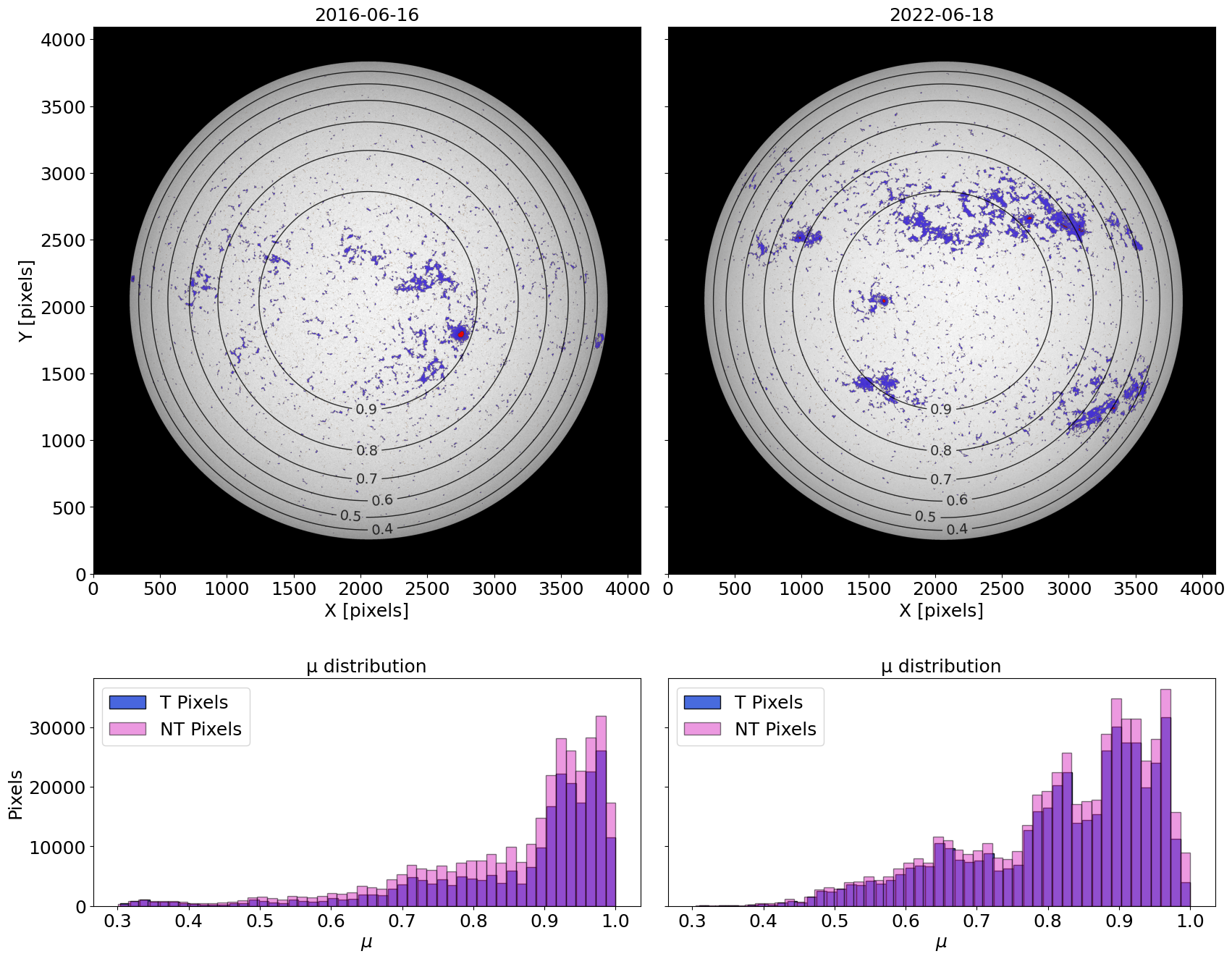}
    \caption{Top row: SDO images of the solar surface for two example days of 2016 June 16 and 2022 June 18. Blue shows pixels defined as faculae with the area threshold applied, orange shows all magnetically active bright points (not visible due to scale), and red shows areas identified as spots. Concentric circles mark limb angles as given by SDO. Bottom row: Histograms of the limb angles calculated by SDO for each pixel defined as faculae, specifically for 2016 June 16 (left) and 2022 June 18 (right). T (blue bars) and NT (pink bars) distinguish between with and without the area threshold, respectively. See text for further definition.}\label{fig:sdo_mu_hist_combined}
\end{figure*}

\subsection{Comparison of Model Estimates and SDO calculations}

We compare the facular filling factors derived from the \phoenix\ and \muram\ models to those determined from SDO images. With \phoenix, we follow the method described in Section \ref{Phoenix} for each day, obtaining a daily estimate for the temperature, radial velocity shift, and faculae areal coverage. The resulting faculae estimates of the 2016 and 2022 periods are presented in the top left and right panels of Figure \ref{fig:plageestimatessdo}, respectively. The bottom two panels of Figure \ref{fig:plageestimatessdo} show the estimates given by the \muram\ models. These estimates are obtained following the procedure outlined in Section~\ref{muram}.
\begin{figure*}
	% To include a figure from a file named example.*
	% Allowable file formats are eps or ps if compiling using latex
	% or pdf, png, jpg if compiling using pdflatex
	\includegraphics[width=\textwidth]{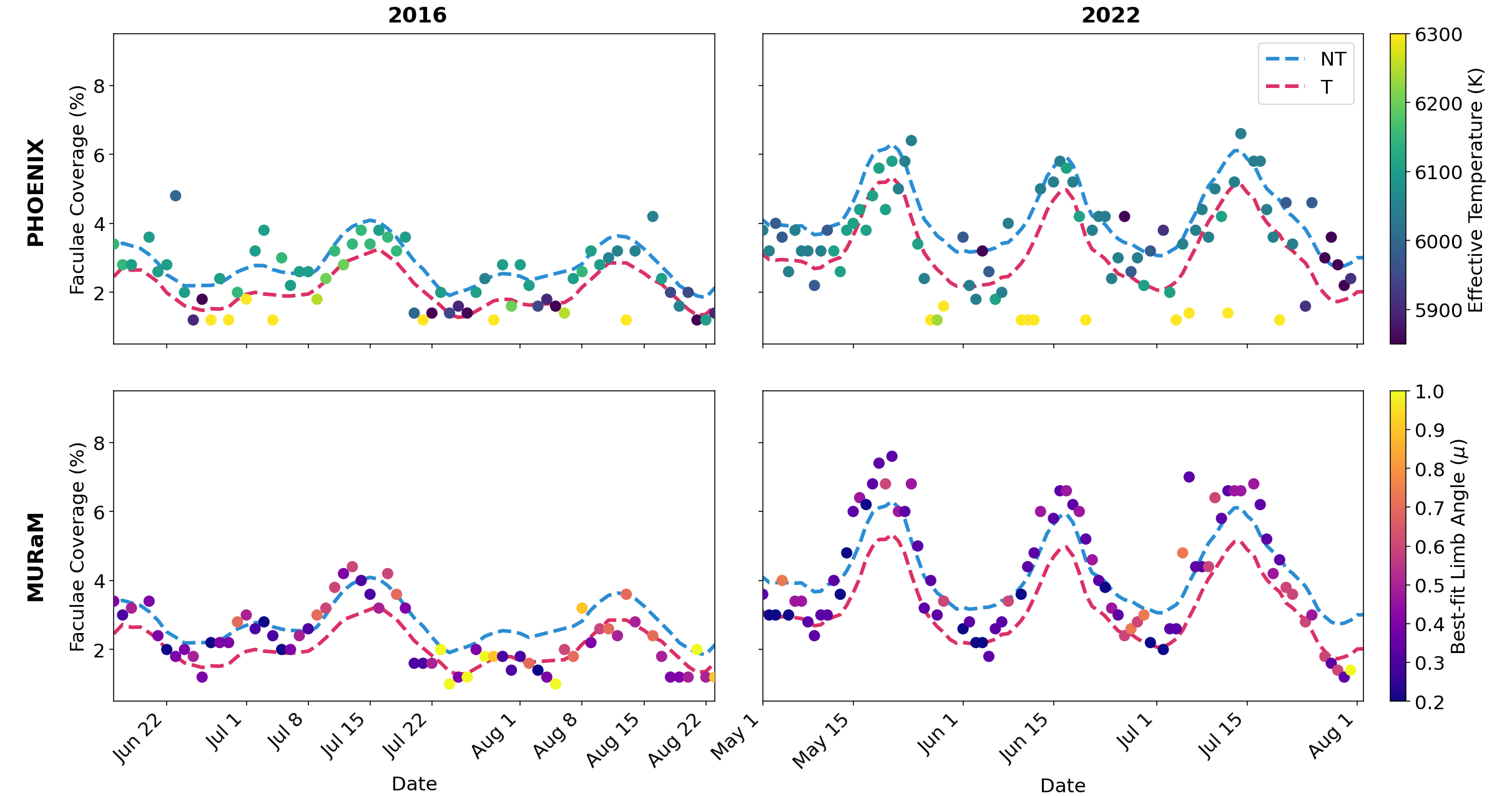}
    \caption{Results of faculae filling-factor estimates using each synthetic stellar model in comparison to HARPS-N solar data. The markers show faculae coverage for each day as predicted by each model, with the top row showing \phoenix\ results in 2016 and 2022, and the bottom row showing the respective results with \muram. The colour bars show the best-fit temperature and best-fit $\mu$ angle for \phoenix\ and \muram, respectively. SDO derived ground truth values are shown in the blue and red dashed lines, representing no area threshold (NT) and an area threshold (T) of 2 microhemispheres, respectively.}
    \label{fig:plageestimatessdo}
\end{figure*}
In Figure~\ref{fig:plageestimatessdo}, we observe that the estimated facular coverage moves between the NT and T curves. During the high-activity phase, the HARPS-N-estimated facular coverage mainly tracks the SDO filling factor derived from all regions (i.e. the NT curve in blue). This includes the facular regions as well as the magnetic network, represented by the bright pixels in the SDO images. Conversely, during lower-activity phase, the facular coverage follows the area-thresholded SDO filling factor, which corresponds only to larger facular regions. These variations could be explained by the fact that, at lower activity levels, the contrast between magnetically active and quiet regions is reduced, thereby limiting our SRA residual fitting sensitivity to smaller-scale structures such as the magnetic network; as a result, the Fe\,{\sc i} 4377 line preferentially traces the more prominent faculae at low activity, and a mix of faculae and network at high activity.

We evaluate the agreement between the model-derived facular filling factors and those obtained from SDO using both definitions of the filling factor, i.e.: including all bright pixels (NT), and applying a 2 $\mu$Hem area threshold (T). Table~\ref{pearsoncorrelations} shows the Pearson's $R$ correlation for \phoenix\ and \muram\ estimates in both time periods, with both NT and T values. Overall, both models reproduce the temporal evolution of facular coverage across both epochs and definitions, however, the \muram\ models provide higher Pearson R values and thus our measurements of faculae filling factor directly from spectra show stronger agreement with the values determined from the disc-resolved SDO images. Additionally, the \phoenix\ model estimates for the 2022 period occasionally fall below the expected values and push to the upper limitations of temperatures. We find that this is due to larger errors bars on these days pulling the reduced \chisq fit into the continuum, rather than properly fitting the peak. Removing these points from the correlation computations, we see an increase in the correlation coefficient for the 2022 interval for both \phoenix\ cases. From 0.640 to 0.795 for $\mathrm{PHOENIX}_{NT}$, and from 0.640 to 0.794 for $\mathrm{PHOENIX}_{T}$. We see no change for $\mathrm{MURaM}_{NT}$m and a slight decrease in $\mathrm{MURaM}_{T}$ from 0.927 to 0.926. 

\begin{table}
\centering
\caption{Pearson correlation coefficients ($R$) between model faculae estimates 
and SDO filling factors for 2016 and 2022. NT = non-thresholded; 
T = area-thresholded.}

\label{tab:correlations}
\begin{tabular}{lcc}
\hline
Model & 2016 & 2022 \\
\hline
$\mathrm{PHOENIX}_{\mathrm{NT}}$ & 0.643 & 0.640 \\
$\mathrm{PHOENIX}_{\mathrm{T}}$  & 0.587 & 0.640 \\
$\mathrm{MURaM}_{\mathrm{NT}}$   & 0.823 & 0.927  \\
$\mathrm{MURaM}_{\mathrm{T}}$    & 0.786 & 0.927 \\
\hline
\end{tabular}
\end{table}\label{pearsoncorrelations}

Since the \phoenix\ models are parametrized by facular temperature, we can directly measure the temperature contrast compared to the quiet Sun with our best-fit values. The best-fit faculae temperatures span approximately 6000–6200 K across the individually modelled observations, corresponding to a temperature contrast of roughly 200 – 400~K (relative to the adopted quiet Sun temperature of 5800~K). This is broadly consistent with the values reported in the literature, where \citet{Meunier2010} and \citet{Unruh1999} find facular temperature contrasts of approximately 50 – 300~K depending on limb angle. 

Likewise, we find that our best fit $\mu$ angle from the \muram\ models is between 0.3 and 0.5 across the individually modelled observations, which corresponds to roughly 73 to 60 degrees from disc centre on the visible disc. While this is a reasonable estimate of where faculae regions might be located, this is not in agreement with the measured average $\mu$ angles from SDO. As an example, on 2016 June 16 and 2022 June 18, the MURaM model fits give an estimated limb-angle of $\sim$ 0.5. However, from the SDO images for these dates, Figure~\ref{fig:sdo_mu_hist_combined} shows that the predominant limb angles of these active regions fall at $\sim$ 0.9-1.0. It is possible that the models favour mid to low limb angles due to the contribution of a variety of active regions at various limb angles across the solar disc. It is also likely that the best fits favour models closer to the limb as these models show higher amplitudes in the residual feature, as seen in Figure \ref{fig:muangle}. We thus conclude that we are currently insensitive to limb angle variations.

As seen in Table~\ref{pearsoncorrelations}, the \muram\ models show stronger agreement with the SDO-derived filling factors. This is likely due to the \muram\ models being based on a more physically-motivated representation of faculae, i.e. by using regions with injected higher magnetic fields. Meanwhile, the \phoenix\ models are based on injection of higher temperature spectra, with no injected magnetic field, and thus likely provide a less accurate representation of the Sun at higher magnetic activity levels. Additionally, the \phoenix\ models are based on models of other stars and different stellar types, while \muram\ is a specifically solar model.  While neither model fully encapsulates the complexity of faculae regions within the method presented here, both provide relative filling factor estimates directly from the HARPS-N spectra.

\subsection{Limitations and Further Work}

Several important caveats should be considered when interpreting these facular filling factors, particularly in terms of their absolute values. First, the SDO-derived filling factors depend on the adopted magnetic field and intensity thresholds; varying these selection criteria would directly change the fraction of pixels classified as faculae. Second, in the \phoenix\ models and the disc-integrated SSD \muram\ model, this approach treats the Sun as a source with uniform surface components and does not account for the spatial distribution of active regions across the solar disc. This simplification neglects centre-to-limb variations and projection effects that influence the observed facular signal. The models themselves also do not fully capture rotational broadening, and the method with which we broaden the synthetic spectra does not capture the exact physics of rotational line broadening. This can affect the inferred line profiles and, in turn, the derived filling factors. A more physically rigorous treatment, in which the stellar surface is tiled with spatially resolved synthetic spectra, each Doppler-shifted according to its projected rotational velocity and then disc-integrated, would better capture these effects and is deferred to future work.

Furthermore, our modelling assumes a fixed magnetic field strength in the \muram\ framework and a limited parametrization of temperature contrast in the \phoenix\ models, whereas real facular regions exhibit a range of magnetic field strengths, temperatures, and geometric structures.
In the \phoenix\ approach, the filling factor and the estimated effective temperature are inversely related (see Figure \ref{fig:chisq}), creating a clear degeneracy where the areal coverage estimate is dependent on the temperature contrast. 
While the PHOENIX-based temperature estimates are consistent with values reported in the literature, this approach is inherently limited. Faculae cannot be fully described by a simple temperature enhancement, as they are magnetically structured regions with more complex temperature and opacity variations. In our method, these regions are approximated using a non-magnetic stellar template with an increased temperature, which does not capture the full underlying physics. A more physically accurate method is the use of MURaM MHD simulations, where faculae are a natural by-product of the magnetic field injected into the simulations themselves. 

Estimates from the \muram\ based models are dependent on magnetic field strength and limb angle. As noted in Section \ref{muram}, we use only 200~G models in our analysis. Activity regions predominantly emerge in belts located at mid-latitudes, as illustrated by the solar butterfly diagram (e.g., \citealt{Hathaway2015}), rather than near the equator. As a result, their closest approach to disc centre is limited by their latitude. For a given heliocentric angle $\mu$, the corresponding latitude of closest approach is given by $\cos^{-1}(\mu)$; thus, features at latitudes of $\pm 25.8^\circ$ only reach $\mu \sim 0.9$. This geometrical effect, combined with the limb brightening of faculae, further reduces the observable contrast of these active regions near disc centre. However, as seen in the histograms in the bottom row of Figure~\ref{fig:sdo_mu_hist_combined}, we find that a large proportion of the pixels we classify as faculae sit between $\mu = 0.8 - 1.0$, which is confirmed visually in the top panel of Figure \ref{fig:sdo_mu_hist_combined}.

In principle, our framework was intended to also provide constraints on the distribution of active regions as a function of limb angle, by comparing line shape changes in the spectra to those from the \muram\ models. However, we are unable to track or make conclusions surrounding the limb angle of active regions within the current modelling setup. It is possible that the disc-integrated spectra lack sufficient spatial information in this method to disentangle limb-dependent variations, and that the non-uniform distribution of facular regions across the solar disc further complicates the interpretation, as different active regions contribute unevenly to the integrated signal, effectively smearing any systematic limb-dependent trends. As a result, the filling factors presented here should be interpreted as relative indicators of solar activity rather than precise absolute measurements.

Finally, in Figure \ref{fig:trailedspectrum}, we see indications of the blue-to-red shift of the 4377 \AA\ feature reported by \citet{Thompson2020magneticactivityharpsnpaper2}. While previous studies (e.g., \citet{Thompson2017}; \citet{Costes2026} have successfully extracted velocity information from multiple lines via SRA residuals, our model-derived estimates for this single feature do not show a significant velocity trend. This is likely because our model is optimized to fit the morphology of an individual spectral line, rather than providing a global estimate of the line centroid.

It is likely that many of the limitations discussed in this paper arise from the reliance on an individual spectral line. Figure \ref{fig:muramlines} shows residual MURaM models with spectral lines 4377 \AA\ and 4382 \AA\, at $\mu$ angles of 1.0, 0.5, and 0.1, spanning from disc centre to the limb. Crucially, the amplitude of these features, as well as their relative amplitudes to one another, shifts significantly as a function of limb angle. This differential behaviour suggests that by analysing a broader suite of spectral lines with similar sensitivities, we may be able to further constrain the parameters of facular regions and improve our areal coverage estimations. Such a multi-feature approach could move these methods beyond relative indication of faculae and toward a more accurate physical representation of the solar atmosphere.

\begin{figure*}
	% To include a figure from a file named example.*
	% Allowable file formats are eps or ps if compiling using latex
	% or pdf, png, jpg if compiling using pdflatex
	\includegraphics[width=\textwidth]{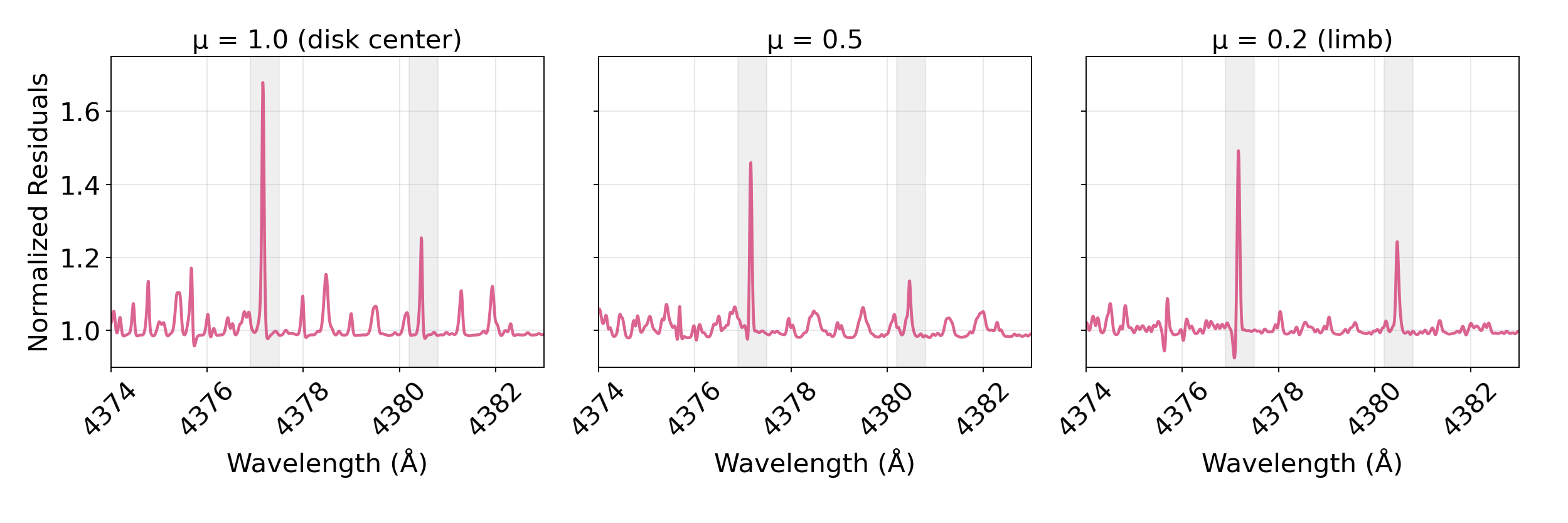}
    \caption{\muram\ model residuals computed with a 1\% faculae filling-factor of the 200~G injected magnetic field model. Left panel shows residuals at disc centre, middle panel shows residuals at $\mu = 0.5$, and the right panel shows residuals at the limb. These residuals highlight the change in amplitudes between the 4377 \AA\ and 4381 \AA\ features, showing how their relative strengths vary considerably with limb angle and, importantly, in different ways to each other.}
    \label{fig:muramlines}
\end{figure*}

\section{Conclusions}\label{sec:conclusions}

In this work, we present a detailed analysis of the Fe\,{\sc i} 4377 \AA\ line using Spectral Ratio Analysis (SRA) in combination with both \phoenix\ and \muram\ synthetic spectra to isolate and model the spectral signatures of solar activity. Following the approach of \citet{Thompson2017}, SRA enables the construction of residual spectra that highlight the impact of activity on individual spectral features. We show that the 4377 \AA\ line tracks solar rotation and is strongly sensitive to variations in faculae, making it a powerful diagnostic of facular activity.

By fitting synthetic \phoenix\ residual spectra -- constructed from combinations of a quiet-Sun (G2) template and a faculae proxy -- to HARPS-N observations, we are able to estimate both facular temperature contrasts and areal filling factors.
Extending this framework to \muram\ models with varying magnetic field strengths and limb angles further strengthens the physical interpretation. We find that the best estimates of faculae filling factor are given by magnetic field strengths of approximately 200\,G, consistent with canonical plage values. 
We conclude that we are insensitive to limb angle and are limited by the use of only one spectral line. Further work should extend to more features to place further constraints on limb angle.
In both approaches, the inferred filling factors show good agreement with Solar Dynamics Observatory (SDO) measurements, demonstrating that this approach can reliably recover the temporal evolution of facular coverage directly from disc-integrated spectra.

The relative performance of the models depends on the level of solar activity. \muram\ models outperform \phoenix\ models at high activity levels, while both models offer consistent estimates in low-activity regimes. These differences likely reflect the underlying physics of each model: \phoenix\ relies primarily on temperature variations and does not explicitly include magnetic fields, whereas \muram\ incorporates magnetic structure and is therefore better suited to reproducing spectra in magnetically active conditions. At the same time, \phoenix\ models may offer a useful approximation in quieter regimes and provide constraints on effective temperature contrasts.

Overall, this work demonstrates that the Fe\,{\sc i} 4377 \AA\ line can be used to trace relative variations in facular filling factors in Sun-as-a-star observations, providing a direct observational pathway to monitoring stellar activity using high-resolution spectra. This line thus provides a diagnostic of activity that can be used in combination with \logrhk\, and provides a better indicator of photospheric variations that imapct RV measurements. This has important implications for radial velocity exoplanet searches, where stellar activity remains a dominant source of noise, as well as for exoplanet atmosphere studies, where unocculted faculae can bias transmission spectra.

However, several limitations remain. Our approach relies on a simplified parametrization of faculae, particularly in the case of \phoenix\ models, where magnetic regions are approximated as temperature perturbations. Even for \muram, the models do not fully reproduce all observed line profile features, indicating that additional physical effects, such as complex temperature gradients, velocity fields, or unresolved magnetic structure, may play a role. Furthermore, this analysis focuses on a single spectral line; extending the method to a broader set of activity-sensitive lines would provide stronger constraints and reduce degeneracies between temperature, magnetic field strength, and filling factor.

Future work will therefore focus on expanding this analysis to multiple spectral features, improving the physical realism of the models, and incorporating more sophisticated treatments of magnetic and radiative transfer effects. Ultimately, developing a comprehensive framework that combines multiple diagnostics will be key to robustly disentangling stellar activity from planetary signals in high-precision spectroscopic observations.

\section*{Acknowledgements}

K.L.H. is supported by a UK Science and Technology Facilities Council (STFC) Studentship (ST/X508706/1). T.R. is supported by an STFC studentship. J.C.C., C.A.W., E.dM., and M.E.Y. would like to acknowledge support from the UK Science and Technology Facilities Council (STFC, grant number ST/X00094X/1). The authors also thank Harry Greatorex and Aisling O'Hare for their assistance with accessing and processing Solar Dynamics Observatory (SDO) data.
AP was supported by grant PI~2102/1-1 from the Deutsche Forschungsgemeinschaft (DFG).
K.S. and A.I.S. acknowledge the support from the European Research Council (ERC) under the European Union’s Horizon 2020 Research and Innovation Programme through grant No.\ 101118581.
The HARPS-N project was funded by the Prodex Program of the Swiss Space Office (SSO), the Harvard University Origin of Life Initiative (HUOLI), the Scottish Universities Physics Alliance (SUPA), the University of Geneva, the Smithsonian Astrophysical Observatory (SAO), the Italian National Astrophysical Institute (INAF), University of St. Andrews, Queen's University Belfast, and University of Edinburgh.
This publication makes use of The Data $\&$ Analysis Center for Exoplanets (DACE), which is a facility based at the University of Geneva (CH) dedicated to extrasolar planets data visualization, exchange and analysis. DACE is a platform of the Swiss National Centre of Competence in Research (NCCR) PlanetS, federating the Swiss expertise in Exoplanet research. The DACE platform is available at \url{https://dace.unige.ch}.
F.R. acknowledges a UK Science and Technology Facilities Council (STFC) small grant ST/Y002334/1.
ACC acknowledges support from STFC consolidated grant number ST/V000861/1
and UKRI/ERC Synergy Grant EP/Z000181/1 (REVEAL).
The Flatiron Institute is a division of the Simons Foundation.

%%%%%%%%%%%%%%%%%%%%%%%%%%%%%%%%%%%%%%%%%%%%%%%%%%
\section*{Data Availability}
HARPS-N data are available at doi: \href{https://doi.org/10.82180/dace-h4s8lp7c}{10.82180/dace-h4s8lp7c}. These data are described in \citep{Dumusque2026}.
SDO data are courtesy of NASA/SDO and the HMI science teams.
PHOENIX models are available from the \href{https://phoenix.astro.physik.uni-goettingen.de/}{G\"{o}ttingen Spectral Library}.

%%%%%%%%%%%%%%%%%%%% REFERENCES %%%%%%%%%%%%%%%%%%

% The best way to enter references is to use BibTeX:

\bibliographystyle{mnras}
\bibliography{zbiblio} % if your bibtex file is called example.bib

@BOOK{Rutten2003RT,
       author = {{Rutten}, Robert J.},
        title = "{Radiative Transfer in Stellar Atmospheres}",
         year = 2003,
       adsurl = {https://ui.adsabs.harvard.edu/abs/2003rtsa.book.....R}
}

@BOOK{Mihalas1978,
       author = {{Mihalas}, Dimitri},
        title = "{Stellar atmospheres}",
         year = 1978,
       adsurl = {https://ui.adsabs.harvard.edu/abs/1978stat.book.....M}
}

@ARTICLE{Gehren2001,
       author = {{Gehren}, T. and {Butler}, K. and {Mashonkina}, L. and {Reetz}, J. and {Shi}, J.},
        title = "{Kinetic equilibrium of iron in the atmospheres of cool dwarf stars. I. The solar strong line spectrum}",
      journal = {\aap},
         year = 2001,
        month = feb,
       volume = {366},
        pages = {981-1002},
          doi = {10.1051/0004-6361:20000287},
       adsurl = {https://ui.adsabs.harvard.edu/abs/2001A&A...366..981G}
}

@ARTICLE{Sowmya2026,
       author = {{Sowmya}, K. and {Shapiro}, A.~I. and {Vasilyev}, V. and {Witzke}, V. and {Collier Cameron}, A. and {Solanki}, S.~K.},
        title = "{Sensitivity of Spectral Lines to Granulation: The Sun}",
      journal = {\apj},
         year = 2026,
        month = may,
       volume = {1003},
       number = {1},
          eid = {28},
        pages = {28},
          doi = {10.3847/1538-4357/ae6102},
archivePrefix = {arXiv},
       eprint = {2509.09824},
 primaryClass = {astro-ph.SR},
       adsurl = {https://ui.adsabs.harvard.edu/abs/2026ApJ..1003...28S}
}

@ARTICLE{VAL1981,
       author = {{Vernazza}, J.~E. and {Avrett}, E.~H. and {Loeser}, R.},
        title = "{Structure of the solar chromosphere. III. Models of the EUV brightness components of the quiet sun.}",
      journal = {\apjs},
         year = 1981,
        month = apr,
       volume = {45},
        pages = {635-725},
          doi = {10.1086/190731},
       adsurl = {https://ui.adsabs.harvard.edu/abs/1981ApJS...45..635V}
}

@ARTICLE{VALD,
       author = {{Piskunov}, N.~E. and {Kupka}, F. and {Ryabchikova}, T.~A. and {Weiss}, W.~W. and {Jeffery}, C.~S.},
        title = "{VALD: The Vienna Atomic Line Data Base.}",
      journal = {\aaps},
         year = 1995,
        month = sep,
       volume = {112},
        pages = {525},
       adsurl = {https://ui.adsabs.harvard.edu/abs/1995A&AS..112..525P}
}

@ARTICLE{Ryabchikova2015,
       author = {{Ryabchikova}, T. and {Piskunov}, N. and {Kurucz}, R.~L. and {Stempels}, H.~C. and {Heiter}, U. and {Pakhomov}, Yu and {Barklem}, P.~S.},
        title = "{A major upgrade of the VALD database}",
      journal = {\physscr},
         year = 2015,
        month = may,
       volume = {90},
       number = {5},
          eid = {054005},
        pages = {054005},
          doi = {10.1088/0031-8949/90/5/054005},
       adsurl = {https://ui.adsabs.harvard.edu/abs/2015PhyS...90e4005R}
}

@misc{Champeau2026,
  author       = {Champeau, C{\'e}dric},
  title        = {{JSol'Ex (astro4j)}},
  year         = {2026},
  note         = {Zenodo, \url{https://doi.org/10.5281/zenodo.19582043}}
}

@ARTICLE{Pietrow2026,
       author = {{Pietrow}, Alexander G.~M.},
        title = "{HelioSpectrotron 5000: an interactive solar atlas}",
      journal = {The Open Journal of Astrophysics},
         year = 2026,
        month = feb,
       volume = {9},
        pages = {58273},
          doi = {10.33232/001c.158273},
archivePrefix = {arXiv},
       eprint = {2602.20101},
 primaryClass = {astro-ph.SR},
       adsurl = {https://ui.adsabs.harvard.edu/abs/2026OJAp....958273P}
}

@ARTICLE{Bhatiaetal2026,
       author = {{Bhatia}, Tanayveer Singh and {Cameron}, Robert H. and {Solanki}, Sami K. and {Przybylski}, Damien F. and {Witzke}, Veronika and {Shapiro}, Alexander and {Kostogryz}, Nadiia},
        title = "{Simulations of facular magnetic fields on cool stars: I. Main-sequence stars with solar metallicity}",
      journal = {\aap},
         year = 2026,
        month = feb,
       volume = {706},
          eid = {A308},
        pages = {A308},
          doi = {10.1051/0004-6361/202555256},
archivePrefix = {arXiv},
       eprint = {2512.22379},
 primaryClass = {astro-ph.SR},
       adsurl = {https://ui.adsabs.harvard.edu/abs/2026A&A...706A.308B}
}

@ARTICLE{SchrijverandHarvey1994,
       author = {{Schrijver}, C.~J. and {Harvey}, K.~L.},
        title = "{The Photospheric Magnetic Flux Budget}",
      journal = {\solphys},
         year = 1994,
        month = mar,
       volume = {150},
        pages = {1},
          doi = {10.1007/BF00712873},
       adsurl = {https://ui.adsabs.harvard.edu/abs/1994SoPh..150....1S}
}

@ARTICLE{Schrijver2020,
       author = {{Schrijver}, Carolus J.},
        title = "{Testing the Solar Activity Paradigm in the Context of Exoplanet Transits}",
      journal = {\apj},
         year = 2020,
        month = feb,
       volume = {890},
       number = {2},
          eid = {121},
        pages = {121},
          doi = {10.3847/1538-4357/ab67c1},
archivePrefix = {arXiv},
       eprint = {2001.01093},
 primaryClass = {astro-ph.SR},
       adsurl = {https://ui.adsabs.harvard.edu/abs/2020ApJ...890..121S}
}

@ARTICLE{Pal2025,
       author = {{V{\'a}radi Nagy}, P. and {Pietrow}, A.~G.~M.},
        title = "{An Atlas of Spectroheliograms from 3641 to 6600 {\r{A}}}",
      journal = {RNAAS},
         year = 2025,
        month = jul,
       volume = {9},
       number = {7},
          eid = {188},
        pages = {188},
          doi = {10.3847/2515-5172/adef50},
archivePrefix = {arXiv},
       eprint = {2507.13025},
 primaryClass = {astro-ph.SR},
       adsurl = {https://ui.adsabs.harvard.edu/abs/2025RNAAS...9..188V}
}

@ARTICLE{Buil2023,
       author = {{Buil}, Christian and {Malherbe}, Jean-Marie and {Maksimovic}, Milan},
        title = "{Sol'Ex et l'imagerie monochromatique solaire}",
      journal = {Photoniques},
         year = 2023,
        month = jul,
       volume = {120},
        pages = {36-40},
          doi = {10.1051/photon/202312036},
       adsurl = {https://ui.adsabs.harvard.edu/abs/2023Phot..120...36B}
}

@ARTICLE{Thompson2020magneticactivityharpsnpaper2,
       author = {{Thompson}, A.~P.~G. and {Watson}, C.~A. and {Haywood}, R.~D. and {Costes}, J.~C. and {de Mooij}, E. and {Collier Cameron}, A. and {Dumusque}, X. and {Phillips}, D.~F. and {Saar}, S.~H. and {Mortier}, A. and {Milbourne}, T.~W. and {Aigrain}, S. and {Cegla}, H.~M. and {Charbonneau}, D. and {Cosentino}, R. and {Ghedina}, A. and {Latham}, D.~W. and {L{\'o}pez-Morales}, M. and {Micela}, G. and {Molinari}, E. and {Poretti}, E. and {Sozzetti}, A. and {Thompson}, S. and {Walsworth}, R.},
        title = "{The spectral impact of magnetic activity on disc-integrated HARPS-N solar observations: exploring new activity indicators}",
      journal = {\mnras},
         year = 2020,
        month = may,
       volume = {494},
       number = {3},
        pages = {4279-4290},
          doi = {10.1093/mnras/staa1010},
archivePrefix = {arXiv},
       eprint = {2004.09830},
 primaryClass = {astro-ph.SR},
       adsurl = {https://ui.adsabs.harvard.edu/abs/2020MNRAS.494.4279T}
}

@INPROCEEDINGS{HARPSN,
       author = {{Cosentino}, Rosario and {Lovis}, Christophe and {Pepe}, Francesco and {Collier Cameron}, Andrew and {Latham}, David W. and {Molinari}, Emilio and {Udry}, Stephane and {Bezawada}, Naidu and {Black}, Martin and {Born}, Andy and {Buchschacher}, Nicolas and {Charbonneau}, Dave and {Figueira}, Pedro and {Fleury}, Michel and {Galli}, Alberto and {Gallie}, Angus and {Gao}, Xiaofeng and {Ghedina}, Adriano and {Gonzalez}, Carlos and {Gonzalez}, Manuel and {Guerra}, Jose and {Henry}, David and {Horne}, Keith and {Hughes}, Ian and {Kelly}, Dennis and {Lodi}, Marcello and {Lunney}, David and {Maire}, Charles and {Mayor}, Michel and {Micela}, Giusi and {Ordway}, Mark P. and {Peacock}, John and {Phillips}, David and {Piotto}, Giampaolo and {Pollacco}, Don and {Queloz}, Didier and {Rice}, Ken and {Riverol}, Carlos and {Riverol}, Luis and {San Juan}, Jose and {Sasselov}, Dimitar and {Segransan}, Damien and {Sozzetti}, Alessandro and {Sosnowska}, Danuta and {Stobie}, Brian and {Szentgyorgyi}, Andrew and {Vick}, Andy and {Weber}, Luc},
        title = "{Harps-N: the new planet hunter at TNG}",
    booktitle = {Ground-based and Airborne Instrumentation for Astronomy IV},
         year = 2012,
       editor = {{McLean}, Ian S. and {Ramsay}, Suzanne K. and {Takami}, Hideki},
       series = {Society of Photo-Optical Instrumentation Engineers (SPIE) Conference Series},
       volume = {8446},
        month = sep,
          eid = {84461V},
        pages = {84461V},
          doi = {10.1117/12.925738},
       adsurl = {https://ui.adsabs.harvard.edu/abs/2012SPIE.8446E..1VC}
}

@INPROCEEDINGS{Phillips2016,
       author = {{Phillips}, David F. and {Glenday}, Alex G. and {Dumusque}, Xavier and {Buchschacher}, Nicolas and {Collier Cameron}, Andrew and {Cecconi}, Massimo and {Charbonneau}, David and {Cosentino}, Rosario and {Ghedina}, Adriano and {Haywood}, Raph{\"a}elle and {Latham}, David W. and {Li}, Chih-Hao and {Lodi}, Marcello and {Lovis}, Christophe and {Molinari}, Emilio and {Pepe}, Francesco and {Sasselov}, Dimitar and {Szentgyorgyi}, Andrew and {Udry}, Stephane and {Walsworth}, Ronald L.},
        title = "{An astro-comb calibrated solar telescope to search for the radial velocity signature of Venus}",
    booktitle = {Advances in Optical and Mechanical Technologies for Telescopes and Instrumentation II},
         year = 2016,
       editor = {{Navarro}, Ram{\'o}n and {Burge}, James H.},
       series = {SPIE},
       volume = {9912},
        month = jul,
          eid = {99126Z},
        pages = {99126Z},
          doi = {10.1117/12.2232452},
       adsurl = {https://ui.adsabs.harvard.edu/abs/2016SPIE.9912E..6ZP}
}

@ARTICLE{Dumusque2015,
       author = {{Dumusque}, Xavier and {Glenday}, Alex and {Phillips}, David F. and {Buchschacher}, Nicolas and {Collier Cameron}, Andrew and {Cecconi}, Massimo and {Charbonneau}, David and {Cosentino}, Rosario and {Ghedina}, Adriano and {Latham}, David W. and {Li}, Chih-Hao and {Lodi}, Marcello and {Lovis}, Christophe and {Molinari}, Emilio and {Pepe}, Francesco and {Udry}, St{\'e}phane and {Sasselov}, Dimitar and {Szentgyorgyi}, Andrew and {Walsworth}, Ronald},
        title = "{HARPS-N Observes the Sun as a Star}",
      journal = {\apjl},
         year = 2015,
        month = dec,
       volume = {814},
       number = {2},
          eid = {L21},
        pages = {L21},
          doi = {10.1088/2041-8205/814/2/L21},
archivePrefix = {arXiv},
       eprint = {1511.02267},
 primaryClass = {astro-ph.EP},
       adsurl = {https://ui.adsabs.harvard.edu/abs/2015ApJ...814L..21D}
}

@ARTICLE{Vogler2005,
       author = {{V{\"o}gler}, A. and {Shelyag}, S. and {Sch{\"u}ssler}, M. and {Cattaneo}, F. and {Emonet}, T. and {Linde}, T.},
        title = "{Simulations of magneto-convection in the solar photosphere.  Equations, methods, and results of the MURaM code}",
      journal = {\aap},
         year = 2005,
        month = jan,
       volume = {429},
        pages = {335-351},
          doi = {10.1051/0004-6361:20041507},
       adsurl = {https://ui.adsabs.harvard.edu/abs/2005A&A...429..335V}
}

@ARTICLE{Oshagh2014,
       author = {{Oshagh}, M. and {Santos}, N.~C. and {Ehrenreich}, D. and {Haghighipour}, N. and {Figueira}, P. and {Santerne}, A. and {Montalto}, M.},
        title = "{Impact of occultations of stellar active regions on transmission spectra. Can occultation of a plage mimic the signature of a blue sky?}",
      journal = {\aap},
         year = 2014,
        month = aug,
       volume = {568},
          eid = {A99},
        pages = {A99},
          doi = {10.1051/0004-6361/201424059},
archivePrefix = {arXiv},
       eprint = {1407.2066},
 primaryClass = {astro-ph.EP},
       adsurl = {https://ui.adsabs.harvard.edu/abs/2014A&A...568A..99O}
}

@ARTICLE{Rackham2018,
       author = {{Rackham}, Benjamin V. and {Apai}, D{\'a}niel and {Giampapa}, Mark S.},
        title = "{The Transit Light Source Effect: False Spectral Features and Incorrect Densities for M-dwarf Transiting Planets}",
      journal = {\apj},
         year = 2018,
        month = feb,
       volume = {853},
       number = {2},
          eid = {122},
        pages = {122},
          doi = {10.3847/1538-4357/aaa08c},
archivePrefix = {arXiv},
       eprint = {1711.05691},
 primaryClass = {astro-ph.EP},
       adsurl = {https://ui.adsabs.harvard.edu/abs/2018ApJ...853..122R}
}

@ARTICLE{Thompson2017,
       author = {{Thompson}, A.~P.~G. and {Watson}, C.~A. and {de Mooij}, E.~J.~W. and {Jess}, D.~B.},
        title = "{The changing face of {\ensuremath{\alpha}} Centauri B: probing plage and stellar activity in K dwarfs}",
      journal = {\mnras},
         year = 2017,
        month = jun,
       volume = {468},
       number = {1},
        pages = {L16-L20},
          doi = {10.1093/mnrasl/slx018},
archivePrefix = {arXiv},
       eprint = {1702.01647},
 primaryClass = {astro-ph.SR},
       adsurl = {https://ui.adsabs.harvard.edu/abs/2017MNRAS.468L..16T}
}

@ARTICLE{Meunier2010,
       author = {{Meunier}, N. and {Desort}, M. and {Lagrange}, A. -M.},
        title = "{Using the Sun to estimate Earth-like planets detection capabilities . II. Impact of plages}",
      journal = {\aap},
         year = 2010,
        month = mar,
       volume = {512},
          eid = {A39},
        pages = {A39},
          doi = {10.1051/0004-6361/200913551},
archivePrefix = {arXiv},
       eprint = {1001.1638},
 primaryClass = {astro-ph.EP},
       adsurl = {https://ui.adsabs.harvard.edu/abs/2010A&A...512A..39M}
}

@ARTICLE{Lakeland2024,
       author = {{Lakeland}, Ben S. and {Naylor}, Tim and {Haywood}, Rapha{\"e}lle D. and {Meunier}, Nad{\`e}ge and {Rescigno}, Federica and {Dalal}, Shweta and {Mortier}, Annelies and {Thompson}, Samantha J. and {Cameron}, Andrew Collier and {Dumusque}, Xavier and {L{\'o}pez-Morales}, Mercedes and {Pepe}, Francesco and {Rice}, Ken and {Sozzetti}, Alessandro and {Udry}, St{\'e}phane and {Ford}, Eric and {Ghedina}, Adriano and {Lodi}, Marcello},
        title = "{The magnetically quiet solar surface dominates HARPS-N solar RVs during low activity}",
      journal = {\mnras},
         year = 2024,
        month = jan,
       volume = {527},
       number = {3},
        pages = {7681-7691},
          doi = {10.1093/mnras/stad3723},
archivePrefix = {arXiv},
       eprint = {2311.16076},
 primaryClass = {astro-ph.SR},
       adsurl = {https://ui.adsabs.harvard.edu/abs/2024MNRAS.527.7681L}
}

@ARTICLE{Cretignier2024,
       author = {{Cretignier}, M. and {Pietrow}, A.~G.~M. and {Aigrain}, S.},
        title = "{Stellar surface information from the Ca II H\&K lines - I. Intensity profiles of the solar activity components}",
      journal = {\mnras},
         year = 2024,
        month = jan,
       volume = {527},
       number = {2},
        pages = {2940-2962},
          doi = {10.1093/mnras/stad3292},
archivePrefix = {arXiv},
       eprint = {2310.15926},
 primaryClass = {astro-ph.SR},
       adsurl = {https://ui.adsabs.harvard.edu/abs/2024MNRAS.527.2940C}
}

@ARTICLE{Haywood2016,
       author = {{Haywood}, R.~D. and {Collier Cameron}, A. and {Unruh}, Y.~C. and {Lovis}, C. and {Lanza}, A.~F. and {Llama}, J. and {Deleuil}, M. and {Fares}, R. and {Gillon}, M. and {Moutou}, C. and {Pepe}, F. and {Pollacco}, D. and {Queloz}, D. and {S{\'e}gransan}, D.},
        title = "{The Sun as a planet-host star: proxies from SDO images for HARPS radial-velocity variations}",
      journal = {\mnras},
         year = 2016,
        month = apr,
       volume = {457},
       number = {4},
        pages = {3637-3651},
          doi = {10.1093/mnras/stw187},
archivePrefix = {arXiv},
       eprint = {1601.05651},
 primaryClass = {astro-ph.EP},
       adsurl = {https://ui.adsabs.harvard.edu/abs/2016MNRAS.457.3637H}
}

@ARTICLE{Milbourne2019,
       author = {{Milbourne}, T.~W. and {Haywood}, R.~D. and {Phillips}, D.~F. and {Saar}, S.~H. and {Cegla}, H.~M. and {Cameron}, A.~C. and {Costes}, J. and {Dumusque}, X. and {Langellier}, N. and {Latham}, D.~W. and {Maldonado}, J. and {Malavolta}, L. and {Mortier}, A. and {Palumbo}, III, M.~L. and {Thompson}, S. and {Watson}, C.~A. and {Bouchy}, F. and {Buchschacher}, N. and {Cecconi}, M. and {Charbonneau}, D. and {Cosentino}, R. and {Ghedina}, A. and {Glenday}, A.~G. and {Gonzalez}, M. and {Li}, C. -H. and {Lodi}, M. and {L{\'o}pez-Morales}, M. and {Lovis}, C. and {Mayor}, M. and {Micela}, G. and {Molinari}, E. and {Pepe}, F. and {Piotto}, G. and {Rice}, K. and {Sasselov}, D. and {S{\'e}gransan}, D. and {Sozzetti}, A. and {Szentgyorgyi}, A. and {Udry}, S. and {Walsworth}, R.~L.},
        title = "{HARPS-N Solar RVs Are Dominated by Large, Bright Magnetic Regions}",
      journal = {\apj},
         year = 2019,
        month = mar,
       volume = {874},
       number = {1},
          eid = {107},
        pages = {107},
          doi = {10.3847/1538-4357/ab064a},
archivePrefix = {arXiv},
       eprint = {1902.04184},
 primaryClass = {astro-ph.SR},
       adsurl = {https://ui.adsabs.harvard.edu/abs/2019ApJ...874..107M}
}

@ARTICLE{Husser2013,
       author = {{Husser}, T. -O. and {Wende-von Berg}, S. and {Dreizler}, S. and {Homeier}, D. and {Reiners}, A. and {Barman}, T. and {Hauschildt}, P.~H.},
        title = "{A new extensive library of PHOENIX stellar atmospheres and synthetic spectra}",
      journal = {\aap},
         year = 2013,
        month = may,
       volume = {553},
          eid = {A6},
        pages = {A6},
          doi = {10.1051/0004-6361/201219058},
archivePrefix = {arXiv},
       eprint = {1303.5632},
 primaryClass = {astro-ph.SR},
       adsurl = {https://ui.adsabs.harvard.edu/abs/2013A&A...553A...6H}
}

@ARTICLE{Fontenla1999,
       author = {{Fontenla}, Juan and {White}, Oran R. and {Fox}, Peter A. and {Avrett}, Eugene H. and {Kurucz}, Robert L.},
        title = "{Calculation of Solar Irradiances. I. Synthesis of the Solar Spectrum}",
      journal = {\apj},
         year = 1999,
        month = jun,
       volume = {518},
       number = {1},
        pages = {480-499},
          doi = {10.1086/307258},
       adsurl = {https://ui.adsabs.harvard.edu/abs/1999ApJ...518..480F}
}

@ARTICLE{Dumusque2014,
       author = {{Dumusque}, X. and {Boisse}, I. and {Santos}, N.~C.},
        title = "{SOAP 2.0: A Tool to Estimate the Photometric and Radial Velocity Variations Induced by Stellar Spots and Plages}",
      journal = {\apj},
         year = 2014,
        month = dec,
       volume = {796},
       number = {2},
          eid = {132},
        pages = {132},
          doi = {10.1088/0004-637X/796/2/132},
archivePrefix = {arXiv},
       eprint = {1409.3594},
 primaryClass = {astro-ph.SR},
       adsurl = {https://ui.adsabs.harvard.edu/abs/2014ApJ...796..132D}
}

@INPROCEEDINGS{Thompson2016,
       author = {{Thompson}, Samantha J. and {Queloz}, Didier and {Baraffe}, Isabelle and {Brake}, Martyn and {Dolgopolov}, Andrey and {Fisher}, Martin and {Fleury}, Michel and {Geelhoed}, Joost and {Hall}, Richard and {Gonz{\'a}lez Hern{\'a}ndez}, Jonay I. and {ter Horst}, Rik and {Kragt}, Jan and {Navarro}, Ram{\'o}n and {Naylor}, Tim and {Pepe}, Francesco and {Piskunov}, Nikolai and {Rebolo}, Rafael and {Sander}, Louis and {S{\'e}gransan}, Damien and {Seneta}, Eugene and {Sing}, David and {Snellen}, Ignas and {Snik}, Frans and {Spronck}, Julien and {Stempels}, Eric and {Sun}, Xiaowei and {Santana Tschudi}, Samuel and {Young}, John},
        title = "{HARPS3 for a roboticized Isaac Newton Telescope}",
    booktitle = {Ground-based and Airborne Instrumentation for Astronomy VI},
         year = 2016,
       editor = {{Evans}, Christopher J. and {Simard}, Luc and {Takami}, Hideki},
       series = {Society of Photo-Optical Instrumentation Engineers (SPIE) Conference Series},
       volume = {9908},
        month = aug,
          eid = {99086F},
        pages = {99086F},
          doi = {10.1117/12.2232111},
archivePrefix = {arXiv},
       eprint = {1608.04611},
 primaryClass = {astro-ph.IM},
       adsurl = {https://ui.adsabs.harvard.edu/abs/2016SPIE.9908E..6FT}
}

@INPROCEEDINGS{Pepe2000,
       author = {{Pepe}, Francesco and {Mayor}, Michel and {Delabre}, Bernard and {Kohler}, Dominique and {Lacroix}, Daniel and {Queloz}, Didier and {Udry}, Stephane and {Benz}, Willy and {Bertaux}, Jean-Loup and {Sivan}, Jean-Pierre},
        title = "{HARPS: a new high-resolution spectrograph for the search of extrasolar planets}",
    booktitle = {Optical and IR Telescope Instrumentation and Detectors},
         year = 2000,
       editor = {{Iye}, Masanori and {Moorwood}, Alan F.},
       series = {Society of Photo-Optical Instrumentation Engineers (SPIE) Conference Series},
       volume = {4008},
        month = aug,
        pages = {582-592},
          doi = {10.1117/12.395516},
       adsurl = {https://ui.adsabs.harvard.edu/abs/2000SPIE.4008..582P}
}

@ARTICLE{Rajpaul2015,
       author = {{Rajpaul}, V. and {Aigrain}, S. and {Osborne}, M.~A. and {Reece}, S. and {Roberts}, S.},
        title = "{A Gaussian process framework for modelling stellar activity signals in radial velocity data}",
      journal = {\mnras},
         year = 2015,
        month = sep,
       volume = {452},
       number = {3},
        pages = {2269-2291},
          doi = {10.1093/mnras/stv1428},
archivePrefix = {arXiv},
       eprint = {1506.07304},
 primaryClass = {astro-ph.EP},
       adsurl = {https://ui.adsabs.harvard.edu/abs/2015MNRAS.452.2269R}
}

@ARTICLE{Klein2024,
       author = {{Klein}, Baptiste and {Aigrain}, Suzanne and {Cretignier}, Michael and {Al Moulla}, Khaled and {Dumusque}, Xavier and {Barrag{\'a}n}, Oscar and {Yu}, Haochuan and {Mortier}, Annelies and {Rescigno}, Federica and {Cameron}, Andrew Collier and {L{\'o}pez-Morales}, Mercedes and {Meunier}, Nad{\`e}ge and {Sozzetti}, Alessandro and {O'Sullivan}, Niamh K.},
        title = "{Investigating stellar activity through eight years of Sun-as-a-star observations}",
      journal = {\mnras},
         year = 2024,
        month = jul,
       volume = {531},
       number = {4},
        pages = {4238-4262},
          doi = {10.1093/mnras/stae1313},
archivePrefix = {arXiv},
       eprint = {2405.12065},
 primaryClass = {astro-ph.EP},
       adsurl = {https://ui.adsabs.harvard.edu/abs/2024MNRAS.531.4238K}
}

@ARTICLE{Rauer2014,
       author = {{Rauer}, H. and {Catala}, C. and {Aerts}, C. and {Appourchaux}, T. and {Benz}, W. and {Brandeker}, A. and {Christensen-Dalsgaard}, J. and {Deleuil}, M. and {Gizon}, L. and {Goupil}, M. -J. and {G{\"u}del}, M. and {Janot-Pacheco}, E. and {Mas-Hesse}, M. and {Pagano}, I. and {Piotto}, G. and {Pollacco}, D. and {Santos}, {\.{C}}. and {Smith}, A. and {Su{\'a}rez}, J. -C. and {Szab{\'o}}, R. and {Udry}, S. and {Adibekyan}, V. and {Alibert}, Y. and {Almenara}, J. -M. and {Amaro-Seoane}, P. and {Eiff}, M. Ammler-von and {Asplund}, M. and {Antonello}, E. and {Barnes}, S. and {Baudin}, F. and {Belkacem}, K. and {Bergemann}, M. and {Bihain}, G. and {Birch}, A.~C. and {Bonfils}, X. and {Boisse}, I. and {Bonomo}, A.~S. and {Borsa}, F. and {Brand{\~a}o}, I.~M. and {Brocato}, E. and {Brun}, S. and {Burleigh}, M. and {Burston}, R. and {Cabrera}, J. and {Cassisi}, S. and {Chaplin}, W. and {Charpinet}, S. and {Chiappini}, C. and {Church}, R.~P. and {Csizmadia}, Sz. and {Cunha}, M. and {Damasso}, M. and {Davies}, M.~B. and {Deeg}, H.~J. and {D{\'\i}az}, R.~F. and {Dreizler}, S. and {Dreyer}, C. and {Eggenberger}, P. and {Ehrenreich}, D. and {Eigm{\"u}ller}, P. and {Erikson}, A. and {Farmer}, R. and {Feltzing}, S. and {de Oliveira Fialho}, F. and {Figueira}, P. and {Forveille}, T. and {Fridlund}, M. and {Garc{\'\i}a}, R.~A. and {Giommi}, P. and {Giuffrida}, G. and {Godolt}, M. and {Gomes da Silva}, J. and {Granzer}, T. and {Grenfell}, J.~L. and {Grotsch-Noels}, A. and {G{\"u}nther}, E. and {Haswell}, C.~A. and {Hatzes}, A.~P. and {H{\'e}brard}, G. and {Hekker}, S. and {Helled}, R. and {Heng}, K. and {Jenkins}, J.~M. and {Johansen}, A. and {Khodachenko}, M.~L. and {Kislyakova}, K.~G. and {Kley}, W. and {Kolb}, U. and {Krivova}, N. and {Kupka}, F. and {Lammer}, H. and {Lanza}, A.~F. and {Lebreton}, Y. and {Magrin}, D. and {Marcos-Arenal}, P. and {Marrese}, P.~M. and {Marques}, J.~P. and {Martins}, J. and {Mathis}, S. and {Mathur}, S. and {Messina}, S. and {Miglio}, A. and {Montalban}, J. and {Montalto}, M. and {Monteiro}, M.~J.~P.~F.~G. and {Moradi}, H. and {Moravveji}, E. and {Mordasini}, C. and {Morel}, T. and {Mortier}, A. and {Nascimbeni}, V. and {Nelson}, R.~P. and {Nielsen}, M.~B. and {Noack}, L. and {Norton}, A.~J. and {Ofir}, A. and {Oshagh}, M. and {Ouazzani}, R. -M. and {P{\'a}pics}, P. and {Parro}, V.~C. and {Petit}, P. and {Plez}, B. and {Poretti}, E. and {Quirrenbach}, A. and {Ragazzoni}, R. and {Raimondo}, G. and {Rainer}, M. and {Reese}, D.~R. and {Redmer}, R. and {Reffert}, S. and {Rojas-Ayala}, B. and {Roxburgh}, I.~W. and {Salmon}, S. and {Santerne}, A. and {Schneider}, J. and {Schou}, J. and {Schuh}, S. and {Schunker}, H. and {Silva-Valio}, A. and {Silvotti}, R. and {Skillen}, I. and {Snellen}, I. and {Sohl}, F. and {Sousa}, S.~G. and {Sozzetti}, A. and {Stello}, D. and {Strassmeier}, K.~G. and {{\v{S}}vanda}, M. and {Szab{\'o}}, Gy. M. and {Tkachenko}, A. and {Valencia}, D. and {Van Grootel}, V. and {Vauclair}, S.~D. and {Ventura}, P. and {Wagner}, F.~W. and {Walton}, N.~A. and {Weingrill}, J. and {Werner}, S.~C. and {Wheatley}, P.~J. and {Zwintz}, K.},
        title = "{The PLATO 2.0 mission}",
      journal = {Experimental Astronomy},
         year = 2014,
        month = nov,
       volume = {38},
       number = {1-2},
        pages = {249-330},
          doi = {10.1007/s10686-014-9383-4},
archivePrefix = {arXiv},
       eprint = {1310.0696},
 primaryClass = {astro-ph.EP},
       adsurl = {https://ui.adsabs.harvard.edu/abs/2014ExA....38..249R}
}

@ARTICLE{Rauer2025,
       author = {{Rauer}, Heike and {Aerts}, Conny and {Cabrera}, Juan and {Deleuil}, Magali and {Erikson}, Anders and {Gizon}, Laurent and {Goupil}, Mariejo and {Heras}, Ana and {Walloschek}, Thomas and {Lorenzo-Alvarez}, Jose and {Marliani}, Filippo and {Martin-Garcia}, C{\'e}sar and {Mas-Hesse}, J. Miguel and {O'Rourke}, Laurence and {Osborn}, Hugh and {Pagano}, Isabella and {Piotto}, Giampaolo and {Pollacco}, Don and {Ragazzoni}, Roberto and {Ramsay}, Gavin and {Udry}, St{\'e}phane and {Appourchaux}, Thierry and {Benz}, Willy and {Brandeker}, Alexis and {G{\"u}del}, Manuel and {Janot-Pacheco}, Eduardo and {Kabath}, Petr and {Kjeldsen}, Hans and {Min}, Michiel and {Santos}, Nuno and {Smith}, Alan and {Suarez}, Juan-Carlos and {Werner}, Stephanie C. and {Aboudan}, Alessio and {Abreu}, Manuel and {Acu{\~n}a}, Lorena and {Adams}, Moritz and {Adibekyan}, Vardan and {Affer}, Laura and {Agneray}, Fran{\c{c}}ois and {Agnor}, Craig and {Aguirre B{\o}rsen-Koch}, Victor and {Ahmed}, Saad and {Aigrain}, Suzanne and {Al-Bahlawan}, Ashraf and {Alcacera Gil}, Ma de los Angeles and {Alei}, Eleonora and {Alencar}, Silvia and {Alexander}, Richard and {Alfonso-Garz{\'o}n}, Julia and {Alibert}, Yann and {Allende Prieto}, Carlos and {Almeida}, Leonardo and {Alonso Sobrino}, Roi and {Altavilla}, Giuseppe and {Althaus}, Christian and {Alvarez Trujillo}, Luis Alonso and {Amarsi}, Anish and {Ammler-von Eiff}, Matthias and {Am{\^o}res}, Eduardo and {Andrade}, Laerte and {Antoniadis-Karnavas}, Alexandros and {Ant{\'o}nio}, Carlos and {Aparicio del Moral}, Beatriz and {Appolloni}, Matteo and {Arena}, Claudio and {Armstrong}, David and {Aroca Aliaga}, Jose and {Asplund}, Martin and {Audenaert}, Jeroen and {Auricchio}, Natalia and {Avelino}, Pedro and {Baeke}, Ann and {Bailli{\'e}}, Kevin and {Balado}, Ana and {Ballber Balaguer{\'o}}, Pau and {Balestra}, Andrea and {Ball}, Warrick and {Ballans}, Herve and {Ballot}, Jerome and {Barban}, Caroline and {Barbary}, Ga{\"e}le and {Barbieri}, Mauro and {Barcel{\'o} Forteza}, Sebasti{\`a} and {Barker}, Adrian and {Barklem}, Paul and {Barnes}, Sydney and {Barrado Navascues}, David and {Barragan}, Oscar and {Baruteau}, Cl{\'e}ment and {Basu}, Sarbani and {Baudin}, Frederic and {Baumeister}, Philipp and {Bayliss}, Daniel and {Bazot}, Michael and {Beck}, Paul G. and {Belkacem}, Kevin and {Bellinger}, Earl and {Benatti}, Serena and {Benomar}, Othman and {B{\'e}rard}, Diane and {Bergemann}, Maria and {Bergomi}, Maria and {Bernardo}, Pierre and {Biazzo}, Katia and {Bignamini}, Andrea and {Bigot}, Lionel and {Billot}, Nicolas and {Binet}, Martin and {Biondi}, David and {Biondi}, Federico and {Birch}, Aaron C. and {Bitsch}, Bertram and {Bluhm Ceballos}, Paz Victoria and {B{\'o}di}, Attila and {Bogn{\'a}r}, Zs{\'o}fia and {Boisse}, Isabelle and {Bolmont}, Emeline and {Bonanno}, Alfio and {Bonavita}, Mariangela and {Bonfanti}, Andrea and {Bonfils}, Xavier and {Bonito}, Rosaria and {Bonomo}, Aldo Stefano and {B{\"o}rner}, Anko and {Boro Saikia}, Sudeshna and {Borreguero Mart{\'\i}n}, Elisa and {Borsa}, Francesco and {Borsato}, Luca and {Bossini}, Diego and {Bouchy}, Francois and {Bou{\'e}}, Gwena{\"e}l and {Boufleur}, Rodrigo and {Boumier}, Patrick and {Bourrier}, Vincent and {Bowman}, Dominic M. and {Bozzo}, Enrico and {Bradley}, Louisa and {Bray}, John and {Bressan}, Alessandro and {Breton}, Sylvain and {Brienza}, Daniele and {Brito}, Ana and {Brogi}, Matteo and {Brown}, Beverly and {Brown}, David J.~A. and {Brun}, Allan Sacha and {Bruno}, Giovanni and {Bruns}, Michael and {Buchhave}, Lars A. and {Bugnet}, Lisa and {Buldgen}, Ga{\"e}l and {Burgess}, Patrick and {Busatta}, Andrea and {Busso}, Giorgia and {Buzasi}, Derek and {Caballero}, Jos{\'e} A. and {Cabral}, Alexandre and {Cabrero Gomez}, Juan-Francisco and {Calderone}, Flavia and {Cameron}, Robert and {Cameron}, Andrew and {Campante}, Tiago and {Campos Gestal}, N{\'e}stor and {Canto Martins}, Bruno Leonardo and {Cara}, Christophe and {Carone}, Ludmila and {Carrasco}, Josep Manel and {Casagrande}, Luca and {Casewell}, Sarah L. and {Cassisi}, Santi and {Castellani}, Marco and {Castro}, Matthieu and {Catala}, Claude and {Catal{\'a}n Fern{\'a}ndez}, Irene and {Catelan}, M{\'a}rcio and {Cegla}, Heather and {Cerruti}, Chiara and {Cessa}, Virginie and {Chadid}, Merieme and {Chaplin}, William and {Charpinet}, Stephane and {Chiappini}, Cristina and {Chiarucci}, Simone and {Chiavassa}, Andrea and {Chinellato}, Simonetta and {Chirulli}, Giovanni and {Christensen-Dalsgaard}, J{\o}rgen and {Church}, Ross and {Claret}, Antonio and {Clarke}, Cathie and {Claudi}, Riccardo and {Clermont}, Lionel and {Coelho}, Hugo and {Coelho}, Joao and {Cogato}, Fabrizio and {Colom{\'e}}, Josep and {Condamin}, Mathieu and {Conde Garc{\'\i}a}, Fernando and {Conseil}, Simon},
        title = "{The PLATO mission}",
      journal = {Experimental Astronomy},
         year = 2025,
        month = jun,
       volume = {59},
       number = {3},
          eid = {26},
        pages = {26},
          doi = {10.1007/s10686-025-09985-9},
archivePrefix = {arXiv},
       eprint = {2406.05447},
 primaryClass = {astro-ph.IM},
       adsurl = {https://ui.adsabs.harvard.edu/abs/2025ExA....59...26R}
}

@ARTICLE{Carlsson2019,
       author = {{Carlsson}, Mats and {De Pontieu}, Bart and {Hansteen}, Viggo H.},
        title = "{New View of the Solar Chromosphere}",
      journal = {\araa},
         year = 2019,
        month = aug,
       volume = {57},
        pages = {189-226},
          doi = {10.1146/annurev-astro-081817-052044},
       adsurl = {https://ui.adsabs.harvard.edu/abs/2019ARA&A..57..189C}
}

@ARTICLE{Meunier2024,
       author = {{Meunier}, Nad{\`e}ge},
        title = "{Impact of stellar variability on exoplanet detectability and characterisation}",
      journal = {Comptes Rendus Physique},
         year = 2024,
        month = jan,
       volume = {24},
       number = {S2},
          eid = {140},
        pages = {140},
          doi = {10.5802/crphys.140},
       adsurl = {https://ui.adsabs.harvard.edu/abs/2024CRPhy..24S.140M}
}

@ARTICLE{herrero2016,
       author = {{Herrero}, Enrique and {Ribas}, Ignasi and {Jordi}, Carme and {Morales}, Juan Carlos and {Perger}, Manuel and {Rosich}, Albert},
        title = "{Modelling the photosphere of active stars for planet detection and characterization}",
      journal = {\aap},
         year = 2016,
        month = feb,
       volume = {586},
          eid = {A131},
        pages = {A131},
          doi = {10.1051/0004-6361/201425369},
archivePrefix = {arXiv},
       eprint = {1511.06717},
 primaryClass = {astro-ph.EP},
       adsurl = {https://ui.adsabs.harvard.edu/abs/2016A&A...586A.131H}
}

@ARTICLE{zhao2023,
       author = {{Zhao}, Y. and {Dumusque}, X.},
        title = "{SOAP-GPU: Efficient spectral modeling of stellar activity using graphical processing units}",
      journal = {\aap},
         year = 2023,
        month = mar,
       volume = {671},
          eid = {A11},
        pages = {A11},
          doi = {10.1051/0004-6361/202244568},
archivePrefix = {arXiv},
       eprint = {2301.04259},
 primaryClass = {astro-ph.SR},
       adsurl = {https://ui.adsabs.harvard.edu/abs/2023A&A...671A..11Z}
}

@ARTICLE{zhao2025,
       author = {{Zhao}, Yinan and {Dumusque}, Xavier and {Cretignier}, Michael and {Al Moulla}, Khaled and {Ellwarth}, Momo and {Reiners}, Ansgar and {Sozzetti}, Alessandro},
        title = "{Precise and efficient modeling of stellar-activity-affected solar spectra using SOAP-GPU}",
      journal = {\aap},
         year = 2025,
        month = jan,
       volume = {693},
          eid = {A262},
        pages = {A262},
          doi = {10.1051/0004-6361/202450993},
archivePrefix = {arXiv},
       eprint = {2412.13500},
 primaryClass = {astro-ph.SR},
       adsurl = {https://ui.adsabs.harvard.edu/abs/2025A&A...693A.262Z}
}

@ARTICLE{petralia2025,
       author = {{Petralia}, Antonino and {Maldonado}, Jes{\'u}s and {Micela}, Giuseppina},
        title = "{PAStar: A model for stellar surface from the Sun to active stars}",
      journal = {\aap},
         year = 2025,
        month = feb,
       volume = {694},
          eid = {A99},
        pages = {A99},
          doi = {10.1051/0004-6361/202450316},
archivePrefix = {arXiv},
       eprint = {2412.10035},
 primaryClass = {astro-ph.SR},
       adsurl = {https://ui.adsabs.harvard.edu/abs/2025A&A...694A..99P}
}

@ARTICLE{hauschildt1999,
       author = {{Hauschildt}, P.~H. and {Baron}, E.},
        title = "{Numerical solution of the expanding stellar atmosphere problem.}",
      journal = {Journal of Computational and Applied Mathematics},
         year = 1999,
        month = sep,
       volume = {109},
       number = {1},
        pages = {41-63},
          doi = {10.48550/arXiv.astro-ph/9808182},
archivePrefix = {arXiv},
       eprint = {astro-ph/9808182},
 primaryClass = {astro-ph},
       adsurl = {https://ui.adsabs.harvard.edu/abs/1999JCoAM.109...41H}
}

@ARTICLE{Huerta2008_starspotref,
       author = {{Huerta}, Marcos and {Johns-Krull}, Christopher M. and {Prato}, L. and {Hartigan}, Patrick and {Jaffe}, D.~T.},
        title = "{Starspot-Induced Radial Velocity Variability in LkCa 19}",
      journal = {\apj},
         year = 2008,
        month = may,
       volume = {678},
       number = {1},
        pages = {472-482},
          doi = {10.1086/526415},
archivePrefix = {arXiv},
       eprint = {0711.2505},
 primaryClass = {astro-ph},
       adsurl = {https://ui.adsabs.harvard.edu/abs/2008ApJ...678..472H}
}

@INPROCEEDINGS{Luhn2024_oscillations,
       author = {{Luhn}, Jacob and {Robertson}, Paul and {Isaacson}, Howard and {Holden}, Bradford},
        title = "{Pushing the (Convective) Envelope: Mitigating Radial Velocity P-mode Oscillations in Subgiants to Reveal Low-amplitude Companions in Evolved Systems}",
    booktitle = {AAS/Division for Extreme Solar Systems Abstracts},
         year = 2024,
       series = {AAS/Division for Extreme Solar Systems Abstracts},
       volume = {56},
        month = apr,
          eid = {601.27},
        pages = {601.27},
       adsurl = {https://ui.adsabs.harvard.edu/abs/2024ESS.....560127L}
}

@misc{Ervin2022,
       author = {{Ervin}, Tamar and {Halverson}, Samuel and {Burrows}, Abigail and {Murphy}, Nei and {Roy}, Arpita and {Haywood}, Raphaelle D. and {Rescigno}, Federica and {Bender}, Chad F. and {Lin}, Andrea S.~J. and {Burt}, Jennifer and {Mahadevan}, Suvrath},
        title = "{SolAster: 'Sun-as-a-star' radial velocity variations}",
 howpublished = {Astrophysics Source Code Library, record ascl:2207.009},
         year = 2022,
        month = jul,
          eid = {ascl:2207.009},
archivePrefix = {ascl},
       eprint = {2207.009},
       adsurl = {https://ui.adsabs.harvard.edu/abs/2022ascl.soft07009E}
}

@ARTICLE{Schou2012,
       author = {{Schou}, J. and {Scherrer}, P.~H. and {Bush}, R.~I. and {Wachter}, R. and {Couvidat}, S. and {Rabello-Soares}, M.~C. and {Bogart}, R.~S. and {Hoeksema}, J.~T. and {Liu}, Y. and {Duvall}, T.~L. and {Akin}, D.~J. and {Allard}, B.~A. and {Miles}, J.~W. and {Rairden}, R. and {Shine}, R.~A. and {Tarbell}, T.~D. and {Title}, A.~M. and {Wolfson}, C.~J. and {Elmore}, D.~F. and {Norton}, A.~A. and {Tomczyk}, S.},
        title = "{Design and Ground Calibration of the Helioseismic and Magnetic Imager (HMI) Instrument on the Solar Dynamics Observatory (SDO)}",
      journal = {\solphys},
         year = 2012,
        month = jan,
       volume = {275},
       number = {1-2},
        pages = {229-259},
          doi = {10.1007/s11207-011-9842-2},
       adsurl = {https://ui.adsabs.harvard.edu/abs/2012SoPh..275..229S}
}

@ARTICLE{Pesnell2012,
       author = {{Pesnell}, W. Dean and {Thompson}, B.~J. and {Chamberlin}, P.~C.},
        title = "{The Solar Dynamics Observatory (SDO)}",
      journal = {\solphys},
         year = 2012,
        month = jan,
       volume = {275},
       number = {1-2},
        pages = {3-15},
          doi = {10.1007/s11207-011-9841-3},
       adsurl = {https://ui.adsabs.harvard.edu/abs/2012SoPh..275....3P}
}

@ARTICLE{Scherrer2012,
       author = {{Scherrer}, P.~H. and {Schou}, J. and {Bush}, R.~I. and {Kosovichev}, A.~G. and {Bogart}, R.~S. and {Hoeksema}, J.~T. and {Liu}, Y. and {Duvall}, T.~L. and {Zhao}, J. and {Title}, A.~M. and {Schrijver}, C.~J. and {Tarbell}, T.~D. and {Tomczyk}, S.},
        title = "{The Helioseismic and Magnetic Imager (HMI) Investigation for the Solar Dynamics Observatory (SDO)}",
      journal = {\solphys},
         year = 2012,
        month = jan,
       volume = {275},
       number = {1-2},
        pages = {207-227},
          doi = {10.1007/s11207-011-9834-2},
       adsurl = {https://ui.adsabs.harvard.edu/abs/2012SoPh..275..207S}
}

@ARTICLE{Couvidat2016,
       author = {{Couvidat}, S. and {Schou}, J. and {Hoeksema}, J.~T. and {Bogart}, R.~S. and {Bush}, R.~I. and {Duvall}, T.~L. and {Liu}, Y. and {Norton}, A.~A. and {Scherrer}, P.~H.},
        title = "{Observables Processing for the Helioseismic and Magnetic Imager Instrument on the Solar Dynamics Observatory}",
      journal = {\solphys},
         year = 2016,
        month = aug,
       volume = {291},
       number = {7},
        pages = {1887-1938},
          doi = {10.1007/s11207-016-0957-3},
archivePrefix = {arXiv},
       eprint = {1606.02368},
 primaryClass = {astro-ph.SR}
}

@ARTICLE{Fischer2016,
       author = {{Fischer}, Debra A. and {Anglada-Escude}, Guillem and {Arriagada}, Pamela and {Baluev}, Roman V. and {Bean}, Jacob L. and {Bouchy}, Francois and {Buchhave}, Lars A. and {Carroll}, Thorsten and {Chakraborty}, Abhijit and {Crepp}, Justin R. and {Dawson}, Rebekah I. and {Diddams}, Scott A. and {Dumusque}, Xavier and {Eastman}, Jason D. and {Endl}, Michael and {Figueira}, Pedro and {Ford}, Eric B. and {Foreman-Mackey}, Daniel and {Fournier}, Paul and {F{\H{u}}r{\'e}sz}, Gabor and {Gaudi}, B. Scott and {Gregory}, Philip C. and {Grundahl}, Frank and {Hatzes}, Artie P. and {H{\'e}brard}, Guillaume and {Herrero}, Enrique and {Hogg}, David W. and {Howard}, Andrew W. and {Johnson}, John A. and {Jorden}, Paul and {Jurgenson}, Colby A. and {Latham}, David W. and {Laughlin}, Greg and {Loredo}, Thomas J. and {Lovis}, Christophe and {Mahadevan}, Suvrath and {McCracken}, Tyler M. and {Pepe}, Francesco and {Perez}, Mario and {Phillips}, David F. and {Plavchan}, Peter P. and {Prato}, Lisa and {Quirrenbach}, Andreas and {Reiners}, Ansgar and {Robertson}, Paul and {Santos}, Nuno C. and {Sawyer}, David and {Segransan}, Damien and {Sozzetti}, Alessandro and {Steinmetz}, Tilo and {Szentgyorgyi}, Andrew and {Udry}, St{\'e}phane and {Valenti}, Jeff A. and {Wang}, Sharon X. and {Wittenmyer}, Robert A. and {Wright}, Jason T.},
        title = "{State of the Field: Extreme Precision Radial Velocities}",
      journal = {\pasp},
         year = 2016,
        month = jun,
       volume = {128},
       number = {964},
        pages = {066001},
          doi = {10.1088/1538-3873/128/964/066001},
archivePrefix = {arXiv},
       eprint = {1602.07939},
 primaryClass = {astro-ph.IM},
       adsurl = {https://ui.adsabs.harvard.edu/abs/2016PASP..128f6001F}
}

@ARTICLE{Crass2021,
       author = {{Crass}, Jonathan and {Gaudi}, B. Scott and {Leifer}, Stephanie and {Beichman}, Charles and {Bender}, Chad and {Blackwood}, Gary and {Burt}, Jennifer A. and {Callas}, John L. and {Cegla}, Heather M. and {Diddams}, Scott A. and {Dumusque}, Xavier and {Eastman}, Jason D. and {Ford}, Eric B. and {Fulton}, Benjamin and {Gibson}, Rose and {Halverson}, Samuel and {Haywood}, Rapha{\"e}lle D. and {Hearty}, Fred and {Howard}, Andrew W. and {Latham}, David W. and {L{\"o}hner-B{\"o}ttcher}, Johannes and {Mamajek}, Eric E. and {Mortier}, Annelies and {Newman}, Patrick and {Plavchan}, Peter and {Quirrenbach}, Andreas and {Reiners}, Ansgar and {Robertson}, Paul and {Roy}, Arpita and {Schwab}, Christian and {Seifahrt}, Andres and {Szentgyorgyi}, Andy and {Terrien}, Ryan and {Teske}, Johanna K. and {Thompson}, Samantha and {Vasisht}, Gautam},
        title = "{Extreme Precision Radial Velocity Working Group Final Report}",
      journal = {arXiv e-prints},
         year = 2021,
        month = jul,
          eid = {arXiv:2107.14291},
        pages = {arXiv:2107.14291},
          doi = {10.48550/arXiv.2107.14291},
archivePrefix = {arXiv},
       eprint = {2107.14291},
 primaryClass = {astro-ph.IM},
       adsurl = {https://ui.adsabs.harvard.edu/abs/2021arXiv210714291C}
}

@ARTICLE{Burt2025,
       author = {{Burt}, Jennifer A. and {Dumusque}, Xavier and {Halverson}, Samuel},
        title = "{Precise Radial Velocities}",
      journal = {arXiv e-prints},
         year = 2025,
        month = nov,
          eid = {arXiv:2511.01954},
        pages = {arXiv:2511.01954},
          doi = {10.48550/arXiv.2511.01954},
archivePrefix = {arXiv},
       eprint = {2511.01954},
 primaryClass = {astro-ph.EP},
       adsurl = {https://ui.adsabs.harvard.edu/abs/2025arXiv251101954B}
}

@ARTICLE{Tabernero2021,
       author = {{Tabernero}, H.~M. and {Zapatero Osorio}, M.~R. and {Allart}, R. and {Borsa}, F. and {Casasayas-Barris}, N. and {Demangeon}, O. and {Ehrenreich}, D. and {Lillo-Box}, J. and {Lovis}, C. and {Pall{\'e}}, E. and {Sousa}, S.~G. and {Rebolo}, R. and {Santos}, N.~C. and {Pepe}, F. and {Cristiani}, S. and {Adibekyan}, V. and {Allende Prieto}, C. and {Alibert}, Y. and {Barros}, S.~C.~C. and {Bouchy}, F. and {Bourrier}, V. and {D'Odorico}, V. and {Dumusque}, X. and {Faria}, J.~P. and {Figueira}, P. and {G{\'e}nova Santos}, R. and {Gonz{\'a}lez Hern{\'a}ndez}, J.~I. and {Hojjatpanah}, S. and {Lo Curto}, G. and {Lavie}, B. and {Martins}, C.~J.~A.~P. and {Martins}, J.~H.~C. and {Mehner}, A. and {Micela}, G. and {Molaro}, P. and {Nunes}, N.~J. and {Poretti}, E. and {Seidel}, J.~V. and {Sozzetti}, A. and {Su{\'a}rez Mascare{\~n}o}, A. and {Udry}, S. and {Aliverti}, M. and {Affolter}, M. and {Alves}, D. and {Amate}, M. and {Avila}, G. and {Bandy}, T. and {Benz}, W. and {Bianco}, A. and {Broeg}, C. and {Cabral}, A. and {Conconi}, P. and {Coelho}, J. and {Cumani}, C. and {Deiries}, S. and {Dekker}, H. and {Delabre}, B. and {Fragoso}, A. and {Genoni}, M. and {Genolet}, L. and {Hughes}, I. and {Knudstrup}, J. and {Kerber}, F. and {Landoni}, M. and {Lizon}, J.~L. and {Maire}, C. and {Manescau}, A. and {Di Marcantonio}, P. and {M{\'e}gevand}, D. and {Monteiro}, M. and {Monteiro}, M. and {Moschetti}, M. and {Mueller}, E. and {Modigliani}, A. and {Oggioni}, L. and {Oliveira}, A. and {Pariani}, G. and {Pasquini}, L. and {Rasilla}, J.~L. and {Redaelli}, E. and {Riva}, M. and {Santana-Tschudi}, S. and {Santin}, P. and {Santos}, P. and {Segovia}, A. and {Sosnowska}, D. and {Span{\`o}}, P. and {Tenegi}, F. and {Iwert}, O. and {Zanutta}, A. and {Zerbi}, F.},
        title = "{ESPRESSO high-resolution transmission spectroscopy of WASP-76 b}",
      journal = {\aap},
         year = 2021,
        month = feb,
       volume = {646},
          eid = {A158},
        pages = {A158},
          doi = {10.1051/0004-6361/202039511},
archivePrefix = {arXiv},
       eprint = {2011.12197},
 primaryClass = {astro-ph.EP},
       adsurl = {https://ui.adsabs.harvard.edu/abs/2021A&A...646A.158T}
}

@ARTICLE{Luhn2023,
       author = {{Luhn}, Jacob K. and {Ford}, Eric B. and {Guo}, Zhao and {Gilbertson}, Christian and {Newman}, Patrick and {Plavchan}, Peter and {Burt}, Jennifer A. and {Teske}, Johanna and {Gupta}, Arvind F.},
        title = "{Impact of Correlated Noise on the Mass Precision of Earth-analog Planets in Radial Velocity Surveys}",
      journal = {\aj},
         year = 2023,
        month = mar,
       volume = {165},
       number = {3},
          eid = {98},
        pages = {98},
          doi = {10.3847/1538-3881/acad08},
archivePrefix = {arXiv},
       eprint = {2204.12512},
 primaryClass = {astro-ph.EP},
       adsurl = {https://ui.adsabs.harvard.edu/abs/2023AJ....165...98L}
}

@ARTICLE{Langellier2021,
       author = {{Langellier}, N. and {Milbourne}, T.~W. and {Phillips}, D.~F. and {Haywood}, R.~D. and {Saar}, S.~H. and {Mortier}, A. and {Malavolta}, L. and {Thompson}, S. and {Collier Cameron}, A. and {Dumusque}, X. and {Cegla}, H.~M. and {Latham}, D.~W. and {Maldonado}, J. and {Watson}, C.~A. and {Buchschacher}, N. and {Cecconi}, M. and {Charbonneau}, D. and {Cosentino}, R. and {Ghedina}, A. and {Gonzalez}, M. and {Li}, C.-H. and {Lodi}, M. and {L{\'o}pez-Morales}, M. and {Micela}, G. and {Molinari}, E. and {Pepe}, F. and {Poretti}, E. and {Rice}, K. and {Sasselov}, D. and {Sozzetti}, A. and {Udry}, S. and {Walsworth}, R.~L.},
        title = "{Detection Limits of Low-mass, Long-period Exoplanets Using Gaussian Processes Applied to HARPS-N Solar Radial Velocities}",
      journal = {\aj},
         year = 2021,
        month = jun,
       volume = {161},
       number = {6},
          eid = {287},
        pages = {287},
          doi = {10.3847/1538-3881/abf1e0},
archivePrefix = {arXiv},
       eprint = {2008.05970},
 primaryClass = {astro-ph.EP},
       adsurl = {https://ui.adsabs.harvard.edu/abs/2021AJ....161..287L}
}

@PHDTHESIS{Gupta2023,
       author = {{Gupta}, Arvind F.},
        title = "{Harnessing the Potential of Radial Velocity Exoplanet Surveys}",
       school = {Pennsylvania State University},
         year = 2023,
        month = aug,
       adsurl = {https://ui.adsabs.harvard.edu/abs/2023PhDT.........4G}
}

@ARTICLE{Pietrow2024,
       author = {{Pietrow}, A.~G.~M. and {Cretignier}, M. and {Druett}, M.~K. and {Alvarado-G{\'o}mez}, J.~D. and {Hofmeister}, S.~J. and {Verma}, M. and {Kamlah}, R. and {Baratella}, M. and {Amazo-G{\'o}mez}, E.~M. and {Kontogiannis}, I. and {Dineva}, E. and {Warmuth}, A. and {Denker}, C. and {Poppenhaeger}, K. and {Andriienko}, O. and {Dumusque}, X. and {L{\"o}fdahl}, M.~G.},
        title = "{A comparative study of two X2.2 and X9.3 solar flares observed with HARPS-N. Reconciling Sun-as-a-star spectroscopy and high-spatial resolution solar observations in the context of the solar-stellar connection}",
      journal = {\aap},
         year = 2024,
        month = feb,
       volume = {682},
          eid = {A46},
        pages = {A46},
          doi = {10.1051/0004-6361/202347895},
archivePrefix = {arXiv},
       eprint = {2309.03373},
 primaryClass = {astro-ph.SR},
       adsurl = {https://ui.adsabs.harvard.edu/abs/2024A&A...682A..46P}
}

@INPROCEEDINGS{John2021,
       author = {{John}, Ancy Anna and {Collier Cameron}, Andrew and {Ford}, Eric and {Shahaf}, Shahar},
        title = "{Mitigating effects of stellar activity in RV using SCALPELS}",
    booktitle = {Posters from the TESS Science Conference II (TSC2)},
         year = 2021,
        month = jul,
          eid = {116},
        pages = {116},
          doi = {10.5281/zenodo.5130046},
       adsurl = {https://ui.adsabs.harvard.edu/abs/2021tsc2.confE.116J}
}

@ARTICLE{Dumusque2026,
       author = {{Dumusque}, X. and {Al Moulla}, K. and {Cretignier}, M. and {Buchschacher}, N. and {Segransan}, D. and {Phillips}, D.~F. and {Affer}, L. and {Aigrain}, S. and {Anna John}, A. and {Bonomo}, A.~S. and {Bourrier}, V. and {Buchhave}, L.~A. and {Collier Cameron}, A. and {Cegla}, H.~M. and {Cort{\'e}s-Zuleta}, P. and {Cosentino}, R. and {Costes}, J. and {Damasso}, M. and {de Beurs}, Z.~L. and {Ehrenreich}, D. and {Ghedina}, A. and {Gonzales}, M. and {Haywood}, R.~D. and {Klein}, B. and {Lakeland}, B.~S. and {Langellier}, N. and {Latham}, D.~W. and {Leleu}, A. and {Lodi}, M. and {Lopez-Morales}, M. and {Lovis}, C. and {Malavolta}, L. and {Maldonado}, J. and {Mantovan}, G. and {Mat{\'\i}nez Fiorenzano}, A.~F. and {Micela}, G. and {Milbourne}, T. and {Molinari}, E. and {Mortier}, A. and {Naponiello}, L. and {Nicholson}, B.~A. and {O'Sullivan}, N.~K. and {Pepe}, F. and {Pinamonti}, M. and {Piotto}, G. and {Rescigno}, F. and {Rice}, K. and {Dimitar}, S. and {Silva}, A.~M. and {Sozzetti}, A. and {Stalport}, M. and {Tavella}, S. and {Udry}, S. and {Vanderburg}, A. and {Vissapragada}, S. and {Watson}, C.~A.},
        title = "{A decade of solar high-fidelity spectroscopy and precise radial velocities from HARPS-N}",
      journal = {\aap},
         year = 2026,
        month = feb,
       volume = {706},
          eid = {A231},
        pages = {A231},
          doi = {10.1051/0004-6361/202557132},
archivePrefix = {arXiv},
       eprint = {2510.27635},
 primaryClass = {astro-ph.SR},
       adsurl = {https://ui.adsabs.harvard.edu/abs/2026A&A...706A.231D}
}

@misc{SILSO_Sunspot_Number,
 author = {{Clette}, F. and {Lefèvre}, L.},
 title = {SILSO Sunspot Number V2.0},
 howpublished = {https://doi.org/10.24414/qnza-ac80},
 month = {07},
 year = {2015},
 note = {Published by WDC SILSO - Royal Observatory of Belgium (ROB)}
}

@ARTICLE{Witzke2021,
       author = {{Witzke}, V. and {Shapiro}, A.~I. and {Cernetic}, M. and {Tagirov}, R.~V. and {Kostogryz}, N.~M. and {Anusha}, L.~S. and {Unruh}, Y.~C. and {Solanki}, S.~K. and {Kurucz}, R.~L.},
        title = "{MPS-ATLAS: A fast all-in-one code for synthesising stellar spectra}",
      journal = {\aap},
         year = 2021,
        month = sep,
       volume = {653},
          eid = {A65},
        pages = {A65},
          doi = {10.1051/0004-6361/202140275},
archivePrefix = {arXiv},
       eprint = {2105.13611},
 primaryClass = {astro-ph.SR},
       adsurl = {https://ui.adsabs.harvard.edu/abs/2021A&A...653A..65W}
}

@ARTICLE{Witzke2022,
       author = {{Witzke}, Veronika and {Shapiro}, Alexander I. and {Kostogryz}, Nadiia M. and {Cameron}, Robert and {Rackham}, Benjamin V. and {Seager}, Sara and {Solanki}, Sami K. and {Unruh}, Yvonne C.},
        title = "{Can 1D Radiative-equilibrium Models of Faculae Be Used for Calculating Contamination of Transmission Spectra?}",
      journal = {\apjl},
         year = 2022,
        month = dec,
       volume = {941},
       number = {2},
          eid = {L35},
        pages = {L35},
          doi = {10.3847/2041-8213/aca671},
archivePrefix = {arXiv},
       eprint = {2211.02860},
 primaryClass = {astro-ph.SR},
       adsurl = {https://ui.adsabs.harvard.edu/abs/2022ApJ...941L..35W}
}

@ARTICLE{Motalebi2015,
       author = {{Motalebi}, F. and {Udry}, S. and {Gillon}, M. and {Lovis}, C. and {S{\'e}gransan}, D. and {Buchhave}, L.~A. and {Demory}, B.~O. and {Malavolta}, L. and {Dressing}, C.~D. and {Sasselov}, D. and {Rice}, K. and {Charbonneau}, D. and {Collier Cameron}, A. and {Latham}, D. and {Molinari}, E. and {Pepe}, F. and {Affer}, L. and {Bonomo}, A.~S. and {Cosentino}, R. and {Dumusque}, X. and {Figueira}, P. and {Fiorenzano}, A.~F.~M. and {Gettel}, S. and {Harutyunyan}, A. and {Haywood}, R.~D. and {Johnson}, J. and {Lopez}, E. and {Lopez-Morales}, M. and {Mayor}, M. and {Micela}, G. and {Mortier}, A. and {Nascimbeni}, V. and {Philips}, D. and {Piotto}, G. and {Pollacco}, D. and {Queloz}, D. and {Sozzetti}, A. and {Vanderburg}, A. and {Watson}, C.~A.},
        title = "{The HARPS-N Rocky Planet Search. I. HD 219134 b: A transiting rocky planet in a multi-planet system at 6.5 pc from the Sun}",
      journal = {\aap},
         year = 2015,
        month = dec,
       volume = {584},
          eid = {A72},
        pages = {A72},
          doi = {10.1051/0004-6361/201526822},
archivePrefix = {arXiv},
       eprint = {1507.08532},
 primaryClass = {astro-ph.EP},
       adsurl = {https://ui.adsabs.harvard.edu/abs/2015A&A...584A..72M}
}

@ARTICLE{Hojjatpanah2019,
       author = {{Hojjatpanah}, S. and {Figueira}, P. and {Santos}, N.~C. and {Adibekyan}, V. and {Sousa}, S.~G. and {Delgado-Mena}, E. and {Alibert}, Y. and {Cristiani}, S. and {Gonz{\'a}lez Hern{\'a}ndez}, J.~I. and {Lanza}, A.~F. and {Di Marcantonio}, P. and {Martins}, J.~H.~C. and {Micela}, G. and {Molaro}, P. and {Neves}, V. and {Oshagh}, M. and {Pepe}, F. and {Poretti}, E. and {Rojas-Ayala}, B. and {Rebolo}, R. and {Su{\'a}rez Mascare{\~n}o}, A. and {Zapatero Osorio}, M.~R.},
        title = "{Catalog for the ESPRESSO blind radial velocity exoplanet survey}",
      journal = {\aap},
         year = 2019,
        month = sep,
       volume = {629},
          eid = {A80},
        pages = {A80},
          doi = {10.1051/0004-6361/201834729},
archivePrefix = {arXiv},
       eprint = {1908.04627},
 primaryClass = {astro-ph.EP},
       adsurl = {https://ui.adsabs.harvard.edu/abs/2019A&A...629A..80H}
}

@ARTICLE{Brewer2020,
       author = {{Brewer}, John M. and {Fischer}, Debra A. and {Blackman}, Ryan T. and {Cabot}, Samuel H.~C. and {Davis}, Allen B. and {Laughlin}, Gregory and {Leet}, Christopher and {Ong}, J.~M. Joel and {Petersburg}, Ryan R. and {Szymkowiak}, Andrew E. and {Zhao}, Lily L. and {Henry}, Gregory W. and {Llama}, Joe},
        title = "{EXPRES. I. HD 3651 as an Ideal RV Benchmark}",
      journal = {\aj},
         year = 2020,
        month = aug,
       volume = {160},
       number = {2},
          eid = {67},
        pages = {67},
          doi = {10.3847/1538-3881/ab99c9},
archivePrefix = {arXiv},
       eprint = {2006.02303},
 primaryClass = {astro-ph.EP},
       adsurl = {https://ui.adsabs.harvard.edu/abs/2020AJ....160...67B}
}

@INPROCEEDINGS{Brewer2022,
       author = {{Brewer}, John and {Zhao}, Lily and {Fischer}, Debra and {Llama}, Joe and {Szymkowiak}, Andrew},
        title = "{Revealing the In-Between: Results from the EXPRES 100 Earths Survey}",
    booktitle = {American Astronomical Society Meeting \#240},
         year = 2022,
       series = {American Astronomical Society Meeting Abstracts},
       volume = {240},
        month = jun,
          eid = {434.02},
        pages = {434.02},
       adsurl = {https://ui.adsabs.harvard.edu/abs/2022AAS...24043402B}
}

@ARTICLE{Gupta2021,
       author = {{Gupta}, Arvind F. and {Wright}, Jason T. and {Robertson}, Paul and {Halverson}, Samuel and {Luhn}, Jacob and {Roy}, Arpita and {Mahadevan}, Suvrath and {Ford}, Eric B. and {Bender}, Chad F. and {Blake}, Cullen H. and {Hearty}, Fred and {Kanodia}, Shubham and {Logsdon}, Sarah E. and {McElwain}, Michael W. and {Monson}, Andrew and {Ninan}, Joe P. and {Schwab}, Christian and {Stef{\'a}nsson}, Gu{\dh}mundur and {Terrien}, Ryan C.},
        title = "{Target Prioritization and Observing Strategies for the NEID Earth Twin Survey}",
      journal = {\aj},
         year = 2021,
        month = mar,
       volume = {161},
       number = {3},
          eid = {130},
        pages = {130},
          doi = {10.3847/1538-3881/abd79e},
archivePrefix = {arXiv},
       eprint = {2101.11689},
 primaryClass = {astro-ph.EP},
       adsurl = {https://ui.adsabs.harvard.edu/abs/2021AJ....161..130G}
}

@INPROCEEDINGS{Gupta2026,
       author = {{Gupta}, Arvind and {Fitzmaurice}, Evan and {Giovinazzi}, Mark and {Robertson}, Paul and {Mahadevan}, Suvrath and {Luhn}, Jacob and {Wright}, Jason and {Bender}, Chad and {Stefansson}, Gudmundur and {The NEID Science Team}},
        title = "{The NEID Earth Twin Survey}",
    booktitle = {American Astronomical Society Meeting Abstracts},
         year = 2026,
       series = {American Astronomical Society Meeting Abstracts},
       volume = {247},
        month = feb,
          eid = {352.03},
        pages = {352.03},
       adsurl = {https://ui.adsabs.harvard.edu/abs/2026AAS...24735203G}
}

@MISC{Hartman2018,
       author = {{Hartman}, Colleen N},
        title = "{The Decadal Survey in Astronomy and Astrophysics 2020 (Astro 2020)}",
 howpublished = {NSF Award Number 1852611. Directorate for Mathematical and Physical Sciences, Division Of Astronomical Sciences. 2018.},
         year = 2018,
        month = nov,
        pages = {52611},
       adsurl = {https://ui.adsabs.harvard.edu/abs/2018nsf....1852611L}
}

@ARTICLE{Zhao2022,
       author = {{Zhao}, Lily L. and {Fischer}, Debra A. and {Ford}, Eric B. and {Wise}, Alex and {Cretignier}, Micha{\"e}l and {Aigrain}, Suzanne and {Barragan}, Oscar and {Bedell}, Megan and {Buchhave}, Lars A. and {Camacho}, Jo{\~a}o D. and {Cegla}, Heather M. and {Cisewski-Kehe}, Jessi and {Collier Cameron}, Andrew and {de Beurs}, Zoe L. and {Dodson-Robinson}, Sally and {Dumusque}, Xavier and {Faria}, Jo{\~a}o P. and {Gilbertson}, Christian and {Haley}, Charlotte and {Harrell}, Justin and {Hogg}, David W. and {Holzer}, Parker and {John}, Ancy Anna and {Klein}, Baptiste and {Lafarga}, Marina and {Lienhard}, Florian and {Maguire-Rajpaul}, Vinesh and {Mortier}, Annelies and {Nicholson}, Belinda and {Palumbo}, Michael L. and {Ramirez Delgado}, Victor and {Shallue}, Christopher J. and {Vanderburg}, Andrew and {Viana}, Pedro T.~P. and {Zhao}, Jinglin and {Zicher}, Norbert and {Cabot}, Samuel H.~C. and {Henry}, Gregory W. and {Roettenbacher}, Rachael M. and {Brewer}, John M. and {Llama}, Joe and {Petersburg}, Ryan R. and {Szymkowiak}, Andrew E.},
        title = "{The EXPRES Stellar Signals Project II. State of the Field in Disentangling Photospheric Velocities}",
      journal = {\aj},
         year = 2022,
        month = apr,
       volume = {163},
       number = {4},
          eid = {171},
        pages = {171},
          doi = {10.3847/1538-3881/ac5176},
archivePrefix = {arXiv},
       eprint = {2201.10639},
 primaryClass = {astro-ph.EP},
       adsurl = {https://ui.adsabs.harvard.edu/abs/2022AJ....163..171Z}
}

@ARTICLE{Unruh1999,
       author = {{Unruh}, Y.~C. and {Solanki}, S.~K. and {Fligge}, M.},
        title = "{The spectral dependence of facular contrast and solar irradiance variations}",
      journal = {\aap},
         year = 1999,
        month = may,
       volume = {345},
        pages = {635-642},
       adsurl = {https://ui.adsabs.harvard.edu/abs/1999A&A...345..635U}
}

@ARTICLE{Dravins1981,
       author = {{Dravins}, D. and {Lindegren}, L. and {Nordlund}, A.},
        title = "{Solar granulation - Influence of convection on spectral line asymmetries and wavelength shifts}",
      journal = {\aap},
         year = 1981,
        month = mar,
       volume = {96},
       number = {1-2},
        pages = {345-364},
       adsurl = {https://ui.adsabs.harvard.edu/abs/1981A&A....96..345D}
}

@ARTICLE{Cameron2019,
       author = {{Collier Cameron}, A. and {Mortier}, A. and {Phillips}, D. and {Dumusque}, X. and {Haywood}, R.~D. and {Langellier}, N. and {Watson}, C.~A. and {Cegla}, H.~M. and {Costes}, J. and {Charbonneau}, D. and {Coffinet}, A. and {Latham}, D.~W. and {Lopez-Morales}, M. and {Malavolta}, L. and {Maldonado}, J. and {Micela}, G. and {Milbourne}, T. and {Molinari}, E. and {Saar}, S.~H. and {Thompson}, S. and {Buchschacher}, N. and {Cecconi}, M. and {Cosentino}, R. and {Ghedina}, A. and {Glenday}, A. and {Gonzalez}, M. and {Li}, C.-H. and {Lodi}, M. and {Lovis}, C. and {Pepe}, F. and {Poretti}, E. and {Rice}, K. and {Sasselov}, D. and {Sozzetti}, A. and {Szentgyorgyi}, A. and {Udry}, S. and {Walsworth}, R.},
        title = "{Three years of Sun-as-a-star radial-velocity observations on the approach to solar minimum}",
      journal = {\mnras},
         year = 2019,
        month = jul,
       volume = {487},
       number = {1},
        pages = {1082-1100},
          doi = {10.1093/mnras/stz1215},
archivePrefix = {arXiv},
       eprint = {1904.12186},
 primaryClass = {astro-ph.SR},
       adsurl = {https://ui.adsabs.harvard.edu/abs/2019MNRAS.487.1082C}
}

@ARTICLE{Ribas2018,
       author = {{Ribas}, I. and {Tuomi}, M. and {Reiners}, A. and {Butler}, R.~P. and {Morales}, J.~C. and {Perger}, M. and {Dreizler}, S. and {Rodr{\'\i}guez-L{\'o}pez}, C. and {Gonz{\'a}lez Hern{\'a}ndez}, J.~I. and {Rosich}, A. and {Feng}, F. and {Trifonov}, T. and {Vogt}, S.~S. and {Caballero}, J.~A. and {Hatzes}, A. and {Herrero}, E. and {Jeffers}, S.~V. and {Lafarga}, M. and {Murgas}, F. and {Nelson}, R.~P. and {Rodr{\'\i}guez}, E. and {Strachan}, J.~B.~P. and {Tal-Or}, L. and {Teske}, J. and {Toledo-Padr{\'o}n}, B. and {Zechmeister}, M. and {Quirrenbach}, A. and {Amado}, P.~J. and {Azzaro}, M. and {B{\'e}jar}, V.~J.~S. and {Barnes}, J.~R. and {Berdi{\~n}as}, Z.~M. and {Burt}, J. and {Coleman}, G. and {Cort{\'e}s-Contreras}, M. and {Crane}, J. and {Engle}, S.~G. and {Guinan}, E.~F. and {Haswell}, C.~A. and {Henning}, Th. and {Holden}, B. and {Jenkins}, J. and {Jones}, H.~R.~A. and {Kaminski}, A. and {Kiraga}, M. and {K{\"u}rster}, M. and {Lee}, M.~H. and {L{\'o}pez-Gonz{\'a}lez}, M.~J. and {Montes}, D. and {Morin}, J. and {Ofir}, A. and {Pall{\'e}}, E. and {Rebolo}, R. and {Reffert}, S. and {Schweitzer}, A. and {Seifert}, W. and {Shectman}, S.~A. and {Staab}, D. and {Street}, R.~A. and {Su{\'a}rez Mascare{\~n}o}, A. and {Tsapras}, Y. and {Wang}, S.~X. and {Anglada-Escud{\'e}}, G.},
        title = "{A candidate super-Earth planet orbiting near the snow line of Barnard's star}",
      journal = {\nat},
         year = 2018,
        month = nov,
       volume = {563},
       number = {7731},
        pages = {365-368},
          doi = {10.1038/s41586-018-0677-y},
archivePrefix = {arXiv},
       eprint = {1811.05955},
 primaryClass = {astro-ph.EP},
       adsurl = {https://ui.adsabs.harvard.edu/abs/2018Natur.563..365R}
}

@INPROCEEDINGS{Lubin2022,
       author = {{Lubin}, Jack and {HPF Science}},
        title = "{Stellar Activity Manifesting at a One-year Alias Explains Barnard b as a False Positive}",
    booktitle = {Bulletin of the American Astronomical Society},
         year = 2022,
       volume = {54},
        month = jun,
          eid = {403.02},
        pages = {403.02},
       adsurl = {https://ui.adsabs.harvard.edu/abs/2022BAAS...54e4302L}
}

@ARTICLE{Robertson2015,
       author = {{Robertson}, Paul and {Roy}, Arpita and {Mahadevan}, Suvrath},
        title = "{Stellar Activity Mimics a Habitable-zone Planet around Kapteyn's Star}",
      journal = {\apjl},
         year = 2015,
        month = jun,
       volume = {805},
       number = {2},
          eid = {L22},
        pages = {L22},
          doi = {10.1088/2041-8205/805/2/L22},
archivePrefix = {arXiv},
       eprint = {1505.02778},
 primaryClass = {astro-ph.EP},
       adsurl = {https://ui.adsabs.harvard.edu/abs/2015ApJ...805L..22R}
}

@INPROCEEDINGS{Ji2019,
       author = {{Ji}, Jinbiao and {Fausey}, Hallie and {Bortle}, Anna and {Dodson-Robinson}, Sarah E. and {Gizis}, John},
        title = "{A Gaussian Process Regression Reveals No Evidence for Planets Orbiting Kapteyn's Star}",
    booktitle = {American Astronomical Society Meeting Abstracts \#234},
         year = 2019,
       series = {American Astronomical Society Meeting Abstracts},
       volume = {234},
        month = jun,
          eid = {109.01},
        pages = {109.01},
       adsurl = {https://ui.adsabs.harvard.edu/abs/2019AAS...23410901J}
}

@ARTICLE{Jenkins2013,
       author = {{Jenkins}, J.~S. and {Tuomi}, M. and {Brasser}, R. and {Ivanyuk}, O. and {Murgas}, F.},
        title = "{Two Super-Earths Orbiting the Solar Analog HD 41248 on the Edge of a 7:5 Mean Motion Resonance}",
      journal = {\apj},
         year = 2013,
        month = jul,
       volume = {771},
       number = {1},
          eid = {41},
        pages = {41},
          doi = {10.1088/0004-637X/771/1/41},
archivePrefix = {arXiv},
       eprint = {1304.7374},
 primaryClass = {astro-ph.EP},
       adsurl = {https://ui.adsabs.harvard.edu/abs/2013ApJ...771...41J}
}

@ARTICLE{Jenkins2014,
       author = {{Jenkins}, J.~S. and {Tuomi}, M.},
        title = "{The Curious Case of HD 41248. A Pair of Static Signals Buried Behind Red Noise}",
      journal = {\apj},
         year = 2014,
        month = oct,
       volume = {794},
       number = {2},
          eid = {110},
        pages = {110},
          doi = {10.1088/0004-637X/794/2/110},
archivePrefix = {arXiv},
       eprint = {1406.3093},
 primaryClass = {astro-ph.EP},
       adsurl = {https://ui.adsabs.harvard.edu/abs/2014ApJ...794..110J}
}

@ARTICLE{Santos2014,
       author = {{Santos}, N.~C. and {Mortier}, A. and {Faria}, J.~P. and {Dumusque}, X. and {Adibekyan}, V. Zh. and {Delgado-Mena}, E. and {Figueira}, P. and {Benamati}, L. and {Boisse}, I. and {Cunha}, D. and {Gomes da Silva}, J. and {Lo Curto}, G. and {Lovis}, C. and {Martins}, J.~H.~C. and {Mayor}, M. and {Melo}, C. and {Oshagh}, M. and {Pepe}, F. and {Queloz}, D. and {Santerne}, A. and {S{\'e}gransan}, D. and {Sozzetti}, A. and {Sousa}, S.~G. and {Udry}, S.},
        title = "{The HARPS search for southern extra-solar planets. XXXV. The interesting case of HD 41248: stellar activity, no planets?}",
      journal = {\aap},
         year = 2014,
        month = jun,
       volume = {566},
          eid = {A35},
        pages = {A35},
          doi = {10.1051/0004-6361/201423808},
archivePrefix = {arXiv},
       eprint = {1404.6135},
 primaryClass = {astro-ph.EP},
       adsurl = {https://ui.adsabs.harvard.edu/abs/2014A&A...566A..35S}
}

@ARTICLE{Feng2017b,
       author = {{Feng}, F. and {Tuomi}, M. and {Jones}, H.~R.~A.},
        title = "{Agatha: disentangling periodic signals from correlated noise in a periodogram framework}",
      journal = {\mnras},
         year = 2017,
        month = oct,
       volume = {470},
       number = {4},
        pages = {4794-4814},
          doi = {10.1093/mnras/stx1126},
archivePrefix = {arXiv},
       eprint = {1705.03089},
 primaryClass = {astro-ph.EP},
       adsurl = {https://ui.adsabs.harvard.edu/abs/2017MNRAS.470.4794F}
}

@ARTICLE{Faria2020,
       author = {{Faria}, J.~P. and {Adibekyan}, V. and {Amazo-G{\'o}mez}, E.~M. and {Barros}, S.~C.~C. and {Camacho}, J.~D. and {Demangeon}, O. and {Figueira}, P. and {Mortier}, A. and {Oshagh}, M. and {Pepe}, F. and {Santos}, N.~C. and {Gomes da Silva}, J. and {Costa Silva}, A.~R. and {Sousa}, S.~G. and {Ulmer-Moll}, S. and {Viana}, P.~T.~P.},
        title = "{Decoding the radial velocity variations of HD 41248 with ESPRESSO}",
      journal = {\aap},
         year = 2020,
        month = mar,
       volume = {635},
          eid = {A13},
        pages = {A13},
          doi = {10.1051/0004-6361/201936389},
archivePrefix = {arXiv},
       eprint = {1911.11714},
 primaryClass = {astro-ph.EP},
       adsurl = {https://ui.adsabs.harvard.edu/abs/2020A&A...635A..13F}
}

@ARTICLE{Spruit1976,
       author = {{Spruit}, H.~C.},
        title = "{Pressure equilibrium and energy balance of small photospheric fluxtubes.}",
      journal = {\solphys},
         year = 1976,
        month = nov,
       volume = {50},
       number = {2},
        pages = {269-295},
          doi = {10.1007/BF00155292},
       adsurl = {https://ui.adsabs.harvard.edu/abs/1976SoPh...50..269S}
}

@ARTICLE{Fontenla1993,
       author = {{Fontenla}, J.~M. and {Avrett}, E.~H. and {Loeser}, R.},
        title = "{Energy Balance in the Solar Transition Region. III. Helium Emission in Hydrostatic, Constant-Abundance Models with Diffusion}",
      journal = {\apj},
         year = 1993,
        month = mar,
       volume = {406},
        pages = {319},
          doi = {10.1086/172443},
       adsurl = {https://ui.adsabs.harvard.edu/abs/1993ApJ...406..319F}
}

@ARTICLE{Hathaway2015,
       author = {{Hathaway}, David H.},
        title = "{The Solar Cycle}",
      journal = {Living Reviews in Solar Physics},
         year = 2015,
        month = dec,
       volume = {12},
       number = {1},
          eid = {4},
        pages = {4},
          doi = {10.1007/lrsp-2015-4},
archivePrefix = {arXiv},
       eprint = {1502.07020},
 primaryClass = {astro-ph.SR},
       adsurl = {https://ui.adsabs.harvard.edu/abs/2015LRSP...12....4H}
}

@INPROCEEDINGS{Boisse2011,
       author = {{Boisse}, Isabelle and {Bouchy}, Fran{\c{c}}ois and {H{\'e}brard}, Guillaume and {Bonfils}, Xavier and {Santos}, Nuno and {Vauclair}, Sylvie},
        title = "{Disentangling between stellar activity and planetary signals}",
    booktitle = {Physics of Sun and Star Spots},
         year = 2011,
       editor = {{Prasad Choudhary}, Debi and {Strassmeier}, Klaus G.},
       series = {IAU Symposium},
       volume = {273},
        month = aug,
        pages = {281-285},
          doi = {10.1017/S1743921311015389},
archivePrefix = {arXiv},
       eprint = {1012.1452},
 primaryClass = {astro-ph.EP},
       adsurl = {https://ui.adsabs.harvard.edu/abs/2011IAUS..273..281B}
}

@ARTICLE{Lienhard2023,
       author = {{Lienhard}, F. and {Mortier}, A. and {Cegla}, H.~M. and {Cameron}, A. Collier and {Klein}, B. and {Watson}, C.~A.},
        title = "{Unsigned magnetic flux proxy from solar optical intensity spectra}",
      journal = {\mnras},
         year = 2023,
        month = jul,
       volume = {522},
       number = {4},
        pages = {5862-5878},
          doi = {10.1093/mnras/stad1343},
archivePrefix = {arXiv},
       eprint = {2305.03522},
 primaryClass = {astro-ph.EP},
       adsurl = {https://ui.adsabs.harvard.edu/abs/2023MNRAS.522.5862L}
}

@ARTICLE{Queloz2001,
       author = {{Queloz}, D. and {Henry}, G.~W. and {Sivan}, J.~P. and {Baliunas}, S.~L. and {Beuzit}, J.~L. and {Donahue}, R.~A. and {Mayor}, M. and {Naef}, D. and {Perrier}, C. and {Udry}, S.},
        title = "{No planet for HD 166435}",
      journal = {\aap},
         year = 2001,
        month = nov,
       volume = {379},
        pages = {279-287},
          doi = {10.1051/0004-6361:20011308},
archivePrefix = {arXiv},
       eprint = {astro-ph/0109491},
 primaryClass = {astro-ph},
       adsurl = {https://ui.adsabs.harvard.edu/abs/2001A&A...379..279Q}
}

@ARTICLE{Anglada2014,
       author = {{Anglada-Escude}, G. and {Arriagada}, P. and {Tuomi}, M. and {Zechmeister}, M. and {Jenkins}, J.~S. and {Ofir}, A. and {Dreizler}, S. and {Gerlach}, E. and {Marvin}, C.~J. and {Reiners}, A. and {Jeffers}, S.~V. and {Butler}, R.~P. and {Vogt}, S.~S. and {Amado}, P.~J. and {Rodriguez-Lopez}, C. and {Berdinas}, Z.~M. and {Morin}, J. and {Crane}, J.~D. and {Shectman}, S.~A. and {Thompson}, I.~B. and {Diaz}, M. and {Rivera}, E. and {Sarmiento}, L.~F. and {Jones}, H.~R.~A.},
        title = "{Two planets around Kapteyn's star: a cold and a temperate super-Earth orbiting the nearest halo red dwarf.}",
      journal = {\mnras},
         year = 2014,
        month = sep,
       volume = {443},
        pages = {L89-L93},
          doi = {10.1093/mnrasl/slu076},
archivePrefix = {arXiv},
       eprint = {1406.0818},
 primaryClass = {astro-ph.EP},
       adsurl = {https://ui.adsabs.harvard.edu/abs/2014MNRAS.443L..89A}
}

@ARTICLE{Liang2024,
       author = {{Liang}, Yan and {Winn}, Joshua N. and {Melchior}, Peter},
        title = "{AESTRA: Deep Learning for Precise Radial Velocity Estimation in the Presence of Stellar Activity}",
      journal = {\aj},
         year = 2024,
        month = jan,
       volume = {167},
       number = {1},
          eid = {23},
        pages = {23},
          doi = {10.3847/1538-3881/ad0e01},
archivePrefix = {arXiv},
       eprint = {2311.18326},
 primaryClass = {astro-ph.EP},
       adsurl = {https://ui.adsabs.harvard.edu/abs/2024AJ....167...23L}
}

@ARTICLE{Gilbertson2024,
       author = {{Gilbertson}, Christian and {Ford}, Eric B. and {Halverson}, Samuel and {Fitzmaurice}, Evan and {Blake}, Cullen H. and {Stef{\'a}nsson}, Gu{\dh}mundur and {Mahadevan}, Suvrath and {Wright}, Jason T. and {Luhn}, Jacob K. and {Ninan}, Joe P. and {Robertson}, Paul and {Roy}, Arpita and {Schwab}, Christian and {Terrien}, Ryan C.},
        title = "{Data-Driven Modeling of Telluric Features and Stellar Variability with StellarSpectraObservationFitting.jl}",
      journal = {arXiv e-prints},
         year = 2024,
        month = aug,
          eid = {arXiv:2408.17289},
        pages = {arXiv:2408.17289},
          doi = {10.48550/arXiv.2408.17289},
archivePrefix = {arXiv},
       eprint = {2408.17289},
 primaryClass = {astro-ph.EP},
       adsurl = {https://ui.adsabs.harvard.edu/abs/2024arXiv240817289G}
}

@BOOK{Gray2005,
       author = {{Gray}, David F.},
        title = "{The Observation and Analysis of Stellar Photospheres}",
         year = 2005,
          doi = {10.1017/CBO9781316036570},
       adsurl = {https://ui.adsabs.harvard.edu/abs/2005oasp.book.....G}
}

@ARTICLE{Claret2000,
       author = {{Claret}, A.},
        title = "{A new non-linear limb-darkening law for LTE stellar atmosphere models. Calculations for -5.0 <= log[M/H] <= +1, 2000 K <= T$_{eff}$ <= 50000 K at several surface gravities}",
      journal = {\aap},
         year = 2000,
        month = nov,
       volume = {363},
        pages = {1081-1190},
       adsurl = {https://ui.adsabs.harvard.edu/abs/2000A&A...363.1081C}
}

@ARTICLE{VoeglerandSchuessler2007,
       author = {{V{\"o}gler}, A. and {Sch{\"u}ssler}, M.},
        title = "{A solar surface dynamo}",
      journal = {\aap},
         year = 2007,
        month = apr,
       volume = {465},
       number = {3},
        pages = {L43-L46},
          doi = {10.1051/0004-6361:20077253},
archivePrefix = {arXiv},
       eprint = {astro-ph/0702681},
 primaryClass = {astro-ph},
       adsurl = {https://ui.adsabs.harvard.edu/abs/2007A&A...465L..43V}
}

@ARTICLE{SchuesslerandVoegler2008,
       author = {{Sch{\"u}ssler}, M. and {V{\"o}gler}, A.},
        title = "{Strong horizontal photospheric magnetic field in a surface dynamo simulation}",
      journal = {\aap},
         year = 2008,
        month = apr,
       volume = {481},
       number = {1},
        pages = {L5-L8},
          doi = {10.1051/0004-6361:20078998},
archivePrefix = {arXiv},
       eprint = {0801.1250},
 primaryClass = {astro-ph},
       adsurl = {https://ui.adsabs.harvard.edu/abs/2008A&A...481L...5S}
}

@ARTICLE{Rempel2014,
       author = {{Rempel}, M.},
        title = "{Numerical Simulations of Quiet Sun Magnetism: On the Contribution from a Small-scale Dynamo}",
      journal = {\apj},
         year = 2014,
        month = jul,
       volume = {789},
       number = {2},
          eid = {132},
        pages = {132},
          doi = {10.1088/0004-637X/789/2/132},
archivePrefix = {arXiv},
       eprint = {1405.6814},
 primaryClass = {astro-ph.SR},
       adsurl = {https://ui.adsabs.harvard.edu/abs/2014ApJ...789..132R}
}

@ARTICLE{Witzke2024,
       author = {{Witzke}, Veronika and {Shapiro}, Alexander I. and {Kostogryz}, Nadiia M. and {Mauviard}, Lucien and {Bhatia}, Tanayveer S. and {Cameron}, Robert and {Gizon}, Laurent and {Przybylski}, Damien and {Solanki}, Sami K. and {Unruh}, Yvonne C. and {Yue}, Li},
        title = "{Testing MURaM and MPS-ATLAS against the quiet solar spectrum}",
      journal = {\aap},
         year = 2024,
        month = jan,
       volume = {681},
          eid = {A81},
        pages = {A81},
          doi = {10.1051/0004-6361/202346099},
archivePrefix = {arXiv},
       eprint = {2310.05652},
 primaryClass = {astro-ph.SR},
       adsurl = {https://ui.adsabs.harvard.edu/abs/2024A&A...681A..81W}
}

@ARTICLE{Kostogryz2024,
       author = {{Kostogryz}, Nadiia M. and {Shapiro}, Alexander I. and {Witzke}, Veronika and {Cameron}, Robert H. and {Gizon}, Laurent and {Krivova}, Natalie A. and {Ludwig}, Hans-G. and {Maxted}, Pierre F.~L. and {Seager}, Sara and {Solanki}, Sami K. and {Valenti}, Jeff},
        title = "{Magnetic origin of the discrepancy between stellar limb-darkening models and observations}",
      journal = {Nature Astronomy},
         year = 2024,
        month = jul,
       volume = {8},
        pages = {929-937},
          doi = {10.1038/s41550-024-02252-5},
       adsurl = {https://ui.adsabs.harvard.edu/abs/2024NatAs...8..929K}
}

@INPROCEEDINGS{Cosentino2012,
       author = {{Cosentino}, Rosario and {Lovis}, Christophe and {Pepe}, Francesco and {Collier Cameron}, Andrew and {Latham}, David W. and {Molinari}, Emilio and {Udry}, Stephane and {Bezawada}, Naidu and {Black}, Martin and {Born}, Andy and {Buchschacher}, Nicolas and {Charbonneau}, Dave and {Figueira}, Pedro and {Fleury}, Michel and {Galli}, Alberto and {Gallie}, Angus and {Gao}, Xiaofeng and {Ghedina}, Adriano and {Gonzalez}, Carlos and {Gonzalez}, Manuel and {Guerra}, Jose and {Henry}, David and {Horne}, Keith and {Hughes}, Ian and {Kelly}, Dennis and {Lodi}, Marcello and {Lunney}, David and {Maire}, Charles and {Mayor}, Michel and {Micela}, Giusi and {Ordway}, Mark P. and {Peacock}, John and {Phillips}, David and {Piotto}, Giampaolo and {Pollacco}, Don and {Queloz}, Didier and {Rice}, Ken and {Riverol}, Carlos and {Riverol}, Luis and {San Juan}, Jose and {Sasselov}, Dimitar and {Segransan}, Damien and {Sozzetti}, Alessandro and {Sosnowska}, Danuta and {Stobie}, Brian and {Szentgyorgyi}, Andrew and {Vick}, Andy and {Weber}, Luc},
        title = "{Harps-N: the new planet hunter at TNG}",
    booktitle = {Ground-based and Airborne Instrumentation for Astronomy IV},
         year = 2012,
       editor = {{McLean}, Ian S. and {Ramsay}, Suzanne K. and {Takami}, Hideki},
       series = {Society of Photo-Optical Instrumentation Engineers (SPIE) Conference Series},
       volume = {8446},
        month = sep,
          eid = {84461V},
        pages = {84461V},
          doi = {10.1117/12.925738},
       adsurl = {https://ui.adsabs.harvard.edu/abs/2012SPIE.8446E..1VC}
}

@INPROCEEDINGS{Jurgenson2016,
       author = {{Jurgenson}, C. and {Fischer}, D. and {McCracken}, T. and {Sawyer}, D. and {Szymkowiak}, A. and {Davis}, A. and {Muller}, G. and {Santoro}, F.},
        title = "{EXPRES: a next generation RV spectrograph in the search for earth-like worlds}",
    booktitle = {Ground-based and Airborne Instrumentation for Astronomy VI},
         year = 2016,
       editor = {{Evans}, Christopher J. and {Simard}, Luc and {Takami}, Hideki},
       series = {Society of Photo-Optical Instrumentation Engineers (SPIE) Conference Series},
       volume = {9908},
        month = aug,
          eid = {99086T},
        pages = {99086T},
          doi = {10.1117/12.2233002},
archivePrefix = {arXiv},
       eprint = {1606.04413},
 primaryClass = {astro-ph.IM},
       adsurl = {https://ui.adsabs.harvard.edu/abs/2016SPIE.9908E..6TJ}
}

@INPROCEEDINGS{Schwab2016,
       author = {{Schwab}, C. and {Rakich}, A. and {Gong}, Q. and {Mahadevan}, S. and {Halverson}, S.~P. and {Roy}, A. and {Terrien}, R.~C. and {Robertson}, P.~M. and {Hearty}, F.~R. and {Levi}, E.~I. and {Monson}, A.~J. and {Wright}, J.~T. and {McElwain}, M.~W. and {Bender}, C.~F. and {Blake}, C.~H. and {St{\"u}rmer}, J. and {Gurevich}, Y.~V. and {Chakraborty}, A. and {Ramsey}, L.~W.},
        title = "{Design of NEID, an extreme precision Doppler spectrograph for WIYN}",
    booktitle = {Ground-based and Airborne Instrumentation for Astronomy VI},
         year = 2016,
       editor = {{Evans}, Christopher J. and {Simard}, Luc and {Takami}, Hideki},
       series = {Society of Photo-Optical Instrumentation Engineers (SPIE) Conference Series},
       volume = {9908},
        month = aug,
          eid = {99087H},
        pages = {99087H},
          doi = {10.1117/12.2234411},
       adsurl = {https://ui.adsabs.harvard.edu/abs/2016SPIE.9908E..7HS}
}

@ARTICLE{Kontogiannis2025,
       author = {{Kontogiannis}, I. and {Zhu}, Y. and {Barczynski}, K. and {Stiefel}, M.~Z. and {Collier}, H. and {McKevitt}, J. and {Castellanos Dur{\'a}n}, J.~S. and {Berdyugina}, S. and {Harra}, L.~K.},
        title = "{Near-continuous tracking of solar active region NOAA 13664 over three solar rotations}",
      journal = {\aap},
         year = 2025,
        month = dec,
       volume = {704},
          eid = {A105},
        pages = {A105},
          doi = {10.1051/0004-6361/202556136},
archivePrefix = {arXiv},
       eprint = {2510.05979},
 primaryClass = {astro-ph.SR},
       adsurl = {https://ui.adsabs.harvard.edu/abs/2025A&A...704A.105K}
}

@ARTICLE{Finley2025,
       author = {{Finley}, A.~J. and {Brun}, A.~S. and {Strugarek}, A. and {Perri}, B.},
        title = "{A prolific solar flare factory: Nearly continuous monitoring of an active region nest with Solar Orbiter}",
      journal = {\aap},
         year = 2025,
        month = may,
       volume = {697},
          eid = {A217},
        pages = {A217},
          doi = {10.1051/0004-6361/202554323},
archivePrefix = {arXiv},
       eprint = {2504.06345},
 primaryClass = {astro-ph.SR},
       adsurl = {https://ui.adsabs.harvard.edu/abs/2025A&A...697A.217F}
}

@ARTICLE{Costes2026,
       author = {{Costes}, Jean C. and {Watson}, Christopher A. and {de Mooij}, Ernst and {Hobbs}, Katlyn L. and {Yaptangco}, Dana Clarice S. and {Unruh}, Yvonne C. and {Bedell}, Megan and {Meunier}, Nad{\'e}ge and {Mitchell}, Thomas D.},
        title = "{Spectral Ratio Analysis: probing of a new suite of stellar activity indicators as a tool for astrophysical noise mitigation}",
      journal = {\mnras},
         year = 2026,
        month = jun,
          doi = {10.1093/mnras/stag1173},
       adsurl = {https://ui.adsabs.harvard.edu/abs/2026MNRAS.tmp.1105C}
}

% Alternatively you could enter them by hand, like this:
% This method is tedious and prone to error if you have lots of references
%\begin{thebibliography}{99}
%\bibitem[\protect\citeauthoryear{Author}{2012}]{Author2012}
%Author A.~N., 2013, Journal of Improbable Astronomy, 1, 1
%\bibitem[\protect\citeauthoryear{Others}{2013}]{Others2013}
%Others S., 2012, Journal of Interesting Stuff, 17, 198
%\end{thebibliography}

%%%%%%%%%%%%%%%%%%%%%%%%%%%%%%%%%%%%%%%%%%%%%%%%%%

%%%%%%%%%%%%%%%%% APPENDICES %%%%%%%%%%%%%%%%%%%%%

\appendix

\section{4377 \AA\ feature at all limb angles}
In Figure~\ref{fig:muangle}, we show the \muram\ models for each limb angle, focused on the 4377\AA\ feature.
\begin{figure}
    \includegraphics[width=0.5\textwidth]{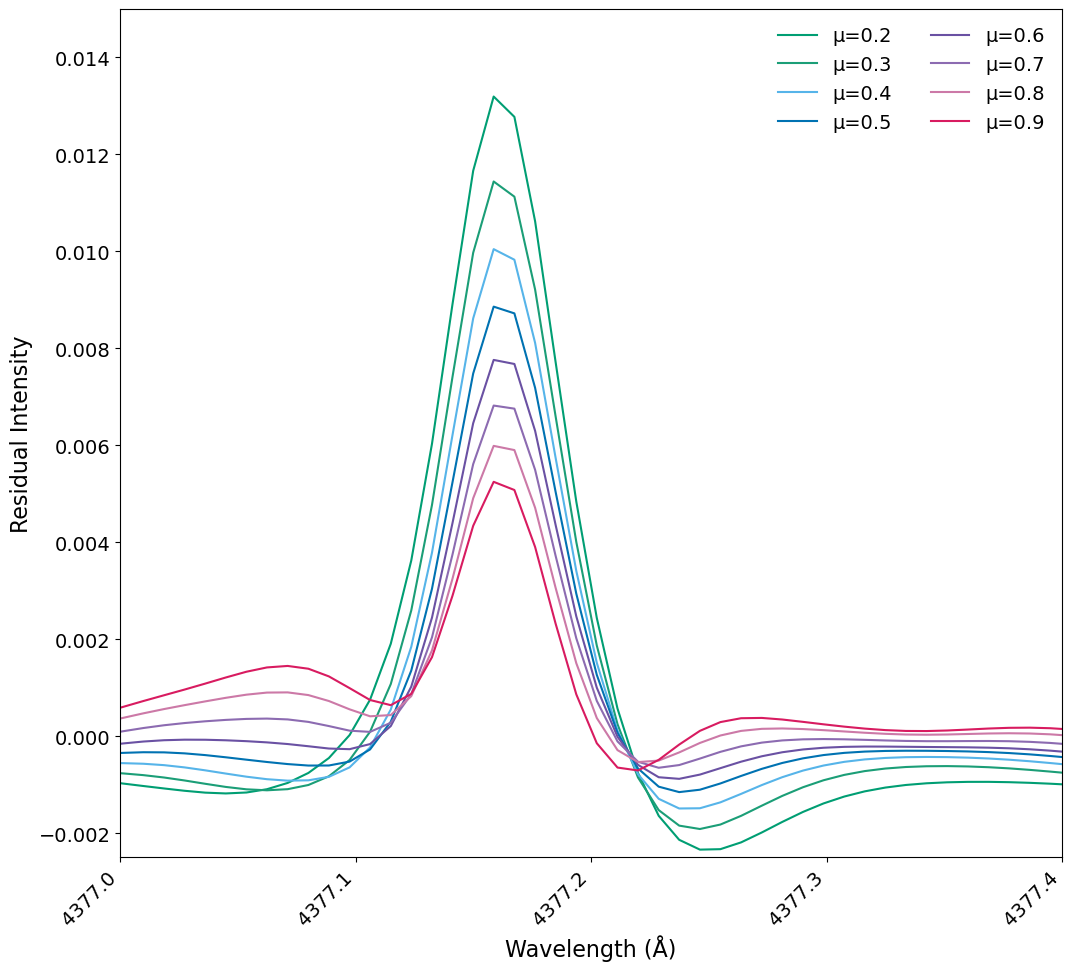}
    \caption{\muram\ models focused on the 4377 \AA\ feature at all limb angles. Angles closer to disc centre give a trough before the feature that is not present moving towards the limb. }
    \label{fig:muangle}
\end{figure}

\section{Faculae coverage versus temperature}

In Figure~\ref{fig:chisq}, we show the reduced \chisq maps for the PHOENIX model fits for two example days.
\begin{figure}
	% To include a figure from a file named example.*
	% Allowable file formats are eps or ps if compiling using latex
	% or pdf, png, jpg if compiling using pdflatex
    % \centering
	\includegraphics[width=0.5\textwidth]{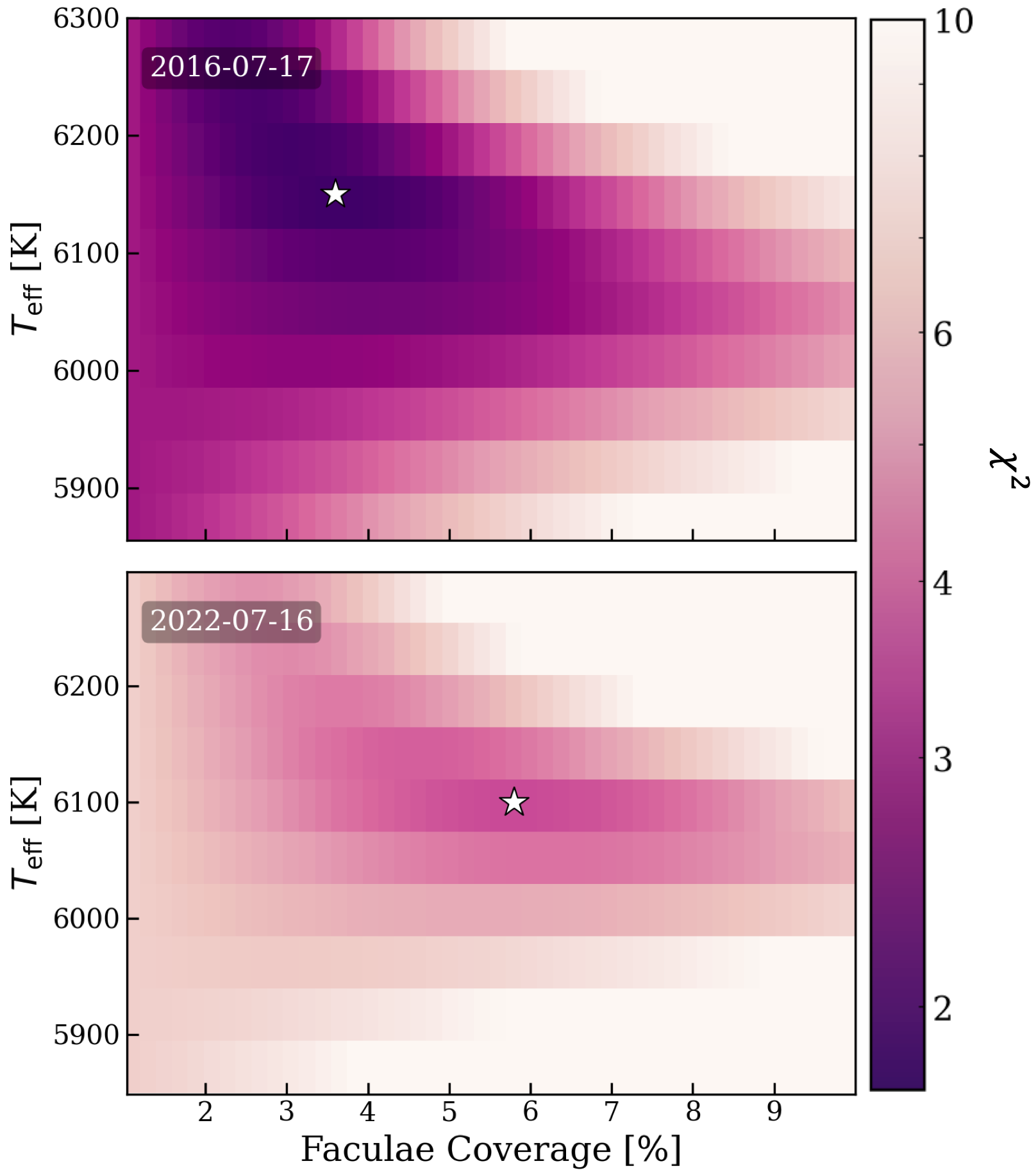}
    \caption{Reduced \chisq\ map of faculae coverage versus temperature derived from 
    \phoenix\ models for example days 2016-07-17 (top) and 2022-07-16 (bottom). 
    The colour bar shows the $\chi^2$ value of each pairing of faculae coverage 
    and temperature, with darker blue representing a lower value and therefore 
    better fit. The best-fitting models for the two days are indicated by the 
    white stars.}
    \label{fig:chisq}
\end{figure}

%%%%%%%%%%%%%%%%%%%%%%%%%%%%%%%%%%%%%%%%%%%%%%%%%%

% Don't change these lines
\bsp	% typesetting comment
\label{lastpage}
\end{document}